\documentclass[fleqn,usenatbib]{mnras}

\usepackage{newtxtext,newtxmath}

\usepackage[T1]{fontenc}
\usepackage{ae,aecompl}

\usepackage{graphicx}	
\usepackage{amsmath}	
\usepackage{amssymb}	
\usepackage{multirow}
\usepackage{latexsym}

\def\prd{Phys. Rev. D}

\def\apj{Astrophys. J.}
\def\apjs{Astrophys. J. Suppl.}
\def\apjl{Astrophys. J. Lett.}

\def\mnras{Mon. Not. R. Astron. Soc.}

\def\Gammaflat{\hat \Gamma}
\def\metricflat{\hat \gamma}
\def\Dflat{\hat {\mathcal D}}
\def\part_n{\partial_\perp}

\def\eul{\mathrm{e}}
\def\ga{\,\,\raise0.14em\hbox{$>$}\kern-0.76em\lower0.28em\hbox
{$\sim$}\,\,}
\def\la{\,\,\raise0.14em\hbox{$<$}\kern-0.76em\lower0.28em\hbox
{$\sim$}\,\,}

\title[General relativistic flux-limited diffusion]{NADA-FLD: A general relativistic, multi-dimensional neutrino-hydrodynamics code employing flux-limited diffusion}

\author[Rahman et al.]{
N.~Rahman,$^{1,2}$\thanks{E-mail: nrahman@mpa-garching.mpg.de}
O.~Just$^{3}$
and H.-Th.~Janka,$^{1}$
\\
$^{1}$Max-Planck-Institut f\"ur Astrophysik, Karl-Schwarzschild-Str. 1, 85748 Garching, Germany \\
$^{2}$Physik Department, Technische Universit\"at M\"unchen, James-Franck-Stra{\ss}e 1, 85748 Garching, Germany \\
$^{3}$Astrophysical Big Bang Laboratory, RIKEN Cluster for Pioneering Research, 2-1 Hirosawa, Wako, Saitama 351-0198, Japan
}

\date{Accepted XXX. Received YYY; in original form ZZZ}

\pubyear{2019}

\begin{document}
\label{firstpage}
\pagerange{\pageref{firstpage}--\pageref{lastpage}}
\maketitle

\begin{abstract}
  We present the new code NADA-FLD to solve multi-dimensional neutrino-hydrodynamics in full general relativity (GR) in spherical polar coordinates. The energy-dependent neutrino transport assumes the flux-limited diffusion (FLD) approximation and evolves the neutrino energy densities measured in the frame comoving with the fluid. Operator splitting is used to avoid multi-dimensional coupling of grid cells in implicit integration steps involving matrix inversions. Terms describing lateral diffusion and advection are integrated explicitly using the Allen-Cheng or the Runge-Kutta-Legendre method, which remain stable even in the optically thin regime. We discuss several toy-model problems in one and two dimensions to test the basic functionality and individual components of the transport scheme. We also perform fully dynamic core-collapse supernova (CCSN) simulations in spherical symmetry. For a Newtonian model we find good agreement with the M1 code ALCAR, and for a GR model we reproduce the main effects of GR in CCSNe already found by previous works.
\end{abstract}




\section{Introduction}

The transport of neutrinos plays a vital role for core-collapse supernovae (CCSNe) and mergers of neutron stars (NSs). According to the standard explosion mechanism of ordinary CCSNe, the stalled accretion shock is revived due to neutrino heating of material below the shock \citep{1966ApJ...143..626C, 1985ApJ...295...14B, 2016ARNPS..66..341J}. In the case of more massive stars undergoing black-hole (BH) formation, neutrino emission has leverage on the time when the central proto-neutron star (PNS) collapses, and subsequently it may regulate the mass-accretion rate onto the BH (e.g. \citealt{1999ApJ...524..262M,2011ApJ...737....6S,2015Natur.528..376M,2017MNRAS.469L..43O}). For any type of CCSN, the nucleosynthesis pattern in material ejected from the central regions depends sensitively on the neutron-to-proton ratio, which is determined by the number of neutrinos and antineutrinos emitted and absorbed during the expansion (e.g. \citealt{1996ApJ...471..331Q,2018ApJ...852...40W,2016MNRAS.459.4174G,2017ApJ...836L..21N}). In NS mergers, neutrino transport is similarly important for setting the nucleosynthesis conditions in ejected matter and for cooling and heating in the central remnant, which typically consists of an accretion disk surrounding either a NS or a BH (e.g. \citealt{2014MNRAS.441.3444M,2015MNRAS.448..541J,2014MNRAS.443.3134P,2017PhRvL.119w1102S}). Furthermore, in NS- or BH-torus systems a gamma-ray burst jet might be produced or enhanced due to annihilation of neutrinos with their anti-particles (e.g. \citealt{1986ApJ...308L..43P,1989Natur.340..126E,2016ApJ...816L..30J,2017ApJ...850L..37P}). In order to better understand all of these scenarios, models including a multi-dimensional, general relativistic treatment of neutrino transport are desirable.

During the last several decades, various approaches have been developed for treating neutrino transport in hydrodynamical simulations. The most sophisticated schemes solve the full Boltzmann equation, e.g. by direct discretization using finite differences (see, e.g. \citealt{2004ApJS..150..263L, 2004ApJ...609..277L, 2012ApJS..199...17S}), or by employing a Monte Carlo treatment (see, e.g. \citealt{1989A&AS...78..375J, 2012ApJ...755..111A, 2015ApJ...813...38R}), or by coupling a somehow simplified Boltzmann solver to an additional system of equations for the lowest angular moments of the Boltzmann equation (see, e.g. \citealt{2002A&A...396..361R, 2018MNRAS.475.4186F}). While these methods have the advantage of providing the full phase-space information of the generally six-dimensional phase-space dependence of the neutrino distribution function, they are still too expensive in terms of computational resources to be used for long-term, high-resolution simulations, or for exhaustive parameter exploration.

A computationally much cheaper alternative of describing neutrino processes comes with so-called neutrino leakage schemes (e.g. \citealt{1996A&A...311..532R,2013PhRvD..88f4009G,2016ApJS..223...22P,2018arXiv180800006A}) that estimate the local matter-neutrino interaction rates just based on the instantaneous fluid configuration without evolving (or evolving only in regions where neutrinos are trapped) conservation equations for neutrino energy and number. Moreover, several additional schemes have been designed for specific application purposes, including light-bulb schemes (e.g. \citealt{1996A&A...306..167J}), the fast-multigroup transport scheme (FMT; \citealt{2015MNRAS.448.2141M}), the M0 scheme \citep{2016MNRAS.460.3255R}, or the isotropic-diffusion-source approximation (IDSA; \citealt{2009ApJ...698.1174L}).

In the class of local-closure moment schemes only the lowest angular moments of the distribution function are dynamically evolved, while all higher-order moments are provided by an approximate closure relation as function of the evolved moments. In flux-limited diffusion (FLD) schemes only the zeroth-order moment (i.e. the energy density) is evolved, while in two-moment (M1) schemes additionally the first-order moment (flux density) is integrated. In the recent years a number of M1 schemes have been developed \citep{2015ApJS..219...24O,2015MNRAS.453.3386J,2015PhRvD..91l4021F,2016ApJS..222...20K,2019ApJS..241....7S,Sekiguchi2012a}.

The concept of FLD was first introduced by \citet{1981JQSRT..26..385P} and \citet{1981ApJ...248..321L} and since then has been used, apart from many applications in the context of photon transport, for neutrino transport in CCSNe and PNS cooling \citep{1982ApJS...50..115B,1985ApJS...58..771B,1986ApJ...307..178B,1987ApJ...318..744M,1989ApJ...339..978B,1990PhDT.......270C,2007ApJ...655..416B,2015ApJ...807L..31L,2018arXiv180905608B}. \citet{1992ApJ...398..531C} derived the FLD equation correct to order $v/c$. The maximum entropy principle was applied by \citet{1989JQSRT..42..603C} in FLD to determine Eddington factors. \citet{1989ApJ...339..978B} generalized the FLD approach to the general relativistic context. An improved flux-limiter was suggested by \citet{1992A&A...256..452J} based on the comparison with a Monte-Carlo study. A two dimensional, multi-group (i.e. energy-dependent) FLD scheme for neutrinos was developed by \citet{2009ApJS..181....1S}.

General relativistic (GR) radiative transfer as a scientific discipline was established with the formulation of the Boltzmann equation in GR by \citet{1966AnPhy..37..487L} and of the corresponding two moment formalism in \citet{1981MNRAS.194..439T}. Until rather recently, most numerical applications were restricted to spherical symmetry \citep[e.g.][]{1989ApJ...339..978B, 2001ApJ...560..326B, 2004ApJS..150..263L}. A few years ago, multi-group GR transport has found its way into multiple dimensions. For instance, \citet{2010ApJS..189..104M} have solved the two-moment equations with a variable Eddington-factor method using the conformal-flatness condition and the ray-by-ray-plus approximation \citep{2006A&A...447.1049B}. Moreover, \citet{Sekiguchi2016a, 2015ApJS..219...24O, 2016ApJS..222...20K}, and \citet{2016ApJ...831...98R} have employed the M1-method to solve neutrino transport in full general relativity.

In this paper, we present the first fully multi-dimensional, multi-group FLD scheme in full general relativity. Although FLD does not necessarily produce more accurate results than an M1 code, we opted for the FLD approach mainly for two reasons. First, FLD evolves only a single equation per neutrino species and energy bin, whereas M1 evolves three additional flux-vector components. Particularly in GR the FLD approach reduces the complexity of the equations and eases the computational work load. Second, the M1 method is currently employed already in several existing codes and has its own short comings, for example with respect to the accuracy in beam-crossing regions (see, e.g. \citealt{2018PhRvD..98f3007F}). Developing different, complementary algorithms therefore enhances the diversity of applied methods and in the long run might help to discriminate numerical artefacts from physical effects.

Our algorithm employs spherical polar coordinates and integrates the GR equations using the partially implicit Runge-Kutta method \citep{2012PhRvD..85l4037M,2013PhRvD..87d4026B}. The transport equations are solved in the comoving (i.e. fluid-rest) frame. In order to avoid multi-directional coupling of grid cells, and therefore the inversion of bigger matrices spanned over the entire grid, we employ operator splitting. The source terms, the radial- and energy-derivatives, as well as the non-radial derivatives are integrated separately, each using an appropriate discretization scheme. In this way, the scheme can be parallelized in a straightforward manner and remains numerically less complex than an unsplit, fully implicit solver.

In Sect.~2, we outline the basic equations of our general relativistic radiation-hydrodynamics scheme. The discretization scheme for solving the transport equations is described in Sect.~3. In Sect.~4, we discuss the results of various toy-model problems and of fully dynamic neutrino-hydrodynamics simulations of the collapse and post-bounce evolution of a massive star in spherical symmetry. In Appendices~\ref{app:der_ene_eqn}, ~\ref{app:der_fld_flux_eqn} and ~\ref{app:evl_R_lambda_D}, we provide the detailed derivation of the main transport equation used in our code, derive and test GR corrections to the FLD flux and outline the numerical computation of the diffusive flux, respectively.

Throughout most of the paper we assume $\mathrm{c}=1$ for the speed of light, except in some microphysics related cases, in which $\mathrm{c}$ appears explicitly. Also gravitational constant, $\mathrm{G}$, Planck constant, $\mathrm{h}$, and Boltzmann constant, $k_\mathrm{b}$, are set to one. We follow the convention that indices or superscripts $a, b, c, \mu,\nu$ run over space-time components $\{0,1,2,3\}$, while $i,j,k,l$ just run over spatial components $\{1,2,3\}$. We denote quantities defined in the comoving orthonormal frame by using an index with a hat (e.g. $\hat i$) and quantities defined in the comoving curvilinear frame index with a bar (e.g. $\bar i$). We denote electron neutrinos and their anti-neutrinos as $\nu_e$, and $\bar\nu_e$, respectively, and we use $\nu_x$ to denote any of the four remaining neutrino types.

\section{Basic equations of general relativistic radiation hydrodynamics}\label{sec:basic_equations}

In this section, we outline the basic equations used in our general relativistic radiation-hydrodynamics scheme, namely those describing the evolution of the space-time metric, hydrodynamics, and radiation transport.

\subsection{Metric equations}\label{sec:metric} 

We use a $3+1$ decomposition in which the space-time manifold is foliated into space-like hyper-surfaces $\Sigma$ (see, e.g., \citealt{Baumgarte:2010:NRS:2019374}). We denote the 4-metric as $g_{ab}$. The time-like future pointing normal vector to $\Sigma$ is $n^a$, and the space-like 3-metric on $\Sigma$ is $\gamma_{ij}$. The line element is then given by:
\begin{eqnarray}
	\mathrm{d}s^2 & \equiv & g_{ab} \mathrm{d}x^a \mathrm{d}x^b \nonumber \\ 
	& = & - \alpha^2 \mathrm{d}t^2 + \gamma_{ij} (\mathrm{d}x^i + \beta^i \mathrm{d}t)(\mathrm{d}x^j + \beta^j \mathrm{d}t)~,
	\label{eq:gr_linele}
\end{eqnarray}
where $\alpha, \beta^i$ are the lapse function and shift-vector, respectively, and
\begin{eqnarray}\label{eq:gr_def} 
	\gamma_{ab} & = & g_{ab} + n_a n_b ~, \nonumber \\
        n^a & = & (1/\alpha, - \beta^i/\alpha)~, \nonumber \\
	n_a & = & (-\alpha,0,0,0) ~ .
\end{eqnarray}
Moreover, 
\begin{eqnarray} 
	\bar \gamma_{ij} & \equiv & \eul^{-4 \phi} \gamma_{ij}~
\end{eqnarray}
is the conformal metric, with the conformal factor $\exp(4 \phi)$ (see, e.g. chapter 3 of \citealt{Baumgarte:2010:NRS:2019374} for a detailed discussion of the conformal transformation).
Furthermore, the extrinsic curvature\footnote{We use parenthesis () to denote the symmetric part $A_{(ij)} \equiv \frac{1}{2}(A_{ij}+A_{ji})$ of any tensor $A_{ij}$.}, $K_{ij}$, the conformal traceless extrinsic curvature, $\bar A_{ij}$, and the trace of the extrinsic curvature, $K$, are defined as:
\begin{eqnarray} 
	K_{ij} & \equiv & - \gamma_i{}^k \gamma_j{}^l \nabla_k n_l \nonumber \\
	& = & - \frac{1}{2 \alpha} \partial_t \gamma_{ij} + D_{(i} \beta_{j)}~, \nonumber \\
    \bar A_{ij} & \equiv & \eul^{-4 \phi} (K_{ij} - \frac{1}{3} K)~, \nonumber \\
    K & \equiv & {K^i}_i~.
	\label{eq:gr_K_def}
\end{eqnarray}
The Minkowski metric in spherical polar coordinates is $\metricflat_{ij} = \mathrm{diag}(1,r^2,r^2 \sin^2 \theta)$. We denote the connection coefficients associated with the metrics $\gamma_{ab}$, $\bar \gamma_{ij}$, and $\metricflat_{ij}$ as $\Gamma^a_{bc}$, $\bar \Gamma^i_{jk}$, and $\Gammaflat^i_{jk}$, respectively. The covariant derivatives associated with $\gamma_{ij}$, $\bar \gamma_{ij}$, and $\metricflat_{ij}$ are denoted by $D$, $\bar D$, and $\Dflat$, respectively. We define the connection vector $\Lambda^i$ as
\begin{eqnarray}\label{eq:Lambda}
	\Lambda^i \equiv \gamma^{jk} \Delta \Gamma^i_{jk}~,
\end{eqnarray}
with
\begin{eqnarray}
	\Delta \Gamma^i_{jk} & \equiv & \bar \Gamma^i_{jk} - \Gammaflat^i_{jk}~,
	\label{delta_Gamma}
\end{eqnarray}
and express the Ricci tensor as
\begin{eqnarray}
	\bar R_{ij} & = & - \frac{1}{2} \bar \gamma^{kl} \Dflat_k \Dflat_l \bar \gamma_{ij} + 
    \bar \gamma_{k(i} \Dflat_{j)} \bar \Lambda^k + \Delta \Gamma^k \Delta \Gamma_{(ij)k} \nonumber \\
	& & + \bar \gamma^{kl} \left( 2 \Delta \Gamma^m_{k(i} \Delta \Gamma_{j)ml} 
	+ \Delta \Gamma^m_{ik} \Delta \Gamma_{mjl} \right)~.
	\label{eq:gr_ricci}
\end{eqnarray}
To evolve the space-time metric we solve the covariant BSSN equations \citep{2013PhRvD..87d4026B}, which are given by:
\begin{eqnarray}
	\part_n \bar \gamma_{ij} & = & - \frac{2}{3} \bar \gamma_{ij} \bar D_k \beta^k - 2 \alpha \bar A_{ij}~, \nonumber \\
	\part_n \bar A_{ij} & = & - \frac{2}{3} \bar A_{ij} \bar D_k \beta^k - 2 \alpha \bar A_{ik} \bar A^k {}_j + \alpha \bar A_{ij} K \nonumber \\
	&& + \eul^{- 4 \phi} \Big[ - 2 \alpha \bar D_i \bar D_j \phi + 4 \alpha \bar D_i \phi \bar D_j \phi \nonumber \\
	& & ~~~~~~~ + 4 \bar D_{(i} \alpha \bar D_{j)} \phi - \bar D_i \bar D_j \alpha \nonumber\\ 
	& & ~~~~~~~ + \alpha (\bar R_{ij} - 8\pi S_{ij}) \Big]^{\rm TF}, \nonumber \\
	\part_n \phi & = & \frac{1}{6} \bar D_k \beta^i - \frac{1}{6} \alpha K~, \nonumber \\
	\part_n K & = & \frac{\alpha}{3} K^2 + \alpha \bar A_{ij} \bar A^{ij}  \nonumber \\ 
	& &- \eul^{- 4 \phi}  ( \bar D^2 \alpha + 2 \bar D^i \alpha \bar D_i \phi ) \nonumber \\
	& & + 4 \pi \alpha (\varrho + S)~, \nonumber \\
	\part_n \bar \Lambda^i & = & \bar \gamma^{jk} \Dflat_j \Dflat_k \beta^i 
	+ \frac{2}{3} \Delta \Gamma^i \bar D_j \beta^j + \frac{1}{3} \bar D^i \bar D_j \beta^j \nonumber \\
	&  & - 2 \bar A^{jk} ( \delta^i{}_j \partial_k \alpha - 6 \alpha \delta^i{}_j \partial_k \phi
	- \alpha \Delta \Gamma^i_{jk} ) \nonumber \\
	& & - \frac{4}{3} \alpha \bar \gamma^{ij} \partial_j K - 16 \pi \alpha \bar \gamma^{ij} S_j~.
	\label{eq:gr_bssn}
\end{eqnarray}
Here,
\begin{eqnarray}
	\part_n \equiv \partial_t - {\mathcal L}_\beta~, \nonumber
	\label{eq:gr_lie}
\end{eqnarray}
where ${\mathcal L}_\beta$ is the Lie derivative along the shift vector $\beta^i$. The superscript TF  denotes the trace-free part of a tensor. The matter-radiation source terms are given by:
\begin{eqnarray}
	\varrho & \equiv & n_a n_b T^{ab}~, \nonumber \\
	S_i & \equiv & \gamma_{ia} n_b T^{ab}~, \nonumber \\
	S_{ij} & \equiv & \gamma_{ia} \gamma_{jb} T^{ab}~, \nonumber \\
	S & \equiv & \gamma^{ij} S_{ij}~,
	\label{eq:gr_source}
\end{eqnarray}
where $T^{ab}$ is the total stress-energy tensor of matter and radiation (see below for more explanation). We use the `1+log' slicing and the gamma driver condition to evolve $\alpha$ and $\beta^i$, respectively, i.e.:
\begin{eqnarray}
	\partial_t \alpha & = & - 2 \alpha K~, \nonumber \\
	\partial_t \beta^i & = & B^i~, \nonumber \\
	\partial_t B^i & = & \frac{3}{4} \partial_t \bar \Lambda^i~.
	\label{eq:gr_gauge}
\end{eqnarray}
The time integration of equations (\ref{eq:gr_bssn}) and (\ref{eq:gr_gauge}) is done using a 2nd-order partially implicit Runge-Kutta method, which is described in detail in \citet{2012PhRvD..85l4037M,2013PhRvD..87d4026B}. The integration time step is given by the Courant-Friedrichs-Lewy condition:
\begin{eqnarray}
	\Delta t = \mathrm{CFL} \min\{\Delta r/2, (\Delta r/2) \Delta \theta, (\Delta r/2) \sin (\Delta \theta/2) \Delta \phi\}~.
	\label{eq:gr_time_step}
\end{eqnarray}
Here, $\Delta r,\Delta \theta,\Delta \phi$ are the minimum widths in the radial, polar and azimuthal directions, respectively, and CFL$\in$(0,1) refers to the chosen Courant number. The minimum is taken over all cells of the computational grid.

\subsection{Hydrodynamics}\label{sec:hydro}

The general relativistic hydrodynamics equations expressing the local conservation of baryonic mass (with current density $J^a$), energy-momentum (with energy-momentum tensor $T^{ab}_\mathrm{h}$), and electron lepton number (with current density $J^a_e$) read (e.g. \citealt{2008LRR....11....7F}):
\begin{eqnarray}
	\nabla_{a} J^a & = & 0~, \nonumber \\    
        \nabla_{a} T^{ab}_\mathrm{h} & = & s^b~, \nonumber \\
    \nabla_{a} J^a_{\mathrm{e}} & = & S_N~,
	\label{eq:hydro_cons}
\end{eqnarray}
where
\begin{eqnarray}\label{eq:hydro_cons_current}
	J^a & \equiv & \rho u^a~, \nonumber \\
	T^{ab}_\mathrm{h} & \equiv & \rho h u^a u^b + p g^{ab}~, \nonumber \\
    J^a_e & \equiv & \rho u^a Y_\mathrm{e}~,
\end{eqnarray}
and
\begin{eqnarray}
    h & = & 1 + e + p / \rho~, \nonumber \\
	u^0 & = & W / \alpha~, \nonumber \\
    u^i & = & W (v^i - \beta^i \alpha^{-1})~.
	\label{eq:stress_energy}
\end{eqnarray}
The symbols $\rho, e, v^i, W, p, h$, and $Y_\mathrm{e}$ denote the baryonic mass density, specific internal energy, 3-velocity, Lorentz factor, gas pressure, specific enthalpy, and electron fraction (equal to the number of protons per nucleon), respectively. In order to obtain explicit expressions of equations (\ref{eq:hydro_cons}), we use the flux-conservative Valencia formulation generalized to curvilinear coordinates, as described in \citet{2014PhRvD..89h4043M}. In this formulation, singular terms proportional to $1/r$ and $\cot \theta$ are scaled out by using the reference metric $\metricflat_{ij}$. The conservative variables $D$, $S_i$, $\tau$, and $D_\mathrm{e}$ that are evolved in time are defined in terms of the primitive variables $\rho$, $e$, $v^i$, $p$, and $Y_\mathrm{e}$ as:
\begin{eqnarray}
    D & \equiv & W \rho~, \nonumber \\
    S_i & \equiv & W^2 \rho h v_i~, \nonumber \\
    \tau & \equiv & W^2 \rho h - p - D~, \nonumber \\
    D_e & \equiv & D Y_\mathrm{e}~.
	\label{eq:cons_var}
\end{eqnarray}
The continuity, Euler, energy, and lepton-number equations in the generalized Valencia formulation read:
\begin{align}
    &\partial_t (\sqrt{\gamma} D) 
    + \; \partial_j (f_D)^j &&= \; 0~, \nonumber\\ 
    &\partial_t (\sqrt{\gamma/\hat \gamma} S_i) 
    + \; \partial_j (f_S)_i{}^j 
    &&= \; (s_S)_i \;+\; (f_S)_k{}^j \Gammaflat^k_{ji} - (f_S)_i{}^k\Gammaflat^j_{kj} \nonumber \\
    & && + \alpha \sqrt{\gamma/\hat \gamma}(S_M)_i~, \nonumber\\ 
    &\partial_t ( \sqrt{ \gamma }\, \tau) 
    + \; \partial_j (f_\tau)^j 
    &&= \; s_\tau \;-\; (f_\tau)^k \Gammaflat^j_{jk} \;+\; \alpha \sqrt{\gamma} S_E~, \nonumber\\
    &\partial_t (\sqrt{\gamma} D_\mathrm{e}) 
    + \; \partial_j (f_{D\mathrm{e}})^j &&= \; \alpha \sqrt{\gamma} S_N~,
    \label{eq:hydro_eqn}
\end{align}
where $\gamma$ is the determinant of the metric $\gamma_{ij}$ and the flux functions are given by:
\begin{eqnarray}
	(f_D)^j & \equiv & \alpha \sqrt{ \gamma } D (v^j - \beta^j \alpha^{-1})~, \nonumber \\
	(f_S)_i{}^j & \equiv & \alpha \eul^{6\phi} \sqrt{\bar \gamma/\hat \gamma} \
	(W^2 \rho h v_i (v^j - \beta^j \alpha^{-1}) + p \delta_i{}^j)~, \nonumber \\
	(f_\tau)^j & \equiv & \alpha \sqrt{ \gamma } \, \left( \tau (v^j - \beta^j \alpha^{-1}) + pv^j \right)~, \nonumber \\
	(f_{D\mathrm{e}})^j & \equiv & \alpha \sqrt{ \gamma } D_\mathrm{e} (v^j - \beta^j \alpha^{-1})~,
	\label{eq:hydro_flux}
\end{eqnarray}
and the source functions are defined by:
\begin{eqnarray}
	(s_S)_i & \equiv &  \alpha \eul^{6 \phi} \sqrt{\bar \gamma/\hat \gamma} 
	\Big( - T^{00} \alpha \partial_i \alpha + T^0{}_k \Dflat_i \beta^k  \nonumber \\
	& + & \frac{1}{2} \big( T^{00} \beta^j \beta^k   + 2 T^{0j} \beta^k + T^{jk} \big) \Dflat_i \gamma_{jk} 
	\Big)~, \nonumber \\
	s_\tau & \equiv & \alpha \eul^{6\phi}\sqrt{\bar \gamma/\hat \gamma}
	\Big(T^{00} (\beta^i \beta^j K_{ij} - \beta^i \partial_i \alpha)  \nonumber \\
 	& + & T^{0i}(2 \beta^j K_{ij} - \partial_i \alpha) + T^{ij} K_{ij} \Big)~.
\label{eq:hydro_source}
\end{eqnarray}
The source terms $S_E$, $(S_M)_i$ and $S_N$ express the change of gas energy, momentum, and lepton number, respectively, due to neutrino-matter interactions and will be quantified in Section \ref{sec:neu_source}. To close the system of equations, an equation of state is required that provides the pressure, temperature and composition as functions of the primitive variables.
The hydrodynamics equations are solved using a finite difference Godunov-type High-Resolution-Shock-Capturing-Method (HRSC) \citep{Toro2009}.
For the reconstruction of primitive variables at the cell interfaces, the PPM \citep{1984JCoPh..54..174C}, CENO \citep{1998JCoPh.142..304L} and MP5 \citep{1997JCoPh.136...83S} methods are implemented. The fluxes at cell interfaces are calculated from primitive variables using the HLL Riemann solver \citep{Harten1983}. The time integration is done with a second-order Runge-Kutta method, where the time step is the same as that used for integrating the BSSN equations (cf. equation (\ref{eq:gr_time_step})).  The numerical implementation and test of the hydrodynamics part of the code is discussed in \citet{2014PhRvD..89h4043M}. The hydrodynamics equations are integrated using the same timestep (given by equation \eqref{eq:gr_time_step}) as used for integrating the GR equations.

\subsection{Neutrino transport}\label{sec:transport}

In this section, we present the evolution equations used in our FLD neutrino transport scheme and their coupling to the evolution of the metric, eqs.~(\ref{eq:gr_bssn}), and of the hydrodynamic quantities, eqs.~(\ref{eq:hydro_eqn}). The formalism of fully general relativistic truncated-moment schemes has been developed and extensively discussed in \citet{2011PThPh.125.1255S, 2012arXiv1212.4064E, 2013PhRvD..87j3004C}, from whom we adopt a great share of our notation. Like in the aforementioned works, all (comoving-frame and lab-frame) angular moments as well as the neutrino stress-energy tensor are expressed as functions of Eulerian (i.e. lab-frame) space-time coordinates, $x^\mu$, and of the neutrino energy measured by a comoving observer, $\epsilon$. One difference of our scheme compared to those of the aforementioned papers is, however, that we evolve the neutrino moments (i.e. the energy densities) as measured in the orthonormal comoving frame, instead of those measured in the lab frame. In this respect, our scheme is similar to that of \citet{2010ApJS..189..104M}.

\subsubsection{Basic definitions}\label{sec:basic-equations}

In terms of the neutrino distribution function, $f$, the comoving-frame 0th-, 1st-, and 2nd-order moments are given by\footnote{We note that our definition of the comoving-frame moments is consistent with \citet{2011PThPh.125.1255S}, but contains an additional factor $\epsilon^2$ compared to those of \citet{2012arXiv1212.4064E, 2013PhRvD..87j3004C}.}:
\begin{eqnarray}
	\mathcal{J}(x^{\mu},\epsilon) & \equiv & \epsilon^3 \int f(x^{\mu},p^{\hat \mu})~\mathrm{d}\Omega~, \nonumber \\
	\mathcal{H}^{\hat i}(x^{\mu},\epsilon) & \equiv & \epsilon^3 \int l^{\hat i} f(x^{\mu},p^{\hat \mu})~\mathrm{d}\Omega~, \nonumber \\
	\mathcal{K}^{\hat i \hat j}(x^{\mu},\epsilon) & \equiv & \epsilon^3 \int l^{\hat i} l^{\hat j} f(x^{\mu},p^{\hat \mu})~\mathrm{d}\Omega~,
	\label{eq:tr_moment}
\end{eqnarray}
where $p^{\hat \mu}\equiv \epsilon (1,l^{\hat i})$ denotes the neutrino momentum-space coordinates, with unit momentum three-vector $l^{\hat i}$, and the angular integration is performed in the comoving-frame momentum space.

The comoving-frame moments in eqs.~(\ref{eq:tr_moment}) are related to the monochromatic lab-frame neutrino stress-energy tensor, $\mathcal{T}^{ab}_\mathrm{r}$, by
\begin{eqnarray}
  \mathcal{T}^{a b}_\mathrm{r}(x^{\mu},\epsilon) &=& {L^a}_{\hat 0} {L^b}_{\hat 0} \mathcal{J} 
    + ({L^a}_{\hat 0} {L^b}_{\hat i} + {L^a}_{\hat i} {L^b}_{\hat 0}) \mathcal{H}^{\hat i} \nonumber \\
  & & + {L^a}_{\hat i} {L^b}_{\hat j} \mathcal{K}^{\hat i \hat j}~,
	\label{eq:tr_stress_energy_decomp}
\end{eqnarray}
from which the corresponding energy-integrated tensor is obtained as
\begin{eqnarray}
    T^{a b}_\mathrm{r}(x^{\mu})  =  \int \mathcal{T}^{a b} \mathrm{d}\epsilon~.
	\label{eq:tr_stress_energy}
\end{eqnarray}
In eq.~(\ref{eq:tr_stress_energy_decomp}), the matrices ${L^a}_{\hat b} \equiv {e^a}_{\bar c}{\Lambda^{\bar c}}_{\hat b}$ are responsible for transforming tensors from the orthonormal comoving frame to the global coordinate (i.e. lab) frame. Here, the Lorentz transformation, ${\Lambda^{\bar c}}_{\hat b}$, converts orthonormal comoving-frame quantities into an orthonormal (i.e. locally Minkowskian) tetrad basis in the lab frame, and the tetrad transformation ${e^a}_{\bar c}$ converts from the orthonormal lab-frame tetrad basis to the basis of global coordinates (which are generally not orthonormal in curved space-time). The tetrad transformation, ${e^a}_{\bar c}$, is given by equation~(57) of \citet{2012arXiv1212.4064E} and we use the Gram-Schmidt orthonormalising process on the spatial metric $\gamma_{ij}$ to evaluate the spatial part of the tetrad transformation, ${e^k}_{\bar i}$.

The lab-frame moments of 0th-, 1st-, and 2nd-order are respectively given in terms of the comoving-frame moments by (cp. equations A18-A20 in \citealt{2012arXiv1212.4064E}):
\begin{eqnarray}
	\mathcal{E}(x^{\mu},\epsilon) &=& W^2 \big( \mathcal{J} + 2 \bar v_{\hat i} \mathcal{H}^{\hat i} 
    + \bar v_{\hat i} \bar v_{\hat j} \mathcal{K}^{\hat i \hat j} \big)~, \nonumber \\
 	\mathcal{F}^i(x^{\mu},\epsilon) &=& W \bigg[ e^i_{\hat i} \mathcal{H}^{\hat i} + W v^i \mathcal{J} 
    + e^i_{\hat i} \bar v_{\hat j} \mathcal{K}^{\hat i \hat j} \nonumber \\
    &+& \frac{W}{W+1} v^i \big\{ (2W+1) \bar v_{\hat i} \mathcal{H}^{\hat i} 
    + W \bar v_{\hat i} \bar v_{\hat j} \mathcal{K}^{\hat i \hat j} \big\} \bigg]~, \nonumber \\
    \mathcal{S}^{ij}(x^{\mu},\epsilon) &=& e^i_{\hat i} e^j_{\hat j} \mathcal{K}^{\hat i \hat j} 
    + W \big( v^i e^j_{\hat j} \mathcal{H}^{\hat j} + v^j e^i_{\hat i} \mathcal{H}^{\hat i} \big) 
    + W^2 v^i v^j \mathcal{J} \nonumber \\
    &+& \frac{W^2}{W+1} \big( \big[ v^i \bar v_{\hat i} e^j_{\hat j} 
    + v^j \bar v_{\hat j} e^i_{\hat i} \big] \mathcal{K}^{\hat i \hat j}
    + 2 W v^i v^j \bar v_{\hat i} \mathcal{H}^{\hat i} \big) \nonumber \\
    &+& \frac{W^4}{(W+1)^2} v^i v^j \bar v_{\hat i} \bar v_{\hat j} \mathcal{K}^{\hat i \hat j}~,  
	\label{eq:tr_moment_lab}
\end{eqnarray}
where $e_{\hat i}^{i}= e_{\bar j}^{i}\delta^{\bar j}_{\hat i}$ and $v^i \equiv e^{i}_{\hat i} 
\bar v^{\hat i}$, with $\bar v^{\hat i}$ being the three-velocity in the orthonormal tetrad basis. Using these lab-frame moments instead of the comoving-frame moments, the monochromatic lab-frame stress-energy tensor (cp. eq.~(\ref{eq:tr_stress_energy_decomp})) reads:
\begin{eqnarray}\label{eq:stress_energy_lab}
  \mathcal{T}^{a b}_\mathrm{r}(x^{\mu},\epsilon) &=& \mathcal{E} n^a n^b
  + \mathcal{F}^a n^b + \mathcal{F}^b n^a + \mathcal{S}^{ab} ~ .
\end{eqnarray}

\subsubsection{Neutrino interaction source terms}\label{sec:neu_source}
The coupling between the transport equations and the equations of hydrodynamics and the Einstein equations is done as follows. We first compute the exchange rates of energy, momentum, and lepton number as measured in the orthonormal comoving frame, given respectivily by:
\begin{eqnarray}
	Q_\mathrm{E} & \equiv & - \sum_{\nu} \int \mathrm{d}\epsilon \kappa_\mathrm{a} (\mathcal{J}^{\mathrm{eq}} - \mathcal{J})~, \nonumber \\
    Q_\mathrm{M}^{\hat i} & \equiv & \sum_{\nu} \int \mathrm{d}\epsilon \kappa_\mathrm{t} \mathcal{H}^{\hat i}~, \nonumber \\
    Q_\mathrm{N} & \equiv & - \mathrm{m_u} \int \mathrm{d}\epsilon \bigg[ 
    \bigg(\frac{\kappa_\mathrm{a} (\mathcal{J}^{\mathrm{eq}} - \mathcal{J})}{\epsilon}\bigg)_{\nu_\mathrm{e}} 
    - \bigg(\frac{\kappa_\mathrm{a} (\mathcal{J}^{\mathrm{eq}} - \mathcal{J})}{\epsilon}\bigg)_{\bar \nu_\mathrm{e}} \bigg]~,\nonumber \\
	\label{eq:tr_en_int}
\end{eqnarray}
where
\begin{eqnarray}\label{eq:jeq}
    \mathcal{J}^\mathrm{eq} & \equiv & \frac{4 \pi \epsilon^3}{\exp((\epsilon-\mu_\nu)/T) + 1}~,
\end{eqnarray}
and $\mathrm{m_u}$, $T$, and $\mu_\nu$ are the atomic mass unit, fluid temperature, and neutrino chemical potential of the equilibrium distribution. We denote the absorption opacity, scattering opacity, transport opacity, and equilibrium energy distribution as $\kappa_\mathrm{a}, \kappa_\mathrm{s}, \kappa_\mathrm{t} \equiv \kappa_\mathrm{a} + 
\kappa_\mathrm{s}$, and $\mathcal{J}^{\mathrm{eq}}$, respectively. 
In eqs.~\eqref{eq:tr_en_int}, the summation, $\sum$, runs over all six neutrino species.
To obtain the lab-frame source terms $S_E$ and
$(S_M)_i$, we consider $q^{\hat b} \equiv (Q_E,Q_M^{\hat
i})$ as four-vector, apply a Lorentz- and tetrad
transformation, resulting in $s^a \equiv L^a_{\hat b} q^{\hat b}$, and perform the projections $S_E =
-s^a n_a$ and $(S_M)_i = s^j \gamma_{ij}$ (see, e.g.
\citealt{2010ApJS..189..104M,2013PhRvD..87j3004C}). The lepton-number exchange rates, $Q_{N}$, are scalar and therefore frame invariant. We end up with:
\begin{eqnarray}
	S_E &=& W (Q_E + \bar v_{\hat i} Q_M^{\hat i})~, \nonumber \\
    (S_M)_i &=& e_{i \hat i}Q_M^{\hat i} + 
    W (v_i - \beta_i/\alpha) Q_E \nonumber \\
    && + W \bigg(\frac{W}{W+1} v_i - \beta_i/\alpha \bigg)
    \bar v_{\hat i} Q_M^{\hat i}~, \nonumber \\
    S_N &=& Q_N~.
	\label{eq:tr_hydro_source}
\end{eqnarray}
The neutrino contributions to the source terms for the Einstein equations (cf. eqs.~(\ref{eq:gr_bssn})) are obtained from the lab-frame neutrino angular moments (cf. eq.~\eqref{eq:tr_moment_lab}) using eq.~(\ref{eq:stress_energy_lab}) as:
\begin{eqnarray}
	\varrho_\mathrm{r} & = & \sum_{\nu} \int \mathrm{d}\epsilon \mathcal{E}~, \nonumber \\
    S^i_\mathrm{r} & = & \sum_{\nu} \int \mathrm{d}\epsilon \mathcal{F}^i~, \nonumber \\
    S^{ij}_\mathrm{r} & = & \sum_{\nu} \int \mathrm{d}\epsilon \mathcal{S}^{ij}~.
	\label{eq:tr_gr_source}
\end{eqnarray}

\subsubsection{Energy equation and flux-limited diffusion approximation}\label{sec:energy-equation-flux}

The evolution equation for the comoving-frame neutrino energies, $\mathcal{J}$, can be derived from the evolution equations for the lab-frame moments, $\mathcal{E}$ and $\mathcal{F}^i$, which are discussed in \citet{2011PThPh.125.1255S, 2012arXiv1212.4064E, 2013PhRvD..87j3004C}. We refer to  Appendix~\ref{app:der_ene_eqn} for the detailed derivation. The resulting evolution equation reads:
\begin{eqnarray}
	&&\frac{1}{\alpha} \frac{\partial}{\partial t} [W(\mathcal{\hat J} + \bar v_{\hat i}\mathcal{\hat H}^{\hat i})]
    + \frac{1}{\alpha} \frac{\partial}{\partial x^j} [\alpha W (v^j-\beta^j \alpha^{-1}) \mathcal{\hat J}] + \nonumber \\
    &&\frac{1}{\alpha} \frac{\partial}{\partial x^j} \Big[\alpha e^j_{\hat i} \mathcal{\hat H}^{\hat i} +
     \alpha W \Big(\frac{W}{W+1}v^j-\beta^j \alpha^{-1} \Big)
    \bar v_{\hat i}\mathcal{\hat H}^{\hat i} \Big] \nonumber + \\
    &&\hat R_\epsilon - \frac{\partial}{\partial \epsilon} (\epsilon \hat R_\epsilon)
    = \kappa_a (\mathcal{\hat J}^{eq}-\mathcal{\hat J})~,
    \label{eq:tr_energy_eqn}
\end{eqnarray}
where we use the notation $(\mathcal{\hat J},\mathcal{\hat J}^\mathrm{eq},\mathcal{\hat H},\mathcal{\hat K}) \equiv \sqrt{\gamma} (\mathcal{J},\mathcal{J}^\mathrm{eq},\mathcal{H},\mathcal{K})$. The terms subsumed in $R_\epsilon-\partial_\epsilon (\epsilon R_\epsilon)$ are responsible for spectral shifts in the energy-density distribution due to Doppler- and gravitational effects. They are functions of the comoving-frame moments by virtue of eqs.~(\ref{eq:5.3}),~(\ref{eq:lab_com_trafo_red}), and~(\ref{eq:tr_moment_lab}). The specific shape of the neutrino source terms on the right-hand side of eq.~(\ref{eq:tr_energy_eqn}) takes account of the fact that the current implementation is restricted to absorption and emission (or formally equivalent) reactions, and iso-energetic scattering processes, i.e. scattering processes without energy exchange between neutrino and target particle.

The flux-limited diffusion (FLD) approximation is implemented as follows. The flux density as measured in the orthonormal comoving frame, $\mathcal{H}^{\hat i}$, is in the diffusion limit approximately given by $\mathcal{H}^{\hat i, \mathrm{diff}} = - e^{k \hat i} \partial_k (\alpha^3 \mathcal{J})/(3\kappa_\mathrm{t}\alpha^3)$. This expression can be obtained from the evolution equation of the neutrino flux densities (shown in eq.~(\ref{eq:2.4})) by neglecting time derivatives and velocity-dependent terms. See Appendix~\ref{app:der_fld_flux_eqn} for details of the derivation of the FLD flux and comments at the end of this section. Going towards lower optical depths, radiation approaches the causality limit, i.e. $\mathcal{H}^{\hat i, \mathrm{free}} \approx \mathcal{J}$. In FLD, a smooth interpolation between these two regimes is accomplished by the use of a scalar flux-limiter, $\lambda\in[0,1/3]$, in terms of which the flux is expressed as:
\begin{eqnarray}
  \mathcal{H}^{\hat i}\longrightarrow - D \frac{e^{k \hat i}}{\alpha^3} \partial_k (\alpha^3 \mathcal{J})~,
	\label{eq:tr_fld_flux}
\end{eqnarray}
where
\begin{eqnarray}
	D \equiv \frac{\lambda}{\kappa_\mathrm{t}}
\end{eqnarray}
is the generalized (scalar) diffusion coefficient that includes the flux-limiter. In doing so, it is implicitly assumed that the partial time derivative of the flux vanishes, i.e. $\partial_t \mathcal{H}^{\hat i} = 0$. In this work, we use the Levermore-Pomraning (LP) limiter \citep{1981JQSRT..26..385P,1981ApJ...248..321L} and the Wilson limiter \citep{1982ApJS...50..115B}, which are computed as:
\begin{eqnarray}\label{eq:limiter}
    \lambda_\mathrm{LP} & \equiv &  \frac{2+R}{6+3R+R^2}~, \nonumber \\
    \lambda_\mathrm{Wilson} & \equiv &  \frac{1}{3+R}~,
\end{eqnarray}
where
\begin{eqnarray}\label{eq:knudsen}
	R & \equiv & \frac{|e^{k \hat i}\partial_{k} (\alpha^3\mathcal{J})|}{\kappa_\mathrm{t}\alpha^3\mathcal{J}}
\end{eqnarray}
is the Knudsen number. To calculate the Knudsen number, diffusion coefficient and the Eddington scalar and tensor, we either follow the procedure described in Appendix \ref{app:evl_R_lambda_D} or employ the method outlined in Appendix H.4 of \citet{2009ApJS..181....1S}. The former ensures that causality is not violated for both the individual flux components, $\mathcal{H}^r,~\mathcal{H}^\theta,~\mathcal{H}^\phi$, and for the total flux, $\mathcal{H}=\sqrt{(\mathcal{H}^r)^2+(\mathcal{H}^\theta)^2+(\mathcal{H}^\phi)^2}$. On the contrary, in the latter method, the Knudsen number along a direction $\hat{i}$ is calculated separately in each direction using the absolute value of the diffusive flux, $\mathcal{H}^{\hat i, \mathrm{diff}}$, in $\hat{i}$ direction. This procedure only ensures that causality is not violated for individual flux components, i.e. $|\mathcal{H}^{\hat i}| \le \mathcal{J}$, but the total flux, $\mathcal{H}=\sqrt{(\mathcal{H}^r)^2+(\mathcal{H}^\theta)^2+(\mathcal{H}^\phi)^2}$, might violate causality. We will come back to this point when discussing our results for 2D test problems and compare both methods of evaluating the diffusion coefficient.

The Eddington tensor, $\chi^{ij}$, which is related to the second moment tensor, $\mathcal{K}^{ij}$, by
\begin{eqnarray}\label{eq:edd_tensor1}
	\chi^{\hat i \hat j} & = &  \frac{\mathcal{K}^{\hat i \hat j}}{\mathcal{J}}~,
\end{eqnarray}
is in the FLD approximation given by (see, e.g. \citealt{1981JQSRT..26..385P, 1981ApJ...248..321L, 2009ApJS..181....1S}):
\begin{eqnarray}\label{eq:edd_tensor_fld}
    \chi^{\hat i \hat j} = \frac{1}{2} [(1-\chi)\delta^{\hat i \hat j} 
    + (3\chi-1)h^{\hat i}h^{\hat j}]~,
\end{eqnarray}
where $h^{\hat i}$ is the unit vector along $\mathcal{H}^{\hat{i}}$, and the (scalar) Eddington factor, $\chi$, is given by
\begin{eqnarray}\label{eq:edd_factor_fld}
    \chi = \lambda + (\lambda R)^2~.
\end{eqnarray}
For future reference, we also define the flux factor as
\begin{eqnarray}\label{eq:flux_fac}
    f^{\hat i} \equiv \frac{\mathcal{H}^{\hat i}}{\mathcal{J}}~.	
\end{eqnarray}
The final FLD equation solved in our code reads:
\begin{eqnarray}
	&&\frac{1}{\alpha} \frac{\partial}{\partial t} (W \mathcal{\hat J})
    + \frac{1}{\alpha} \frac{\partial}{\partial x^j} [\alpha W (v^j-\beta^j/\alpha) \mathcal{\hat J}] \nonumber \\
    &&- \frac{1}{\alpha} \frac{\partial}{\partial x^j} \Big[\alpha^{-2} \sqrt{\gamma} \Big\{ \gamma^{i k}
    + W \Big(\frac{W}{W+1}v^j-\beta^j/\alpha \Big) v^i \Big\} D \partial_k (\alpha^3 \mathcal{J}) \Big] \nonumber \\
    && - \frac{e^{k \hat i}}{\alpha^4} \frac{\partial}{\partial t}
    (W \sqrt{\gamma} \bar v_{\hat i}) D \partial_k(\alpha^3 \mathcal{J})
    +R_\epsilon - \frac{\partial}{\partial \epsilon} (\epsilon R_\epsilon) \nonumber \\
    &&= \kappa_\mathrm{a} (\mathcal{\hat J}^{eq}-\mathcal{\hat J})~.
    \label{eq:tr_fld_energy_eqn}
\end{eqnarray}
The second, third and fourth terms in the above equation describe advection, diffusion, and aberration due to fluid acceleration, respectively. We simplify the equation by neglecting all spatial cross derivatives, which appear due to off-diagonal metric components $\gamma^{r\theta}, \gamma^{r\phi}$ and $\gamma^{\theta \phi}$. These off-diagonal components, become non-vanishing when two components of the velocity vector have non-zero values, for example, because of convection inside the PNS or the contraction of a rotating PNS. Since they are typically strongly subdominant $\sim$$\mathcal{O}(v^4/\mathrm{c}^4)$ (order of magnitude of the components of 3-metric, $\gamma_{ij}$, does not depend on the choice of gauge conditions) compared to the diagonal components $\sim$$\mathcal{O}(1)$ (see, e.g. \citealt{1969ApJ...158...55C,1996PThPh..96...81A}), the corresponding error should remain small. We leave the development of numerical techniques to treat the spatial cross derivatives resulting from off-diagonal metric components to future works.

We end this section by commenting on some shortcomings of present FLD scheme. As standard in FLD schemes, the temporal derivatives of the radiation flux are ignored in the 0th- and 1st-moment equations, and all the velocity-dependent terms are neglected in the 1st-moment equation. In most cases, the impact of velocity-dependent terms in the 1st-moment equation is not as significant as that of the corresponding terms in the 0th-moment equation\footnote{See, e.g., Mihalas 1984 for the relevance of velocity-dependent terms in the 0th- and 1st-moment equations.}. Hence, one motivation for neglecting these terms is that their impact on the solution is assumed to be small compared to the error already introduced by the approximate flux-limiter (eq. 31). This assumption breaks down in a flow with high velocities or in highly time-dependent cases. Another motivation why these terms are neglected here (and, in fact, in all time-dependent, multidimensional FLD schemes to our knowledge) simply is that some of the additional terms are not straightforward to include numerically. \citet{1992A&A...256..428D} and \citet{1993A&A...273..338D} already identified the problem of the missing terms in the 1st-moment equation, coining it the "missing opacity problem", and they suggested an algorithm to include the additional terms in a 1D FLD scheme. However, we decided not to adopt any such receipes in the present code development for reasons of numerical stability and the unavailability of ``artificial opacity'' corrections in multidimensions.

In the 0th-moment equation of our FLD scheme, we include all velocity-dependent and GR related terms, except for the cross derivative terms coming from the off-diagonal components of the metric, which, however, have subdominant effects of order $\sim\mathcal{O}(v^4/\mathrm{c}^4)$ as is discussed before. Moreover, in the 1st-moment equation we include the most dominant GR corrections in deriving the FLD flux, which are related to effects such as gravitational redshifting, time dilation, and ray bending, but we neglected several subdominant GR correction terms because they have vanishing impact on the value of the FLD flux (see the discussion at the end of Appendix~\ref{app:der_fld_flux_eqn} and Fig.~\ref{fig:test_ccsn_flux_correction}).

A general shortcoming of FLD transport solvers is their parabolic nature; they are not covariant and they allow for information to propagate at an infinite speed. In contrast, the hyperbolic BSSN formalism of Einstein's equations restricts information propagation to a finite speed. Moreover, a final problem is connected to the fact that various radiation moments enter as source terms in the BSSN equations and such a coupling between the FLD radiation moments and the general relativistic metric could lead to causality violation and numerical instabilities. However, in applications considered so far (see, e.g., Sect. 4.3) we have not experienced instability problems, and the main source of error may rather be the FLD closure and not the effects connected to covariance violation.

\section{Numerical treatment of the transport}\label{sec:numer-treatm-transp}

\begin{figure}
	\includegraphics[width=0.5\textwidth]{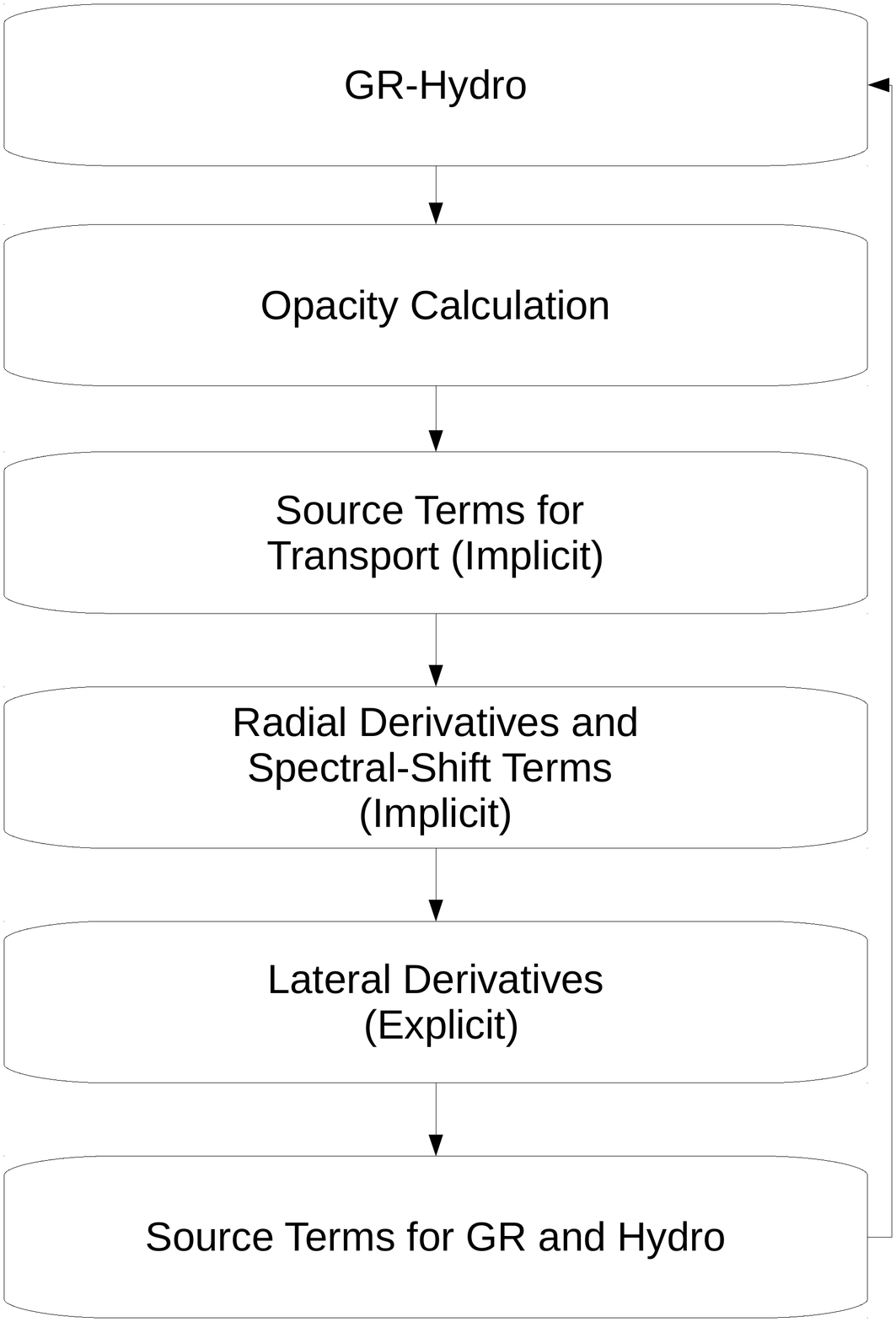}
	\vspace*{-7mm}
    \caption{Flow chart illustrating the steps performed in the evolution scheme.}
	\label{fig:tr_flowchart}
\end{figure}

In this section, we describe the numerical method used to solve the neutrino transport equations together with the Einstein and hydrodynamics equations. The neutrino energy space is discretized into energy groups, and for each of these and for each neutrino species we solve the evolution equation for $\mathcal{J}$, eq.~\eqref{eq:tr_fld_energy_eqn}, which generally depends on three spatial dimensions. We use finite-difference methods for the spatial discretization on the same spatial grid as for the GR and hydrodynamics steps.

The flow chart of our evolution algorithm is depicted in Fig.~\ref{fig:tr_flowchart}. After advancing the GR and hydrodynamics equations by one integration step, we calculate the opacity using updated hydrodynamics quantities as well as transport quantities from the previous time step. Next, we evolve the neutrino energy densities. During the transport steps, all hydrodynamics and GR quantities are kept fixed. Since the FLD equations are generally parabolic and the propagation speed of information is in principle infinity, many existing FLD codes employ a fully implicit time integration. However, with the computational cost roughly increasing with the number of grid points to the third power, unsplit, fully implicit integration schemes become particularly expensive in multi-dimensional applications, and they tend to scale poorly on large numbers of computational cores\footnote{Iterative implicit methods on domains decomposed using the Message Passing Interface (MPI) require several collective MPI communications per time step, whereas explicit methods only require a single point-to-point communication per time step and domain boundary.}. In the present scheme we avoid this inconvenience by using operator splitting and treating parts of the equation explicitly. In the following subsections, we first estimate the relevant timescales to motivate the time-integration steps, and then we present the detailed discretization procedure employed at each step.

For the calculation of the generalized diffusion coefficient (including flux-limiter) and the corresponding Eddington tensor and scalar according to eqs.~(\ref{eq:edd_tensor1})--(\ref{eq:edd_factor_fld}), we implemented two methods: we either follow \citet{2009ApJS..181....1S} (see their Appendix H.4) or use the method outlined in Appendix~\ref{app:evl_R_lambda_D}. In what follows, $D_1,D_2,D_3$ will denote the diffusion coefficients in the radial, polar, and azimuthal coordinate direction, respectively.

\subsection{Relevant timescales and motivation of the integration scheme}\label{sec:relevant-timescales}

Using simple dimensional estimates, we first identify the characteristic timescales on which the different terms in the FLD equation induce a change of $\mathcal{J}$. We denote the grid spacing for simplicity by $\Delta x$, keeping in mind that this quantity generally depends on the grid location. For clarity, in this section we explicitly include the speed of light, $\mathrm{c}$.

Ignoring the energy derivatives, the FLD equation, eq.~(\ref{eq:tr_fld_energy_eqn}), is an advection-diffusion-reaction equation \citep[e.g.][]{daleanderson2011}. The velocity-dependent terms of eq.~(\ref{eq:tr_fld_energy_eqn}) are in this sense advection terms, the characteristic timescale of which is bounded from below by the light-crossing time of a grid cell,
\begin{eqnarray}\label{eq:tlight}
  t_{\mathrm{light}}&=& \frac{\Delta x}{\mathrm{c}} \, .
\end{eqnarray}
The reaction (i.e. neutrino source) terms are associated with timescales \begin{eqnarray}\label{eq:tsource}
  t_{\mathrm{source}}&=& \frac{1}{\mathrm{c}\kappa_{\mathrm{a}}}
\end{eqnarray}
that are typically much shorter than $t_{\mathrm{light}}$ inside a hot PNS and practically infinity far away from any neutrino sources. Finally, the characteristic timescale of the FLD-related terms can be estimated by
\begin{eqnarray}\label{eq:tdiff}
  t_{\mathrm{diff}}&= &\frac{\Delta x^2}{D} ~.
\end{eqnarray}
The time step for an explicit treatment of the advection terms, $\Delta t$, must always be less than or equal to the light-crossing timescale of a grid cell, i.e. $\Delta t\la t_{\mathrm{light}}$. Now, a useful quantity to assess the performance of any method used to integrate the diffusion terms is
\begin{eqnarray}
  r_{\mathrm{diff}}&\equiv& \frac{\Delta t}{t_{\mathrm{diff}}} \, .
\end{eqnarray}
For conventional explicit integration schemes the condition for numerical stability is $r_{\mathrm{diff}}\la 0.5-1$. In order to get some idea about typical values of $r_{\mathrm{diff}}$ encountered in post-bounce configurations, we can use (assuming $\Delta t\sim t_{\mathrm{light}}$ and recalling that $\lambda$ and $\kappa_{\mathrm{t}}\Delta x$ denote the flux-limiter and the optical depth per grid cell, respectively)
\begin{eqnarray}
  r_{\mathrm{diff}}&\sim& \frac{\lambda}{\kappa_{\mathrm{t}}\Delta x} \, \,
\end{eqnarray}
and consider the (simplified) case of constant grid width of $\Delta x\sim \mathcal{O}(100\,\mathrm{m})$: Inside the hot PNS we have $\lambda \approx 1/3$ and $\kappa_{\mathrm{t}}\Delta x \gg 1$, and therefore we expect $r_{\mathrm{diff}}\ll 1$. Far away from any neutrino source the Knudsen number roughly scales as $R\sim (\kappa_{\mathrm{t}}\Delta x)^{-1}$, giving $\lambda \sim R^{-1}\sim \kappa_{\mathrm{t}}\Delta x$ and hence $r_{\mathrm{diff}}\sim \mathcal{O}(1)$. In other words, both the PNS center and the region far away from the PNS do not necessarily demand an implicit integration. Nevertheless, $r_{\mathrm{diff}}$ may still attain high values in the intermediate,  semi-transparent region. However, estimates based on our 1D simulations indicate (cf. Fig.~\ref{fig:test_ccsn_r_diff} and the corresponding discussion in Sect.~\ref{sec:CCSN}) that $r_{\mathrm{diff}}$ may reach values greater than unity only close to shock. At such large distances lateral neutrino fluxes are strongly subdominant compared to radial fluxes.

Backed by these considerations, we decompose eqs.~(\ref{eq:tr_fld_energy_eqn}) into three parts and integrate each part in its own operator-split step: In the first step, we integrate the neutrino source terms implicitly (because $t_\mathrm{source}$ may be $\ll t_\mathrm{light}$) using a Newton-Raphson scheme. Then we solve for the contributions from the radial derivatives and the spectral-shift terms (i.e. $R_\epsilon - \partial_\epsilon (\epsilon R_\epsilon$)) using an implicit Crank-Nicolson scheme. Finally, we obtain the contribution from the non-radial derivatives using one of the following explicit methods, namely the Allen-Cheng method \citep{Allen1970} or the multi-stage Runge-Kutta-Legendre super-time stepping method (RKL2; \citealt{2012MNRAS.422.2102M}). Although these methods are explicit, they have the appealing property that they remain stable for large values of $r_{\mathrm{diff}}$; see Sect.~\ref{sec:test-problems} for exemplary tests. By using an explicit
compared to an implicit scheme for the non-radial terms, not only the
single-core efficiency is improved but, even more importantly, the
scheme can be parallelized very efficiently using MPI decomposition in
the polar and azimuthal directions. The trade-off for using the explicit Allen-Cheng scheme is some loss of accuracy at high values of $r_\mathrm{diff} \ga 1$, while for the explicit RKL2 method, to obtain higher maximum allowed values of $r_\mathrm{diff}$, the needed computational cost increases. For this reason, we apply an explicit method only to the non-radial fluxes. Since the non-radial fluxes tend to be subdominant compared to the radial fluxes in near-shock regions where $r_{\mathrm{diff}}$ peaks, the error introduced by this integration method should remain manageable.

\subsection{Neutrino source terms}\label{sec:neutr-source-terms}

In the first step, we compute the contribution from the neutrino source terms in an implicit manner. We solve the following equations:
\begin{eqnarray}
	\frac{W}{\alpha} \partial_t \mathcal{J}_{\nu,\xi} &=&
    \bigg[ \kappa_\mathrm{a} (\mathcal{J}^{\mathrm{eq}} - \mathcal{J}) \bigg]_{\nu,\xi}, \nonumber \\
    \frac{W}{\alpha} \rho \partial_t e (T,Y_\mathrm{e}) &=&
    - \sum_{\nu,\xi} \bigg[ \kappa_\mathrm{a} (\mathcal{J}^{\mathrm{eq}} - \mathcal{J} ) \Delta \epsilon_\xi \bigg]_{\nu,\xi}, \nonumber \\
    \frac{W}{\alpha} \rho \partial_t Y_\mathrm{e} &=&
    - m_u \sum_{\xi} \bigg[ \big[\kappa_\mathrm{a} (\mathcal{J}^{\mathrm{eq}} - \mathcal{J}) \Delta \epsilon_\xi \big]_{\nu_\mathrm{e}} \nonumber \\
    &&- \big[\kappa_\mathrm{a} (\mathcal{J}^{\mathrm{eq}} - \mathcal{J}) \Delta \epsilon_\xi \big]_{\bar \nu_\mathrm{e}} \bigg]_{\xi}.
	\label{eq:ns_source_term}
\end{eqnarray}
The subscripts $\nu$ and $\xi$ indicate the neutrino species and energy bin, respectively, and $\Delta \epsilon_\xi$ is the width of the energy bin centered at $\epsilon_\xi$.

We discretize eq.~\eqref{eq:ns_source_term} in time employing a backward Euler scheme and solve the resulting system of equations for the neutrino energy densities, $\mathcal{J}_{\nu,\xi}$, temperature, $T$ and electron fraction, $Y_\mathrm{e}$, using the Newton-Raphson method. We keep $\alpha$, $W$, $\rho$, and $\kappa_\mathrm{a}$ constant during this step at values obtained after the GR-hydro step. The Jacobian of eq.~\eqref{eq:ns_source_term} is determined numerically, and a direct matrix solver from the LAPACK library \citep{laug} is used for inverting the Jacobian. The values of neutrino energy densities, $\mathcal{J}_{\nu,\xi}$, obtained in this step are used as initial values in the next step.

\subsection{Radial derivatives and spectral-shift terms}\label{sec:radi-deriv-spectr}

In the next operator-split step, the following equation is solved:
\begin{eqnarray}
	&&W\partial_t \mathcal{\hat J} + \mathcal{R}_r = 0~,
  \label{eq:2nd_step_equation}
\end{eqnarray}
where
\begin{eqnarray}
	&& \mathcal{R}_r \equiv \partial_t (W) \mathcal{\hat J} +
    \partial_r [\alpha W(v^r-\beta^r \alpha^{-1})\mathcal{\hat J}] \nonumber \\
    &&- \partial_r \Big[ \alpha^{-2} \sqrt{\gamma} \Big\{ \gamma^{rr} 
    + W \Big(\frac{W}{W+1}v^r-\beta^r \alpha^{-1}\Big) v^r \Big\} 
    D_1 \partial_r(\alpha^3 \mathcal{J}) \Big] \nonumber \\
    &&- \alpha^{-3} e^{r \hat i} \partial_t
    (W \sqrt{\gamma} \bar v_{\hat i})D_1 \partial_r (\alpha^3 \mathcal{J})
    + \alpha \Big[\hat R_\epsilon - \frac{\partial}{\partial \epsilon} (\epsilon \hat R_\epsilon) \Big]~.
    \label{eq:tr_rad_red_term}
\end{eqnarray}
contains the radial advection and diffusion terms, the radial acceleration term, and the spectral-shift terms. The diffusion coefficient in radial direction is denoted by $D_1$ and $\hat{\mathcal{J}} \equiv \sqrt{\gamma} \mathcal{J}$. Equation~(\ref{eq:2nd_step_equation}) is integrated by using the implicit Crank-Nicolson method. The old time is denoted as $t^n$ and the new time as $t^{n+1}$. The time indices for all GR and hydrodynamics quantities are omitted as they are kept fixed in all transport steps. Using superscripts $n$ and $n+1$ to label quantities defined before and after this partial integration step, respectively, the discretized equation reads:
\begin{eqnarray}
	&& (W \sqrt{\gamma}) \frac{\mathcal{J}^{n+1}_i-\mathcal{J}^{n}_i}{\Delta t}
	= -\frac{1}{2} (\mathcal{R}^{n+1}_{r,i}+\mathcal{R}^{n}_{r,i})~.    
    \label{eq:tr_rad_red_term_discrete}
\end{eqnarray}
Here, $\Delta t \equiv t^{n+1} - t^n$ and $i$ denotes quantities measured at the cell center in the radial direction. In the following, we provide the constituents of $\mathcal{R}_{r,i}^{n+1}$, while the corresponding expressions for $\mathcal{R}_{r,i}^{n}$ are obtained by replacing $n+1$ with $n$. For simplicity, we assume a uniform radial grid with constant cell size $\Delta r$; the generalization to non-uniform grids is straightforward.

The diffusion term is spatially discretized as:
\begin{eqnarray}
	&&\Big[ \partial_r \{A^r D_1 \partial_r(\alpha^3 \mathcal{J})\} \Big]^{n+1}_{i} = \nonumber \\
    &&\frac{1}{\Delta r} \Big[A^{r}_{i+1/2}D^{n}_{1,i+1/2} 
    \frac{\alpha^3_{i+1} \mathcal{J}^{n+1}_{i+1}-\alpha^3_{i} \mathcal{J}^{n+1}_{i}}{\Delta r} \nonumber \\
    &&- A^{r}_{i-1/2}D^{n}_{1,i-1/2}
    \frac{\alpha^3_{i} \mathcal{J}^{n+1}_i-\alpha^3_{i-1} \mathcal{J}^{n+1}_{i-1}}{\Delta r} \Big]~,
    \label{eq:tr_rad_red_diff_discrete0}
\end{eqnarray}
where
\begin{eqnarray}
    A^r &\equiv& \alpha^{-2} \sqrt{\gamma} \Big\{ \gamma^{rr} 
    + W \Big(\frac{W}{W+1}v^r-\beta^r \alpha^{-1}\Big) v^r \Big\}~. 
    \label{eq:tr_rad_red_diff_discrete}
\end{eqnarray}
Indices $i+1/2$ and $i-1/2$ denote the right and left cell interface of the $i$-th cell, respectively. If not mentioned otherwise, all cell interface values of hydrodynamic quantities and metric terms (contained in $A^r$ and in other terms below) are calculated by linear interpolation of the cell centered values.

The fluid-acceleration term (fourth term in eq.~\eqref{eq:tr_rad_red_term}) is computed as:
\begin{eqnarray}
	&&\Big[B^r D_{1} \partial_r(\alpha^3 \mathcal{J})\Big]^{n+1}_{i} = \nonumber \\
    &&\frac{B^{r}_i}{2} \Big[ D^{n}_{1,i+1/2} 
    \frac{\alpha^3_{i+1} \mathcal{J}^{n+1}_{i+1}-\alpha^3_{i}\mathcal{J}^{n+1}_{i}}{\Delta r} \nonumber \\
    &&+ D^{n}_{1,i-1/2} \frac{\alpha^3_{i}\mathcal{J}^{n+1}_i-\alpha^3_{i-1}\mathcal{J}^{n+1}_{i-1}}{\Delta r} \Big]~, 
    \label{eq:tr_rad_red_aber_discrete}
\end{eqnarray}
where
\begin{eqnarray}
	B^r &\equiv&  \alpha^{-3} e^{r \hat i} \partial_t(W \sqrt{\gamma} \bar v_{\hat i})~.
    \label{eq:tr_rad_red_aber_discrete_B}
\end{eqnarray}
The time derivative in eq.~(\ref{eq:tr_rad_red_aber_discrete_B}) is calculated using values of the hydrodynamic and metric quantities before and after the initial GR-hydro step.

The advection term is discretized using an upwind-type method (see, e.g. \citealt{A.Dorfi1998,2002A&A...396..361R}) as:
\begin{align}
	\Big[ \partial_r (C^r {\mathcal{J}}) \Big]^{n+1}_{i} &=&
    \frac{1}{\Delta r} \Big[C^{r}_{i+1/2} 
    \mathcal{J}^{n+1}_{\iota(i+1/2)}
    - C^{r}_{i-1/2}
    \mathcal{J}^{n+1}_{\iota(i-1/2)} \Big]~,
    \label{eq:tr_rad_red_adv_discrete}
\end{align} 
where 
\begin{eqnarray}
    C^r &\equiv& \alpha \sqrt{\gamma} W(v^r-\beta^r \alpha^{-1})
    \label{eq:tr_rad_red_adv_discrete_C}
\end{eqnarray}
and
\begin{eqnarray}
    \iota(i+1/2) &\equiv&  
	\begin{cases}
    i, & \text{if } v^r_{i+1/2} > 0  ~ ,\\
    i+1, & \text{otherwise}~.
	\end{cases}
    \label{eq:tr_rad_red_adv_discrete_jota}
\end{eqnarray}
The spectral-shift term, $R_\epsilon - \partial_\epsilon (\epsilon R_\epsilon)$, is discretized using the number-conservative scheme developed in \citet{2010ApJS..189..104M}. The terms with $\mathcal{H}^{\hat j}$ and $\mathcal{K}^{\hat i \hat j}$ that appear in $R_\epsilon$ are replaced by $f^{\hat j} \mathcal{J}$ and $\chi^{\hat i \hat j} \mathcal{J}$, respectively, and the flux factor, $f^{\hat j}$, and Eddington tensor, $\chi^{\hat i \hat j}$, are defined at instance $t^n$, while only $\mathcal{J}$ is defined at $t^{n+1}$.

The Crank-Nicolson method requires to solve a linear system of equations. Direct methods for solving linear systems are relatively expensive, we therefore use the iterative ``Generalized Minimal Residual Method with Restart'' (GMRES) along with the incomplete LU decomposition as a preconditioner from the NAG library\footnote{\textsf{www.nag.co.uk}} for this purpose. The values of neutrino energy densities, $\mathcal{J}$, obtained in this step are used as initial values in the next step.

\subsection{Non-radial derivatives}\label{sec:non-radi-deriv}

Finally, we include the contribution from the remaining lateral advection and diffusion terms by integrating the equation
\begin{align}
    &&W \sqrt{\gamma} \partial_t (\mathcal{J})
	= \mathcal{R}(\mathcal{J})~,
	\label{eq:3rd_step_equation}
\end{align}
where
\begin{eqnarray}
    &&\mathcal{R}(\mathcal{J})
	\equiv -\partial_\theta [\alpha W (v^\theta-\beta^\theta \alpha^{-1}) \mathcal{\hat J}] 
    - \partial_\phi [\alpha W (v^\phi-\beta^\phi \alpha^{-1}) \mathcal{\hat J}] \nonumber \\
    &&+ \partial_\theta \Big[ \alpha^{-2} \sqrt{\gamma} \Big\{ \gamma^{\theta \theta} 
    + W \Big(\frac{W}{W+1}v^\theta-\beta^\theta \alpha^{-1}\Big) v^\theta \Big\} 
    D_{2} \partial_\theta(\alpha^3 \mathcal{J}) \Big] \nonumber \\
    &&+ \partial_\phi \Big[ \alpha^{-2} \sqrt{\gamma} \Big\{ \gamma^{\phi \phi} 
    + W \Big(\frac{W}{W+1}v^\phi-\beta^\phi \alpha^{-1}\Big) v^\phi \Big\} 
    D_{3} \partial_\phi(\alpha^3 \mathcal{J}) \Big] \nonumber \\
    &&+ \alpha^{-3} e^{\theta \hat i} \partial_t(W \sqrt{\gamma} \bar v_{\hat i}) D_2 \partial_\theta(\alpha^3 \mathcal{J}) \nonumber \\
    &&+ \alpha^{-3} e^{\phi \hat i} \partial_t(W \sqrt{\gamma} \bar v_{\hat i}) D_3 \partial_\phi(\alpha^3 \mathcal{J}).
	\label{eq:ns_lat_term}
\end{eqnarray}
using the explicit Allen-Cheng method (\citealt{Allen1970}, \citealt{daleanderson2011}) or the explicit RKL2 method \citep{2012MNRAS.422.2102M}, where $D_{2}$ and $D_{3}$ are the diffusion coefficients in polar and azimuthal direction, respectively.

\subsubsection{Allen-Cheng method}\label{sec:non-radi-deriv-allen-cheng}
The discretized version of eq.~(\ref{eq:3rd_step_equation}) is presented below exemplarily for a single dimension (representative of the $\theta$- or $\phi$-direction) and a uniform grid, whose points are labeled by $k$ and spaced apart by $\Delta y$. The method consists of two steps, a predictor step and a corrector step. We again use $n$ and $n+1$ to label quantities before and after the two substeps. The value of $\mathcal{J}$ obtained after the predictor step, $\mathcal{J}^*$, is used in the corrector step to determine $\mathcal{J}^{n+1}$. The predictor step is given by
\begin{eqnarray}
    &&\frac{(W\sqrt{\gamma})}{\Delta t}(\mathcal{J}^{*}_k - \mathcal{J}^{n}_k)
	= - \frac{1}{2 \Delta y} (F_{k+1} \mathcal{J}^n_{k+1} - F_{k-1} \mathcal{J}^n_{k-1}) \nonumber \\
    &&+ \frac{1}{\Delta y^2} 
    [E_{k+1/2} (\alpha^3_{k+1} \mathcal{J}^n_{k+1}-\alpha^3_{k} \mathcal{J}^{*}_k) \nonumber \\
    &&- E_{k-1/2} (\alpha^3_{k} \mathcal{J}^{*}_k-\alpha^3_{k-1} \mathcal{J}^n_{k-1})]~ \nonumber \\
    && \frac{G_{k}}{2\Delta y} \big[ D_{k+1/2} (\alpha^3_{k+1} \mathcal{J}^{n}_{k+1}-\alpha^3_{k} \mathcal{J}^{*}_{k}) \nonumber \\
    &&+ D_{k-1/2} (\alpha^3_{k} \mathcal{J}^{*}_{k}-\alpha^3_{k-1} \mathcal{J}^{n}_{k-1}) \big]~
    \label{eq:ns_allen_cheng_pred}
\end{eqnarray}
and the corrector step by
\begin{eqnarray}
    &&\frac{(W\sqrt{\gamma})}{\Delta t}(\mathcal{J}^{n+1}_k - \mathcal{J}^{n}_k)
	= - \frac{1}{2 \Delta y} (F_{k+1} \mathcal{J}^{*}_{k+1} - F_{k-1} \mathcal{J}^{*}_{k-1}) \nonumber \\
    &&+ \frac{1}{\Delta y^2} 
    [E_{k+1/2} (\alpha^3_{k+1} \mathcal{J}^{*}_{k+1}-\alpha^3_{k} \mathcal{J}^{n+1}_k) \nonumber \\
    &&- E_{k-1/2} (\alpha^3_{k} \mathcal{J}^{n+1}_k-\alpha^3_{k-1} \mathcal{J}^{*}_{k-1})] \nonumber \\
    &&\frac{G_{k}}{2\Delta y} \big[ D_{k+1/2} (\alpha^3_{k+1} \mathcal{J}^{*}_{k+1}-\alpha^3_{k} \mathcal{J}^{n+1}_{k}) \nonumber \\
    &&+ D_{k-1/2} (\alpha^3_{k} \mathcal{J}^{n+1}_{k}-\alpha^3_{k-1} \mathcal{J}^{*}_{k-1}) \big]~,
    \label{eq:ns_allen_cheng_cor}
\end{eqnarray}
where we used
\begin{eqnarray}
    &&E \equiv \alpha^{-2} \sqrt{\gamma} \Big\{ \gamma^{jj} 
    + W \Big(\frac{W}{W+1}v^j-\beta^j \alpha^{-1}\Big) v^j \Big\} D~, \nonumber \\
    &&F \equiv \alpha \sqrt{\gamma} W (v^j-\beta^j \alpha^{-1})~, \nonumber \\
    &&G \equiv \alpha^{-3} e^{j \hat i} \partial_t(W \sqrt{\gamma} \bar v_{\hat i}),
	\label{eq:ns_allen_cheng_FE}
\end{eqnarray}
with $j$ denoting the considered direction, $\theta$ or $\phi$. The values $\mathcal{J}^{n+1}$ obtained in this step are the final values at the new time $t^{n+1}$. These values are used to calculate the neutrino source terms for the hydrodynamics equations (cf. eqs.~(\ref{eq:tr_hydro_source})) and for the metric equations (\ref{eq:tr_gr_source}), which are used in the next GR-Hydro step.

\subsubsection{Runge-Kutta-Legendre super-time stepping method (RKL2)}\label{sec:non-radi-deriv-runge-kutta-legendre}
We also implemented the second-order accurate explicit Runge-Kutta-Legendre super-time stepping method with four stages ($s=4$) \citep{2012MNRAS.422.2102M}. The RKL2 method is a conditionally stable method for solving the diffusion equation with maximum allowed values of $r_\mathrm{diff}$ of $s^2/8$ (see \citealt{2012MNRAS.422.2102M} for a detailed discussion of the stability properties). According to this stability criterion, the RKL2 method with four stages allows for a maximum value of $r_\mathrm{diff}$ of $\sim$2. However, if needed, a higher number of stages, $s$, can be implemented in the future. The four stage RKL2 scheme is (again assuming a constant grid width $\Delta y$):
\begin{eqnarray}
    \mathcal{J}_{0} &=& \mathcal{J}^{n}~, \nonumber \\
    \mathcal{J}_{1} &=& \mathcal{J}_{0}
	+ \frac{2}{27} \frac{\Delta t}{W\sqrt{\gamma}} \mathcal{R}(\mathcal{J}_{0})~, \nonumber \\
    \mathcal{J}_{2} &=& \frac{3}{2} \mathcal{J}_{1} - \frac{1}{2} \mathcal{J}_{0}
	+ \frac{\Delta t}{W\sqrt{\gamma}} \Bigg( \frac{1}{3} \mathcal{R}(\mathcal{J}_{1})
	- \frac{2}{9} \mathcal{R}(\mathcal{J}_{0}) \Bigg)~, \nonumber \\
    \mathcal{J}_{3} &=& \frac{25}{12} \mathcal{J}_{2} - \frac{5}{6} \mathcal{J}_{1} - \frac{1}{4} \mathcal{J}_{0}
	+ \frac{\Delta t}{W\sqrt{\gamma}} \Bigg( \frac{25}{54} \mathcal{R}(\mathcal{J}_{2})
	- \frac{25}{81} \mathcal{R}(\mathcal{J}_{0}) \Bigg)~, \nonumber \\
    \mathcal{J}_{4} &=& \frac{189}{100} \mathcal{J}_{3} - \frac{81}{80} \mathcal{J}_{2} + \frac{49}{400} \mathcal{J}_{0} \nonumber \\
	&& + \frac{\Delta t}{W\sqrt{\gamma}} \Bigg( \frac{21}{50} \mathcal{R}(\mathcal{J}_{3})
	- \frac{49}{200} \mathcal{R}(\mathcal{J}_{0}) \Bigg)~, \nonumber \\
	\mathcal{J}^{n+1} &=& \mathcal{J}_{4}~.
    \label{eq:ns_runge_kutta_legendre}
\end{eqnarray}
The quantity $\mathcal{R}(\mathcal{J})$ at stage ``$s$'' and grid point ``$k$'' is discretized as:
\begin{eqnarray}
    &&\mathcal{R}_k(\mathcal{J}_s)
	= - \frac{1}{2 \Delta y} (F_{k+1} \mathcal{J}_{s,k+1} - F_{k-1} \mathcal{J}_{s,k-1}) \nonumber \\
	&& + \frac{1}{\Delta y^2} 
    (E_{k+1/2} (\alpha^3_{k+1} \mathcal{J}_{s,k+1}-\alpha^3_{k} \mathcal{J}_{s,k}) \nonumber \\
    &&- E_{k-1/2} (\alpha^3_{k} \mathcal{J}_{s,k}-\alpha^3_{k-1} \mathcal{J}_{s,k-1})) \nonumber \\
    && \frac{G_{k}}{2\Delta y} \big[ D_{k+1/2} (\alpha^3_{k+1} \mathcal{J}_{s,k+1}-\alpha^3_{k} \mathcal{J}_{s,k}) \nonumber \\
    &&+ D_{k-1/2} (\alpha^3_{k} \mathcal{J}_{s,k}-\alpha^3_{k-1} \mathcal{J}_{s,k-1}) \big], \nonumber \\
    \label{eq:ns_runge_kutta_legendre_R}
\end{eqnarray}
where $E$, $F$ and $G$ are given by equation~(\ref{eq:ns_allen_cheng_FE}). Here, again we assumed a uniform grid spacing $\Delta y$ and have shown the discretization only for a single dimension.

\subsection{Boundary conditions}\label{sec:boundary-conditions}

For our spherical polar coordinate system, we use the standard boundary conditions in angular directions, namely reflecting boundary conditions at the polar axis and periodic boundary conditions in azimuthal direction. For the outer radial boundary, we typically use the ``free'' boundary condition, meaning that the flux is set according to free-streaming conditions, $D\partial_r \mathcal{J} = \mathcal{J}$. For the inner radial boundary, the user may choose a ``flat'' boundary condition, given by $D\partial_r \mathcal{J} = 0$ (adequate, e.g., at the coordinate center for symmetry reasons), or a ``fixed'' boundary condition, for which $\mathcal{J}$ is set to some predefined value (e.g.,  if the inner boundary is placed at a nonzero radius). We set the lower boundary of the neutrino energy grid at $\epsilon = 0$ and, therefore, $\epsilon R_\epsilon=0$. At the boundary of the highest energy bin, we exponentially extrapolate the neutrino energy density, $\mathcal{J}$.

\section{Test Problems}\label{sec:test-problems}

In this section, we discuss various setups for testing and validating the transport scheme. In Sects.~\ref{sec:1d-test-problems} and~\ref{sec:2d-test-problems}, we will consider 1D and 2D tests with simplified radiation-matter interactions, and in Sect.~\ref{sec:CCSN} we examine fully
dynamic 1D core-collapse supernova simulations with a microphysical equation of state and more realistic neutrino-matter interactions. For future reference, we define the L1 and L2 error norms as
\begin{eqnarray}
\mathrm{L1-error} &\equiv& \frac{1}{N} \sum_i \frac{|\mathcal{J}^{\mathrm{num}}_i-\mathcal{J}^{\mathrm{an}}_i|}{\mathcal{J}^{\mathrm{an}}_i}~, \nonumber \\
\mathrm{L2-error} &\equiv& \frac{1}{N} \sqrt{\sum_i \Big(\frac{\mathcal{J}^{\mathrm{num}}_i-\mathcal{J}^{\mathrm{an}}_i}{\mathcal{J}^{\mathrm{an}}_i}\Big)^2}~,
\label{eq:test_error_norm}
\end{eqnarray}
where the sums run over all $N$ spatial grid cells, and $\mathcal{J}^\mathrm{num}$ and $\mathcal{J}^\mathrm{an}$ denote the numerical and analytical solution for the radiation energy density, respectively. If not explicitly mentioned otherwise, we will employ the Allen-Cheng method for the explicit part of the time integration.

\subsection{1D test problems}\label{sec:1d-test-problems}

We first consider 1D toy-model problems, namely the diffusion of a Gaussian pulse and a differentially expanding isothermal atmosphere.

\subsubsection{Diffusion of Gaussian pulse with Crank-Nicolson}\label{sec:diff-gauss-pulse}

We set up a well-known test problem consisting of a Gaussian-shaped pulse of radiation that diffuses through a medium with constant scattering opacity, $\kappa_\mathrm{s}$. This problem is chosen to test the basic working capability of the code, in particular the correct implementation of the implicit Crank-Nicolson method used for the radial diffusion terms. Diffusion of a Gaussian-shaped pulse with constant scattering opacity has the analytical solution \citep[e.g.][]{2009ApJS..181....1S, 2016ApJS..222...20K}:
\begin{eqnarray}
	\mathcal{J}^\mathrm{an}(\tilde{r}) = \bigg(\frac{\kappa_\mathrm{s}}{t}\bigg)^{d/2}
    \exp\bigg(\frac{-3\kappa_\mathrm{s} \tilde{r}^2}{4t}\bigg)
	\label{eq:test_1d_gaussian_analytic}
\end{eqnarray}
in $d=1,2,3$ dimensions, where $\tilde{r}$ is the distance to the center of the pulse. In the present 1D case, a constant scattering opacity of $\kappa_\mathrm{s} = 10^3$ is used. The pulse is initialized at time $t=10^{-9}$ such that its peak coincides with the center of our computational domain, which has a total length of 2. In our spherical polar coordinate system, we mimic the 1D Cartesian grid (plane geometry) by locating the computational domain at some very large radius $r\sim 10^4$. 
The domain is divided into $N=\{128,256,512\}$ cells and a single radiation energy bin is evolved. We employ a ``flat" boundary condition for the inner boundary and a ``free" boundary condition for the outer boundary, following \citet{2009ApJS..181....1S}. We consider two choices for the CFL value (cp. eq.~(\ref{eq:gr_time_step})), 1 and 10. The problem is stopped at $t=2\times10^{-9}$.
\begin{table}
	\centering
	\caption{Gaussian pulse test using the implicit Crank-Nicolson method (cf. Sect.~\ref{sec:diff-gauss-pulse}). For each CFL factor and grid resolution the L2-error is given together with the ratio of the current L2-error to that resulting with half the resolution. The L2-error decreases quadratically with the number of grid points, consistent with the 2nd-order accuracy of the Crank-Nicolson method.}
	\begin{tabular}{lccr} 
		\hline
		CFL & resolution, $N$ & L2 error & error ratio\\
		\hline
		1.0 & 128 & 0.258 & \\
		& 256 & 0.054 & 4.751\\
		& 512 & 0.013 & 4.023\\
		10.0 & 128 & 0.259 & \\
		& 256 & 0.055 & 4.711\\
		& 512 & 0.013 & 4.062\\
		\hline
	\end{tabular}
	\label{tab:test_1d_gaussian_accuracy}
\end{table}
In Table \ref{tab:test_1d_gaussian_accuracy}, the L2-error and the ratio of the L2-error for two consecutive resolutions are shown. We obtain a second-order accuracy for both CFL values, which is in agreement with the formal accuracy of the Crank-Nicolson Method. The test confirms the basic functionality of the code and validates the correct implementation of the Crank-Nicolson scheme used for the diffusion terms.

\subsubsection{Diffusion of Gaussian pulse with Allen-Cheng and RKL2}\label{sec:diff-gauss-pulse-2}
\begin{table}
  \centering
  \caption{Gaussian pulse test using the explicit Allen-Cheng method (cf. Sect.~\ref{sec:diff-gauss-pulse-2}). For each value of $r_\mathrm{diff}$ the L2-error is given together with the ratio of the current L2-error to that resulting with twice the value of $r_\mathrm{diff}$. The L2-error decreases linearly with $r_\mathrm{diff}$, consistent with the 1st-order temporal accuracy of the Allen-Cheng method.}
  \begin{tabular}{lccr} 
    \hline
    $r_\mathrm{diff}$ & resolution, $N$ & L2 error & error ratio\\
    \hline
    3.2 & 200 & 0.0885 & \\     
    1.6 & 200 & 0.0583 & 1.518\\
    0.8 & 200 & 0.0359 & 1.623\\
    0.4 & 200 & 0.0206 & 1.742\\
    0.2 & 200 & 0.0089 & 2.314\\
    0.1 & 200 & 0.0037 & 2.405\\
    \hline
  \end{tabular}
  \label{tab:test_1d_gaussian_accuracy_allen}
\end{table}
As described in Sect.~\ref{sec:numer-treatm-transp}, we use the implicit Crank-Nicolson scheme only for the radial diffusion and advection terms, while for all lateral terms we employ the explicit Allen-Cheng method. In this test problem, we want to check if the Allen-Cheng method is implemented correctly and produces reasonable results, and if it remains stable at conditions where conventional explicit schemes crash.

The setup is again a pure scattering-medium similar as in Sect.~\ref{sec:diff-gauss-pulse}, except now we use a diffusion coefficient of $D = 10^{-3}$ and fix the flux-limiter value to 1/3. We use a single spatial resolution of 200 points and we initialize the problem at $t=1$ using equation \eqref{eq:test_1d_gaussian_analytic}. We evolve the problem from time $t=1$ to $t=2$. The only characteristic timescale of this problem is the diffusion timescale $t_\mathrm{diff}={\Delta x}^2/D$, hence the performance of the integration method can be characterized entirely by the ratio $r_\mathrm{diff}=\Delta t/t_\mathrm{diff}$, where $\Delta t$ is the employed time step. We conduct several simulations varying the value of $r_\mathrm{diff}$ by changing $\Delta t$.

As can be seen in Table~\ref{tab:test_1d_gaussian_accuracy_allen}, the L2-error decreases roughly linearly with decreasing time step in agreement with the formal temporal accuracy of the Allen-Cheng method. Moreover, the test demonstrates that for $r_\mathrm{diff}>0.5$ the Allen-Cheng method indeed remains stable, and that, as expected, the accuracy decreases for higher values of $r_\mathrm{diff}$. We also conducted the test with the RKL2 method with $r_\mathrm{diff}=1.6$ and $2.3$ corresponding to CFL values of 16 and 23, respectively. The calculation with the RKL2 method and $r_\mathrm{diff}=1.6$ have an L2-error of $0.0013$, an order of magnitude smaller than the Allen-Cheng method. However, in the case of $r_\mathrm{diff}=2.3$, the RKL2 solution becomes unstable, which is consistent with its formal stability criterion (see section~\ref{sec:non-radi-deriv-runge-kutta-legendre}). In summary, as expected, the RKL2 method has better accuracy compared to the Allen-Cheng method, but becomes unstable at a critical $r_\mathrm{diff}$, whereas the Allen-Cheng method is unconditionally stable for all values of $r_\mathrm{diff}$. In Sect.~\ref{sec:diff-gauss-pulse-1}, we will consider a similar test in two dimensions.

\subsubsection{Differentially expanding atmosphere}\label{sec:diff-expand-atmosph}

\begin{figure*}
  \begin{minipage}{1.01\textwidth}
  \includegraphics[width=\textwidth]{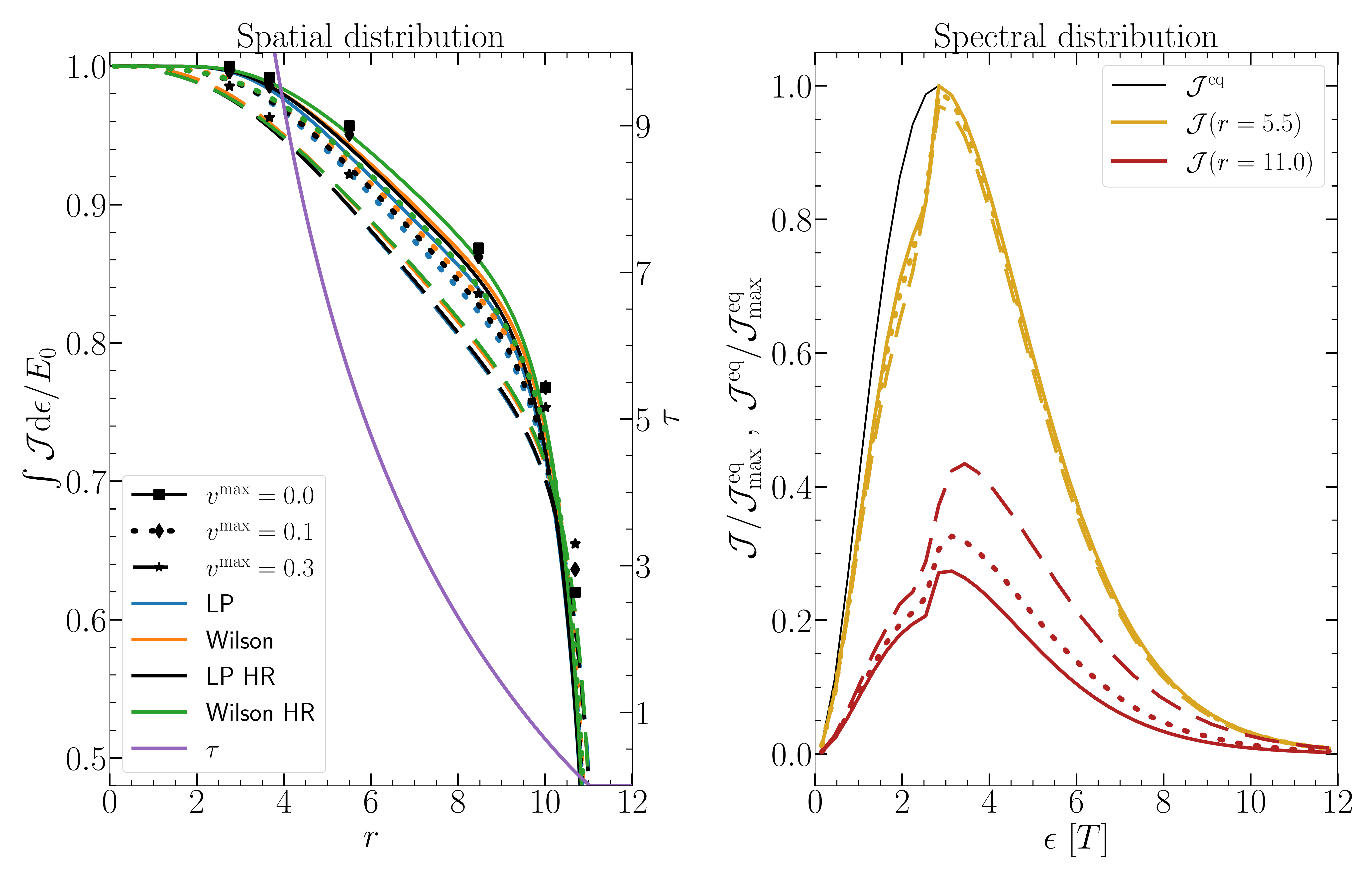}
  \end{minipage}
  \vspace*{-5mm}
  \caption{Differentially expanding atmosphere (cf. Sect.~\ref{sec:diff-expand-atmosph}). In the left plot, the energy-integrated energy density, $E(r) \equiv \int \mathcal{J}(r,\epsilon)~\mathrm{d}\epsilon$, normalized by $E_0=\int \mathcal{J}(r=0,\epsilon)~\mathrm{d}\epsilon$, is shown against radius for $v^\mathrm{max}=0$ (solid lines), 0.1 (dotted lines), 0.3 (dashed lines), where blue (black) lines denote solutions obtained with the LP limiter (LP limiter and higher grid resolution, LP HR), and orange (green) lines denote the solutions obtained with the Wilson limiter (Wilson HR). The markers show the reference solution by \citet{1980ApJ...237..574M}. The mean optical depth $\tau(r) \equiv \int_{\infty}^{r} \mathrm{d}r'\,\big(\int \mathrm{d}\epsilon \,\kappa_\mathrm{a}(r',\epsilon) \mathcal{J}(r',\epsilon)\epsilon^{-1} / \int \mathrm{d}\epsilon\,\mathcal{J}(r',\epsilon)\epsilon^{-1}\big)$ is shown by the violet line (scale on the right side). In the right plot, the radiation spectra, normalized by the maximum of the equilibrium distribution function, $\mathcal{J}^\mathrm{eq}_\mathrm{max}$, resulting with the Wilson limiter are plotted at radius $r=5.5$ (golden lines) and 11 (red lines), together with the equilibrium distribution function (see eq.~\eqref{eq:test_1d_DEA_Jeq}, normalized  by $\mathcal{J}^\mathrm{eq}_\mathrm{max}$) at a temperature of $T=1$ (black line).}
  \label{fig:test_1d_DEA}
\end{figure*}

Next, we consider a differentially expanding, isothermal atmosphere in spherical symmetry having a temperature of $T=1$ \citep{1980ApJ...237..574M, 2002A&A...396..361R, 2015MNRAS.453.3386J} in order to check the correct implementation of the energy-bin coupling and velocity-dependent terms in our code. The velocity profile is given by
\begin{eqnarray}
  v_r(r) = v^{\mathrm{max}}\frac{r-r_{\mathrm{min}}}{r_{\mathrm{max}}-r_{\mathrm{min}}}~.
	\label{eq:test_1d_DEA_vel}
\end{eqnarray}
in the region $[r_\mathrm{min},r_\mathrm{max}]$ and by $v_r=0$ elsewhere. We consider three cases with $v^\mathrm{max}=\{0.0,0.1,0.3\}$. The radius- and energy-dependent absorption opacity is given by:
\begin{equation}
  \kappa_\mathrm{a} =
  \begin{cases}
    \frac{10a}{r^2}\exp\Big(-\frac{(\epsilon-\epsilon_0)^2}{\Delta^2}\Big)
    + \frac{a}{r^2}\left(1-\exp\Big(-\frac{(\epsilon-\epsilon_0)^2}{\Delta^2}\Big)\right) & , \, \epsilon \leq \epsilon_0 ~, \\
    \frac{10a}{r^2} & , \, \epsilon > \epsilon_0 ~, 
  \end{cases}
\end{equation}
and the equilibrium distribution by:
\begin{eqnarray}\label{eq:test_1d_DEA_Jeq}
\mathcal{J}^\mathrm{eq} = \frac{8 \pi \epsilon^3}{\exp(\epsilon/T) - 1}~. 
\end{eqnarray}
The parameters in the aforementioned prescriptions are given by $\{r_\mathrm{min},r_\mathrm{max},\epsilon_0,\Delta,a\} = \{1.0,11.0,3.0~T,0.2~T,10.9989\}$. We use 400 grid points to discretize the simulation domain uniformly within $[0.1,15]$, and employ 40 energy bins to cover the radiation energy range $[0,11.8~T]$. At $r=0.1$ the ``flat'' boundary condition is applied and at $r=15$ the free-streaming boundary condition. Each simulation is performed with the Crank-Nicolson scheme using a CFL value of 0.5 and is stopped once stationarity is reached. We run a simulation for each of the three values of $v^\mathrm{max}$ as well as for both the LP and the Wilson limiters (cf. eqs.~(\ref{eq:limiter}).

In the left plot of Fig.~\ref{fig:test_1d_DEA} we show radial profiles of the energy-integrated radiation energy density in the comoving frame, $E(r) \equiv \int \mathcal{J}(r,\epsilon)~\mathrm{d}\epsilon$, normalized by $E_0=\int \mathcal{J}^\mathrm{eq}(r=0,\epsilon)~\mathrm{d}\epsilon$. In agreement with the reference solution (taken from \citealp{1980ApJ...237..574M} and indicated by markers), $E$ shows a gradual decrease with growing expansion velocities at each given radius $r \la 10$, which is because of Doppler redshifting in the comoving frame. At higher radii, $r \ga 10$, cases with higher velocities show, again in agreement with the reference solution, higher values of $E$, mainly because of the cumulative effect of reduced absorption rates in the underlying layers where $E$ is reduced.

We notice that radiation in the FLD solutions departs from equilibrium and transitions into free-streaming conditions at somewhat lower radii than radiation in the reference solution. However, the L1-error of the FLD solution with respect to the reference solution is still rather small, namely $4\%$ for the LP limiter and $3\%$ for the Wilson limiter. In this test, the Wilson limiter reproduces the reference solution slightly better than the LP limiter.

We test the resolution dependence of the results by running higher resolution tests with 800 grid points with the LP and Wilson limiters, models LP HR and Wilson HR, respectively. The high resolution grid has 200 grid points uniformly distributed within  $r \le 7.5$, i.e. the region where the optical depth is greater than 3 (diffusive regime), and 400 grid points uniformly placed between $7.5 < r < 10.5$ and another 200 grid points uniformly distributed outside of $r = 10.5$. Repeating the runs with even higher resolution, namely 1400 radial grid points, does not lead to any visible differences compared to results with 800 radial zones (see, Fig. 1 of the supplementary material). The results of the test are shown by black lines (LP HR) and green lines (Wilson HR) in Fig.~\ref{fig:test_1d_DEA}. In the $v^\mathrm{max}=0$ case, the Wilson HR model shows better agreement with the reference solutions. However, for higher velocities the solutions do hardly converge better to the reference solutions than the simulations with 400 radial zones. This may be due to the fact that the flux-limiter is agnostic to the velocity-dependent terms in the first moment equation (see Appendix \ref{app:der_fld_flux_eqn} for the derivation of the FLD flux and approximations involved). We observe that the FLD result depends both on the choice of the flux limiter and the resolution. A better flux limiter which optimally incorporates velocity effects could maybe improve the numerical results.

In the right plot of Fig.~\ref{fig:test_1d_DEA}, the radiation energy density spectra, normalized by the maximum of equilibrium distribution function, $\mathcal{J}^\mathrm{eq}_\mathrm{max}$, are shown at radii $r = 5.5$ and $11.0$, representative of optically thick and thin conditions, respectively, along with the equilibrium spectrum at $r=5.5$ (see Fig. 2 of \citealt{2015MNRAS.453.3386J} for comparison). The jump in the spectra is associated with the jump in the absorption opacity at energy $\epsilon = \epsilon_0$. Due to radiation being redshifted (in the frame comoving with the background fluid) on its way to the surface, the jump in the spectra around $\epsilon_0$ is smeared out, all the more for higher values of $v^\mathrm{max}$.

The overall satisfactory results of this test prove that our FLD code can handle the transition of radiation from diffusion to free-streaming, and they indicate that the velocity-dependent terms describing Doppler effects are implemented properly.

\subsection{2D test problems}\label{sec:2d-test-problems}

In this section, we have a look at two-dimensional (2D) toy-model problems in order to check basic multi-dimensional features of our transport solver.

\subsubsection{Hemispheric difference test}\label{sec:hemisph-diff-test}

\begin{figure*}
    \begin{minipage}{1.01\textwidth}
	\includegraphics[width=\textwidth]{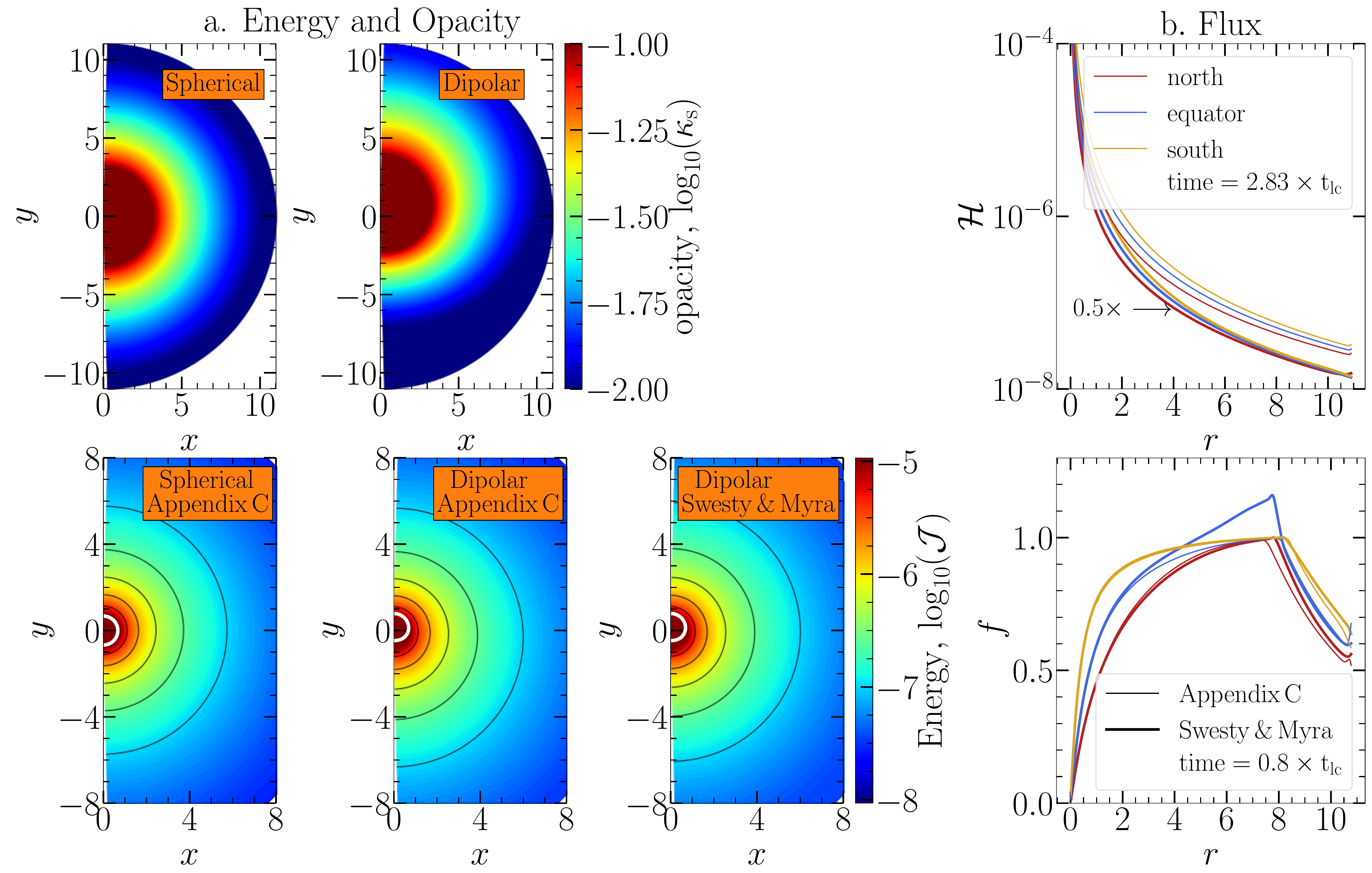}
    \end{minipage}
    \vspace*{-4mm}
    \caption{Hemispheric difference test (cf. Sect.~\ref{sec:hemisph-diff-test}). Panel (a): The top (bottom) row shows color maps of the logarithm of the  scattering opacity (radiation energy density). The angle-dependent locations $r_0$, at which the optical depth $\tau(r_0,\theta) =-\int_\infty^{r_0} \kappa_\mathrm{s} (r,\theta) dr$ reaches $2/3$, are marked by white lines in the bottom plots.
    Panel (b): In the top plot, the radiation energy flux, $\mathcal{H}^r$, are shown at the north pole (red lines), equator (blue lines) and south pole (yellow lines) for the dipolar opacity case at 2.83 light crossing times ($t_\mathrm{lc}$). The thin and thick lines show the results when the flux-limiter is evaluated according to Appendix~\ref{app:evl_R_lambda_D} or \citet{2009ApJS..181....1S}, respectively. For clarity, the radiation energy fluxes obtained by using \citet{2009ApJS..181....1S} have been shifted downward by a factor of 0.5. In the bottom plot, the flux factors are shown at a time of 0.8 light-crossing times for the dipolar case with the mentioned methods of the flux-limiter evaluation.}
	\label{fig:test_2d_HSD}
\end{figure*}

We first discuss a simple configuration to test the basic ability of the code to deal with multiple dimensions without becoming unstable or producing numerical artefacts. We consider radiation diffusing out of a static scattering atmosphere. The absorption opacity vanishes everywhere.

In the first of two versions of this test, the scattering opacity, $\kappa_\mathrm{s}$, has a spherically symmetric profile, given by
\begin{eqnarray}
  \kappa_\mathrm{s}(r) =
  \begin{cases}
    \frac{1}{r^2} & , \, r \leq r_{\mathrm{max}}, \\
    10^{-10} & , \, r > r_{\mathrm{max}} \, ,
  \end{cases}
\end{eqnarray}
with $r_\mathrm{max} = 10$, while in the second version we consider a dipole-shaped opacity profile by multiplying the opacity with the factor $(1+0.5\cos\theta)$. We use 600 grid points to cover the radial domain of $r \in [0,11]$, with 200 grid points uniformly distributed between radii 0 and 1 (optically thick region) and 400 grid points uniformly distributed between radii 1 and 11 (optically thin region). Adequate resolution is required by the FLD scheme in the optically thin region, otherwise the propagating radiation front losses its sharp profile (see, e.g. \citealt{2009ApJS..181....1S}). We use $64$ uniformly spaced grid points in polar direction with $\theta \in [0,\pi]$. A single energy group is used and the CFL value is set to 0.5. At $r=0.01$ the ``fixed'' boundary condition is applied with $\mathcal{J}(r=0.01,t)=1$. The problem is initialized with a constant value of $\mathcal{J}(r,t=0)=10^{-10}$.

In panel~(a) of Fig.~\ref{fig:test_2d_HSD}, the two top plots show the scattering opacity for spherical (left) and dipolar (right) opacity distributions, while the bottom plots depict the radiation energy densities after the numerical solutions have converged to a steady-state at a time of $t=2.83 \times t_\mathrm{lc}$ (93000 iterations), where $t_\mathrm{lc}$ is the light crossing timescale of the computational domain. Tests for longer run times do not show any further evolution. In the bottom row, the left and middle plots show the spherical and dipolar cases, respectively, when the flux-limiter is evaluated using the method described in Appendix~\ref{app:evl_R_lambda_D}, and the right plot shows the dipolar case with the flux-limiter evaluated according to the method from \citet{2009ApJS..181....1S}. We see that for the spherically symmetric opacity configuration the solution remains spherically symmetric, i.e. our mixed-type integration scheme combining the Crank-Nicolson and Allen-Cheng methods does not lead to spurious asphericities. The relative pole-to-equator and pole-to-pole differences of $\mathcal{J}$ are $< 0.1\,\%$.

In the case of the dipole-shaped opacity, in which the southern hemisphere has lower scattering opacity than the northern hemisphere, we observe, as expected, also a hemispheric difference in the radiation energy density: A greater amount of radiation is able to escape from the southern hemisphere compared to the northern hemisphere.

In panel~(b) of Fig.~\ref{fig:test_2d_HSD}, in the top plot, radial profiles of the radiation energy density flux, $\mathcal{H}^r$ are shown along the $\theta=0,~\pi/2,~\pi$ directions. For higher $\theta$, we observe enhanced fluxes and energies, as well as a transition to free-streaming (i.e. $\mathcal{H}/\mathcal{J} \rightarrow 1$) at smaller radii. Both methods of evaluation of the flux-limiter lead to hardly distinguishable steady-state solutions as we can see from Fig.~\ref{fig:test_2d_HSD}.

We checked the causality violation of the total flux (see the comments after eq.~(\ref{eq:knudsen})). In panel~(b) of Fig.~\ref{fig:test_2d_HSD}, in the bottom plot, we show the total flux factor at a time of $t=0.8 \times t_\mathrm{lc}$, when the causality violation is largest for the \citet{2009ApJS..181....1S} case. We obtained a maximum total flux factor of 1.2 in this test when the flux-limiter was evaluated using the method described in \citet{2009ApJS..181....1S}, whereas the flux factor remained $\le$ 1.0 with the improved method detailed in Appendix~\ref{app:evl_R_lambda_D}. In the upper plot of panel (b) one can notice an enhanced spherisization of the solution at large radii with the \citet{2009ApJS..181....1S} prescription.

Overall, the stability of the conducted simulations and the plausible physics results demonstrate the basic functionality of the multidimensional version of our transport solver.

\subsubsection{Diffusion of Gaussian pulse}\label{sec:diff-gauss-pulse-1}

\begin{figure*}
  \begin{minipage}{\textwidth}
  \includegraphics[width=\textwidth]{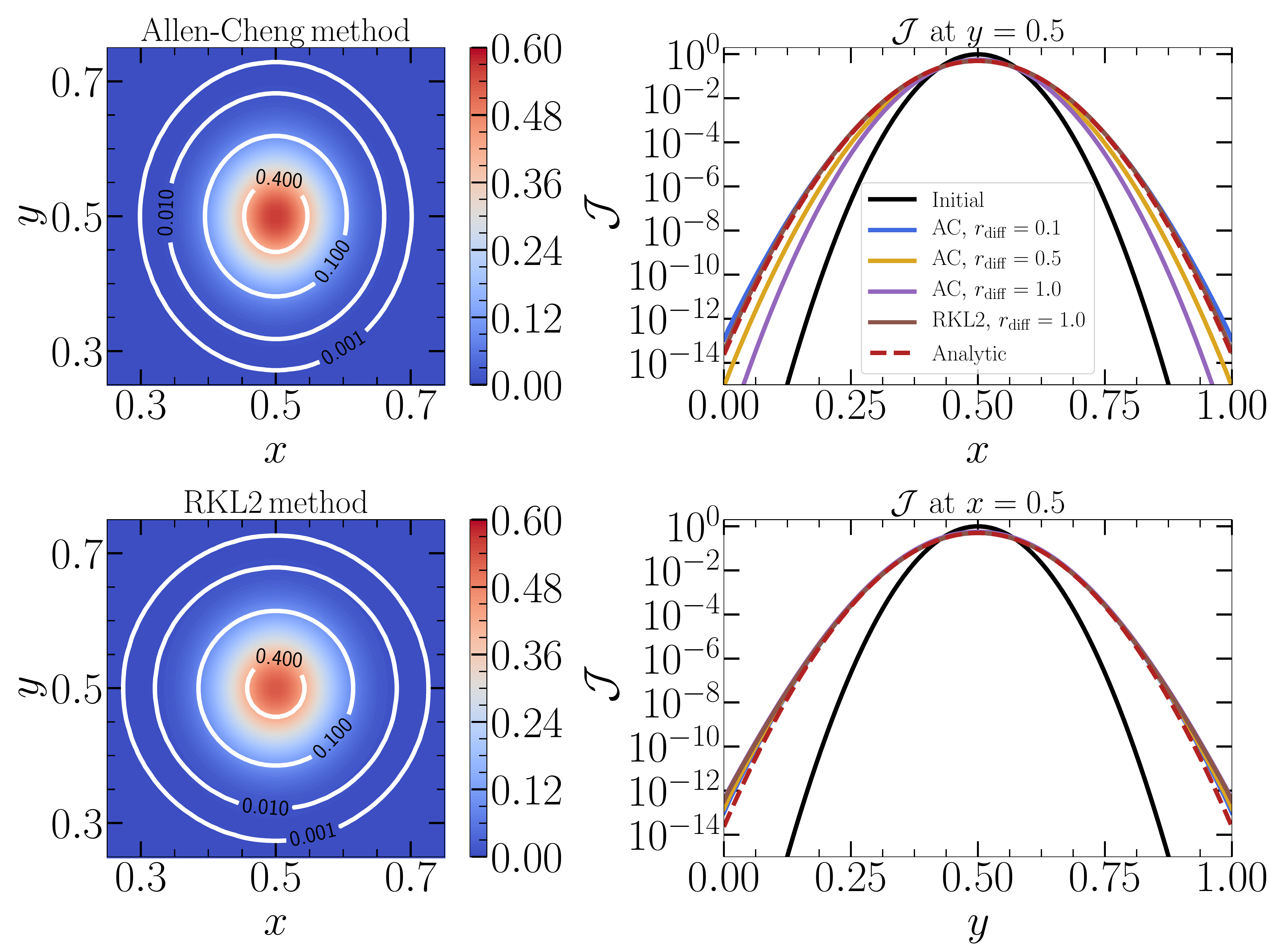}
  \end{minipage}
  \vspace*{-4mm}
  \caption{Two-dimensional diffusion of a Gaussian pulse (cf. Sect.~\ref{sec:diff-gauss-pulse-1}). The Allen-Cheng method (AC; with $r_{\mathrm{diff}}=0.1,0.5,1.0$) or the RKL2 method (with $r_{\mathrm{diff}}=1.0$) are applied along the $x$-direction and the Crank-Nicolson method along the $y$-direction. On the left side, contour plots of the radiation energy density are shown for the case of $r_{\mathrm{diff}}=1.0$ with the Allen-Cheng method (top) and the RKL2 method (bottom) at $t=1.995$. On the right side, profiles of the radiation energy density are plotted along the $x$-direction at $y=0.5$ (top) and along the $y$-direction at $x=0.5$ (bottom). The numerical solutions for different $r_{\mathrm{diff}}$ are shown by solid lines and the analytical solution by dashed lines. The initial condition is shown by solid black lines.}
  \label{fig:test_2d_gauss}
\end{figure*}

We now investigate two-dimensional diffusion of a Gaussian pulse, which has been considered already in 1D in Sects.~\ref{sec:diff-gauss-pulse} and~\ref{sec:diff-gauss-pulse-2}. In contrast to the test in Sect.~\ref{sec:hemisph-diff-test}, the diffusion test allows to compare with an analytical solution and, hence, we are now able to check also on a quantitative level the proper functionality of the multi-dimensional transport, with a particular focus on the impact of the dimensional splitting with mixed explicit-implicit treatments.

We use Cartesian coordinates in a domain of size $1 \times 1$. A uniform grid with 100 points is employed in each direction, and one energy bin is used. The diffusion coefficient is set to $D=10^{-3}$ and the problem is initialized at $t=1$ using equation \eqref{eq:test_1d_gaussian_analytic} with $d = 2$ and $\tilde{r}^2 = (x-0.5)^2 + (y-0.5)^2$. We again define the characteristic time-step parameter $r_\mathrm{diff}=D \Delta t/{\Delta x}^2$, where $\Delta t$ is the integration time step and $\Delta x$ the grid spacing. The values of $r_\mathrm{diff}$ are varied between $\{0.1,0.5,1.0\}$, corresonding to CFL values of $\{1,5,10\}$, respectively. The Allen-Cheng scheme is applied along the $x$-direction and Crank-Nicolson scheme is applied along the $y$-direction. The simulation is stopped at $t=1.995$. We also conducted the same test with the RKL2 method applied along the $x$-direction and the Crank-Nicolson method along the $y$-direction with $r_\mathrm{diff} = 1.0$.

The left panels in Fig. 4 show contour plots of the radiation energy
density for $r_\mathrm{diff} = 1.0$ with the Allen-Cheng method (top) and the RKL2 method (bottom) at the end of the simulation at time $t = 1.995$. The right column compares profiles along the lines at $y = 0.5$ (top) and $x = 0.5$ (bottom) of the numerical
solution with that of the analytical solution, which is given by eq.
\eqref{eq:test_1d_gaussian_analytic}. We first note that
the integration remains well-behaved and numerically stable, which is
indicated by the absence of spurious numerical features
in the plotted data. The Gaussian pulse
retains a circular shape up to a good degree, even for $r_\mathrm{diff}=0.5$,
although a non-circular deformation is visible and becomes stronger for
values of $r_\mathrm{diff} \ga 0.5$. The deformation is a result of the fact that
for high values of $r_\mathrm{diff}$ the diffusion rates are somewhat reduced in
$x$-direction, along which the explicit Allen-Cheng method is used. The
error for higher values of $r_\mathrm{diff}$ increases much more
strongly in $x$-direction than in $y$-direction. This is expected, because the
Allen-Cheng method is only first-order accurate while the Crank-
Nicolson method is second-order accurate. However, large relative errors
only appear at energy densities that are orders of magnitude smaller
than the peak energy, for which reason the global error is still small. The
test confirms that the dimensional splitting of our algorithm works well
and that the Allen-Cheng method remains stable and reasonably accurate
even for values $r_\mathrm{diff} \sim 0.5-1$. In the case of the RKL2 method, the Gaussian pulse retains its spherical shape even for $r_\mathrm{diff} = 1$ because of its higher accuracy compared to the Allen-Cheng method.

\subsection{Spherically symmetric stellar core collapse}\label{sec:CCSN}

\begin{figure*}
    \begin{minipage}{1.01\textwidth}
    \includegraphics[width=\textwidth]{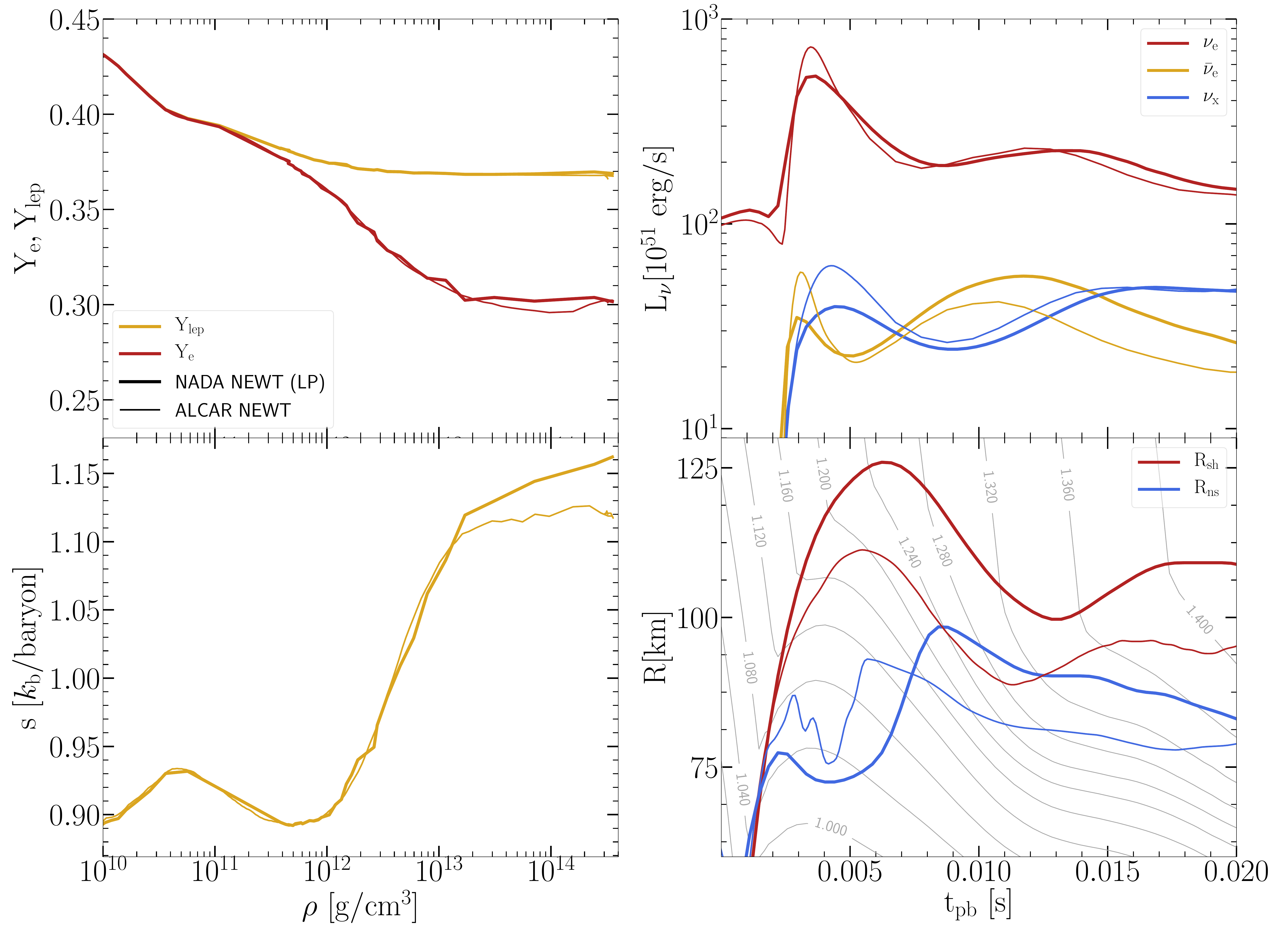}
    \end{minipage}
    \vspace*{-4mm}
    \caption{Properties obtained during the collapse (left) and shortly after bounce (right) in the NADA and ALCAR CCSN simulations of a 20 solar mass progenitor, both with Newtonian gravity (see Sect.~\ref{sec:CCSN}). NADA (ALCAR) results are marked by thick (thin) lines. The top left plot shows the central electron fraction (red) and total lepton fraction (yellow) and the bottom left plot shows the central entropy per baryon as functions of central density. The top right plot depicts as functions of time the neutrino luminosities of $\nu_e$ (red), $\bar{\nu}_e$ (yellow), and an individual species of $\nu_x$ (blue), measured in the comoving frame at a radius of 500\,km. The bottom right plot shows the shock radii (red) and proto-neutron star radii (blue) together with contours of constant enclosed mass in units of the solar mass (gray). The mass shells are depicted for the NADA NEWT model.}
    \label{fig:test_ccsn_nada_alcar_comp_1}
\end{figure*}

\begin{figure}
    \includegraphics[width=0.49\textwidth]{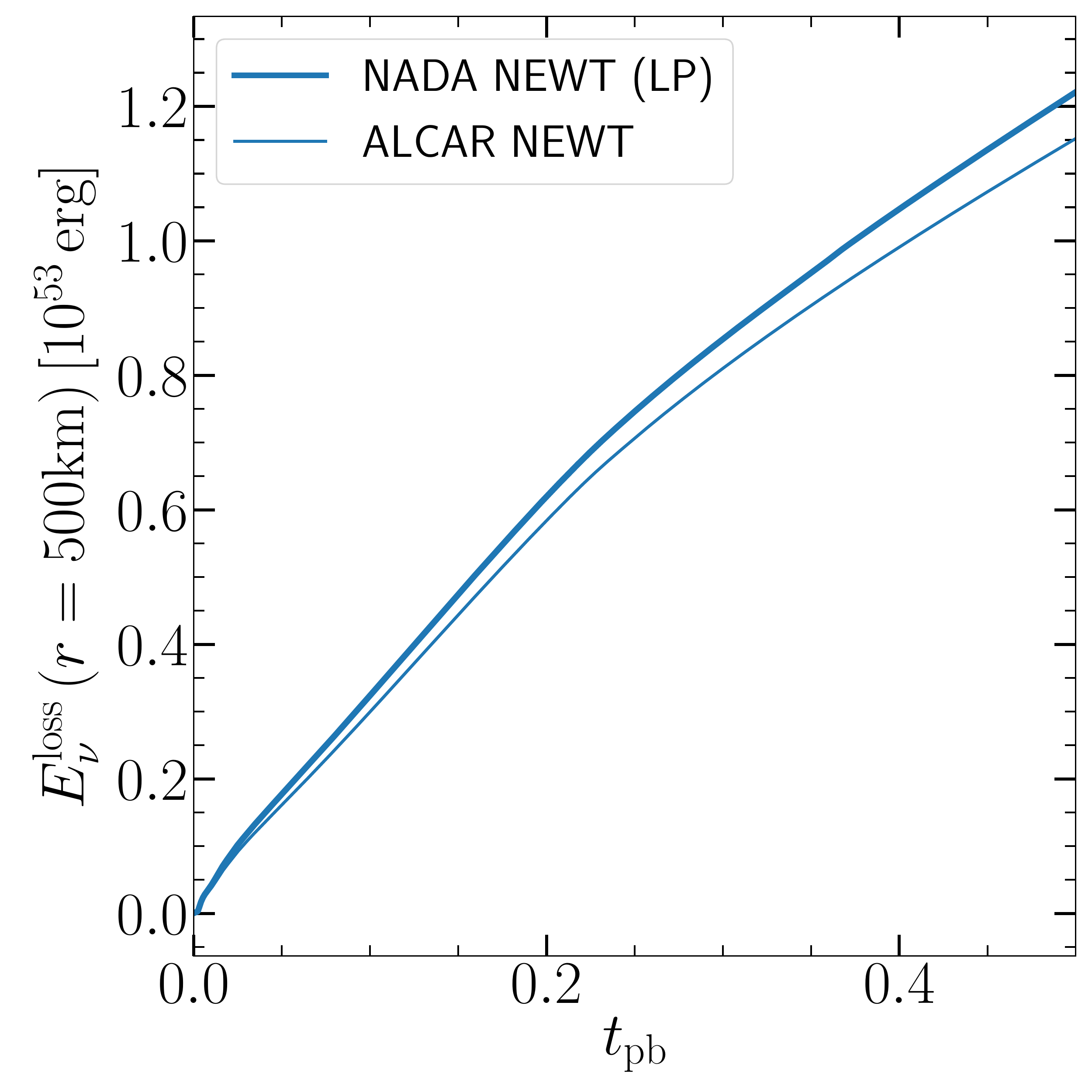}
    \vspace*{-6mm}
    \caption{Time evolution of the cumulative energy radiated in neutrinos after core bounce measured at a radial distance of 500\,km. The cumulative radiated energy is higher by 6\% at 500\,ms post-bounce time in the NADA NEWT (LP) simulation (thick line) than in the ALCAR NEWT simulation (thin line).}
    \label{fig:test_ccsn_cumulative_energy_loss}
\end{figure}

\begin{figure*}
  \includegraphics[width=0.99\textwidth]{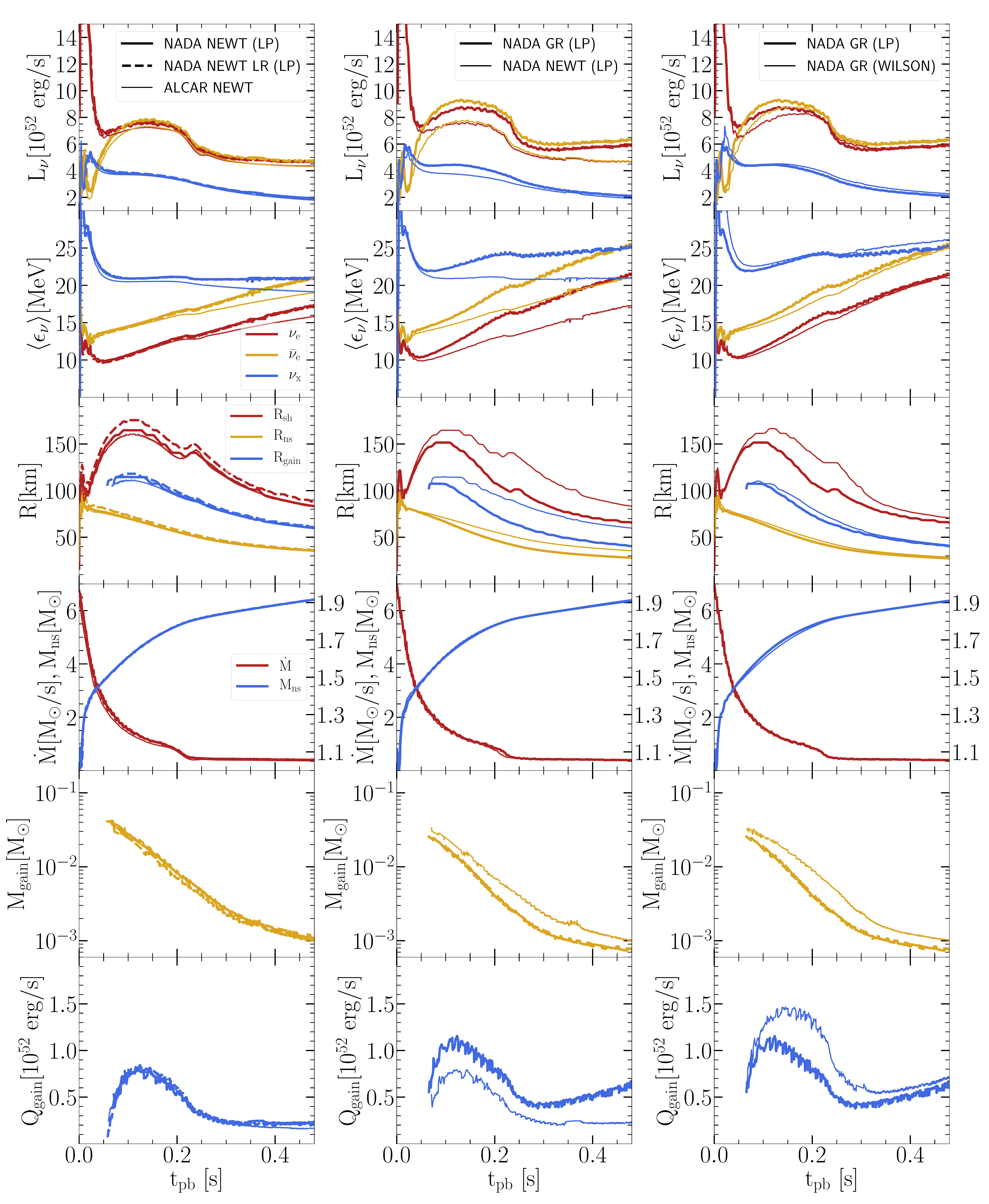}
  \vspace*{-4mm}
  \caption{Comparison of global properties as functions of time for several CCSN simulations. Shown are results in the left column for the NADA (thick lines), NADA low resolution (dashed thick lines) and ALCAR (thin lines) simulations that use Newtonian gravity, in the middle column for NADA simulations with GR (thick lines), and Newtonian (thin lines) treatment of gravity, and in the right column for NADA simulations with GR gravity using different flux-limiters (cf. Sect.~\ref{sec:energy-equation-flux}), namely the LP flux-limiter (thick lines) and the Wilson limiter (thin lines). From top to bottom the panels display the neutrino luminosities, the mean neutrino energies, the shock-, PNS-, and gain radii, the mass accretion rate measured at 500\,km, the mass in the gain layer, and the total neutrino-heating rate in the gain layer. The maximum difference in the PNS masses between models NADA NEWT and ALCAR NEWT is $4 \times 10^{-3}~\mathrm{M_{\odot}}$, with the NADA NEWT model having heavier mass. The differences between the corresponding lines are hardly visible.}
  \label{fig:test_ccsn_nada_alcar_comp_2}
\end{figure*}

\begin{figure*}
  \includegraphics[width=\textwidth]{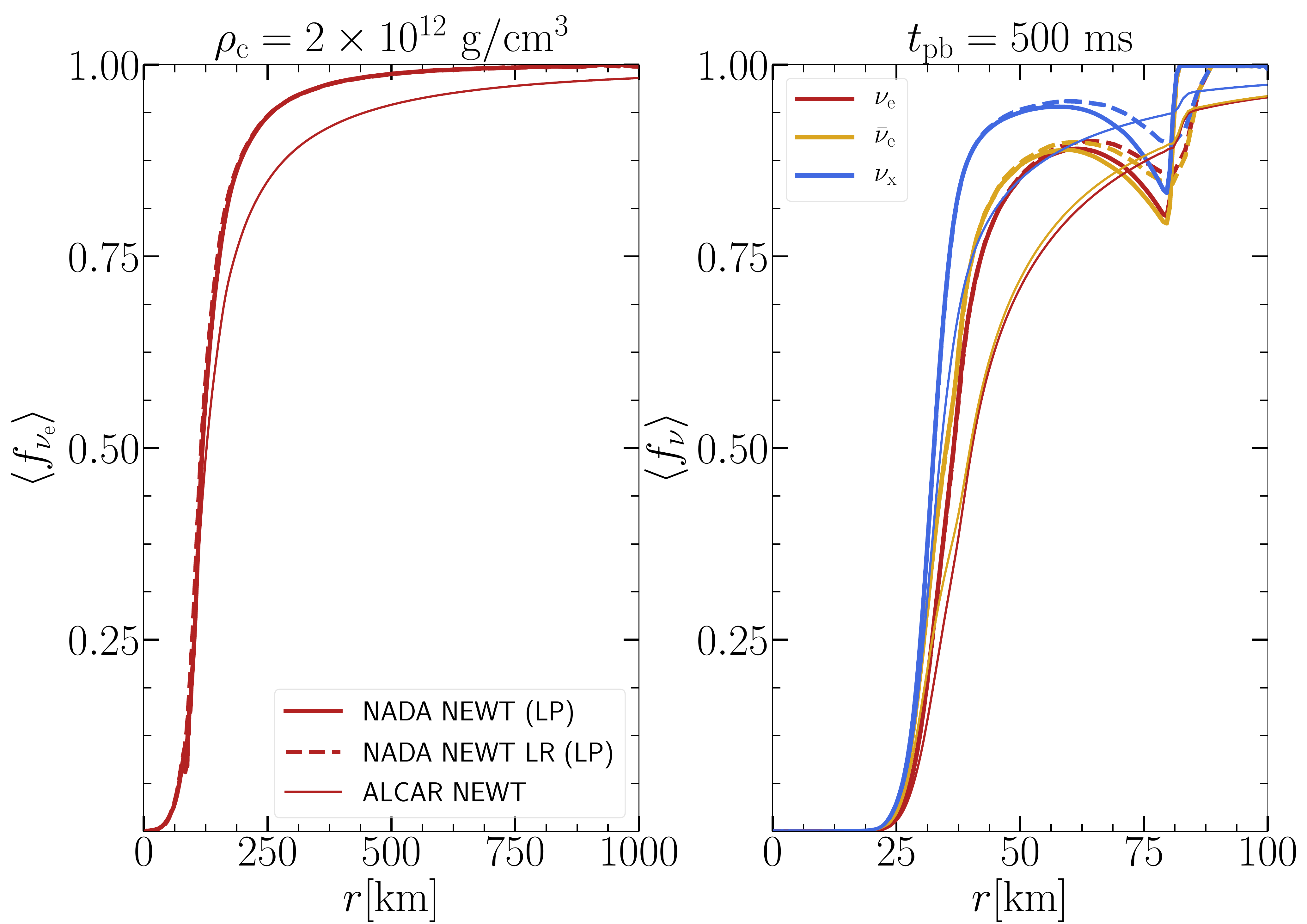}
  \vspace*{-6mm}
  \caption{Comparison of the energy-averaged flux factors, eq.~(\ref{eq:ccsn_mean_ff}), as functions of radius between the NADA (thick lines) and ALCAR (thin lines) CCSN simulations using Newtonian gravity for different neutrino species $\nu_e$ (red), $\bar{\nu}_e$ (yellow), and $\nu_x$ (blue). The mean flux factors from the NADA NEWT LR (LP) model are shown by the dashed lines. The left panel (only for $\nu_e$) shows the mean flux factors at a time during collapse when the central density reaches $2 \times 10^{12}$\,g\,cm$^{-3}$, and the right plot at a post-bounce time of 500 ms. For all cases the transition to free-streaming (i.e. to high flux factors, $\langle f_\nu\rangle \ga 0.5$), takes place at somewhat smaller radii and higher densities for the NADA simulations, which employ the FLD approximation, compared to the ALCAR simulations, which make use of the M1 approximation.}
  \label{fig:test_ccsn_nada_alcar_comp_3}
\end{figure*}

\begin{figure*}
  \includegraphics[width=\textwidth]{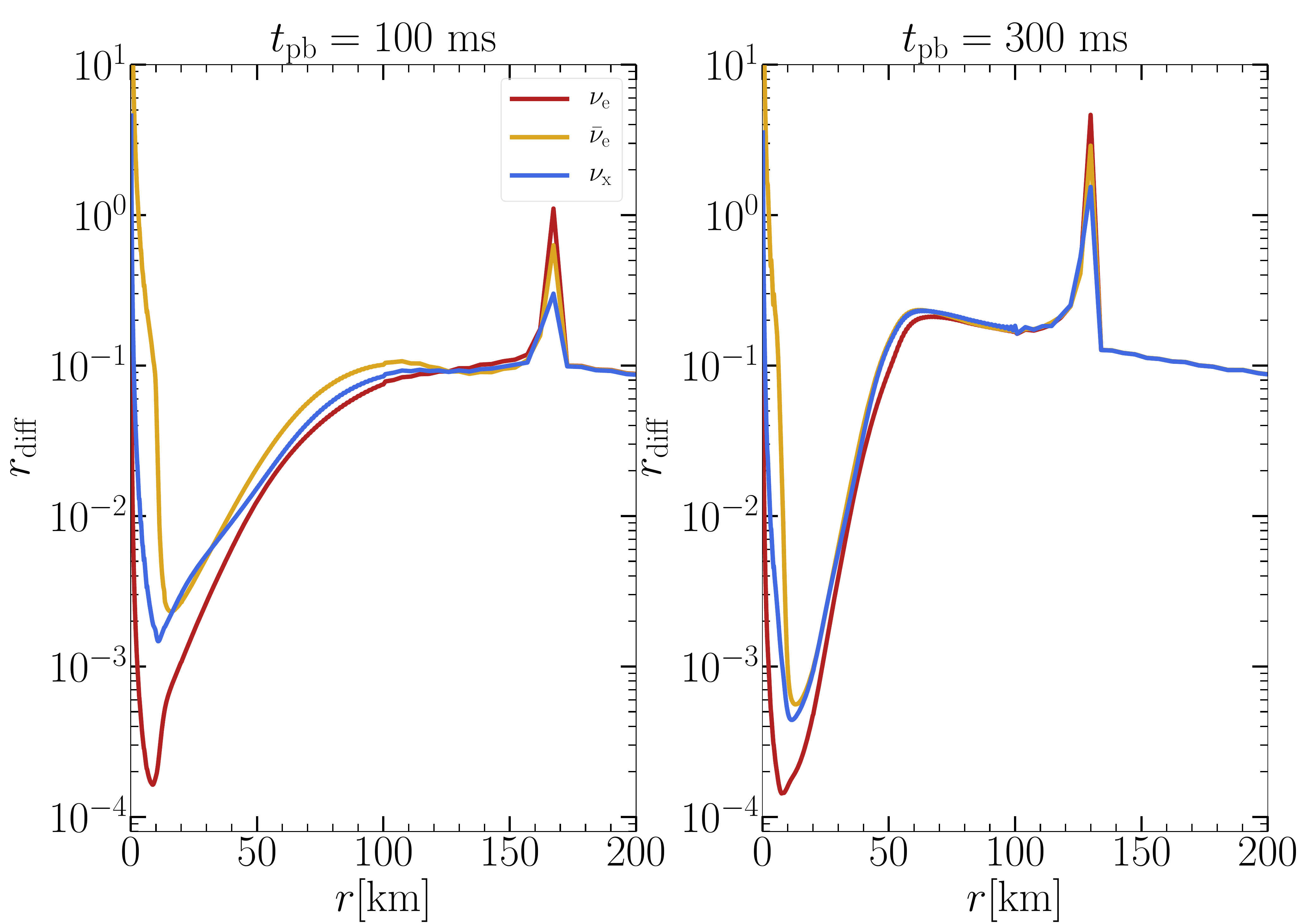}
  \vspace*{-6mm}
  \caption{Estimate of the time-step parameter $r_\mathrm{diff}$ characterizing the accuracy of the explicit integration of lateral fluxes in the future case of a 2D axisymmetric simulation. We use data from the one-dimensional, Newtonian NADA simulation with LP limiter, employ eqs.~(\ref{eq:ccsn_rdiff}) and~(\ref{eq:ccsn_rdiff_D}), and assume an angular resolution of 1.4 degrees. The resulting $r_\mathrm{diff}$ is shown at post-bounce times of 100\,ms (left) and 300\,ms (right) for species $\nu_e$ (red), $\bar{\nu}_e$ (yellow), and $\nu_x$ (blue)}
  \label{fig:test_ccsn_r_diff}
\end{figure*}

\begin{figure*}
    \includegraphics[width=0.49\textwidth]{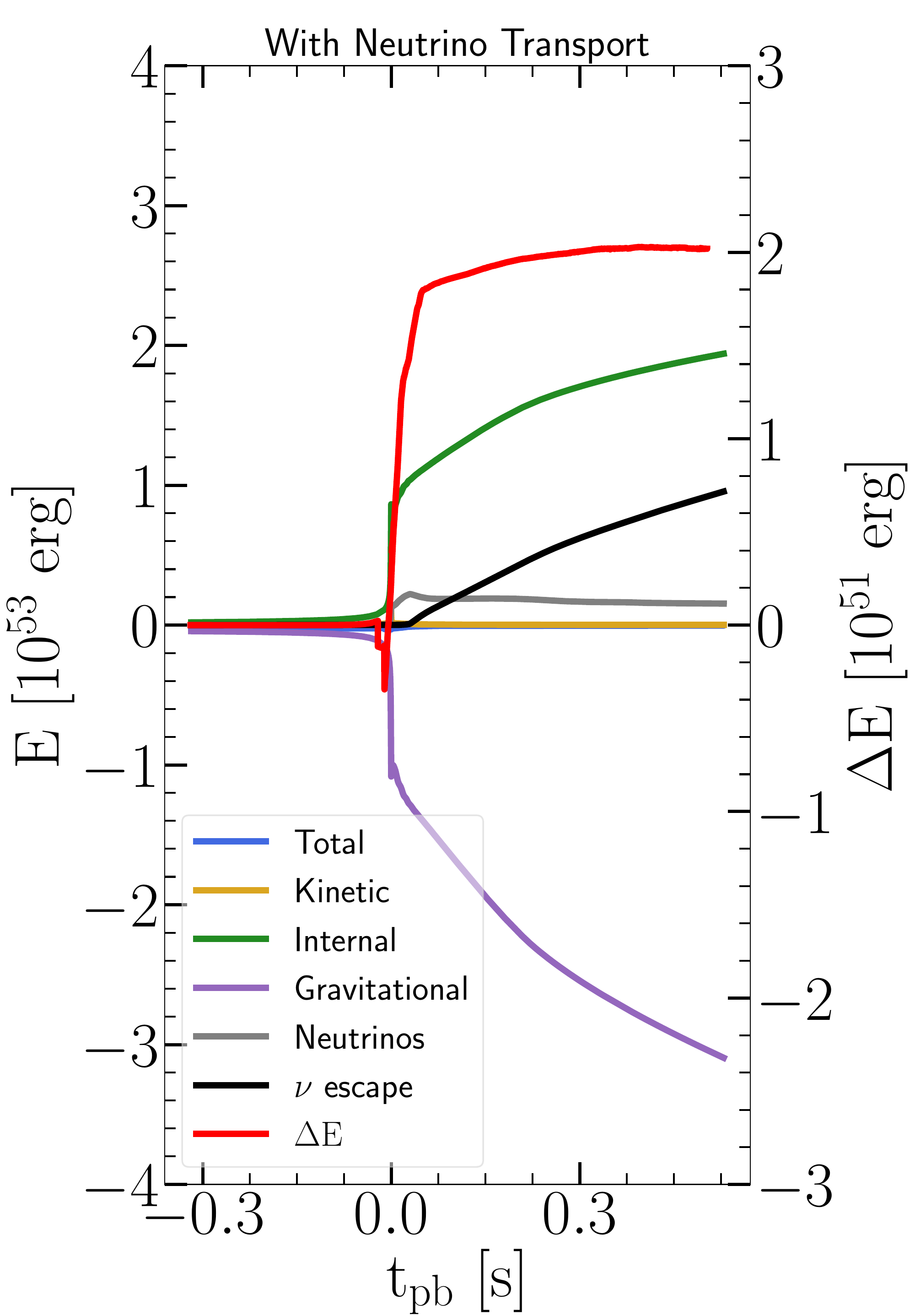}
    \includegraphics[width=0.49\textwidth]{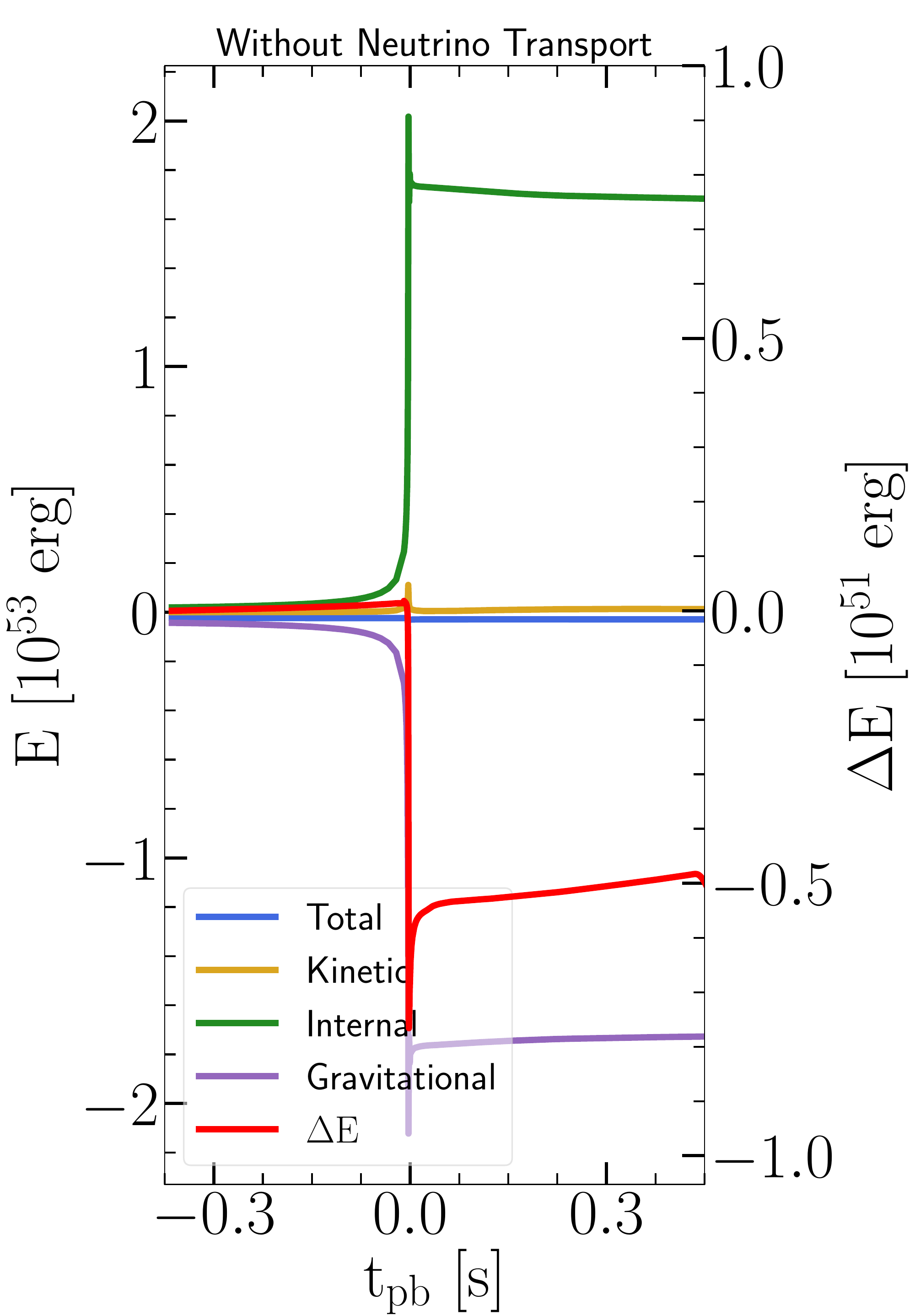}
    \vspace*{-3mm}
    \caption{Time evolution of different components of the energy and energy violation, $\Delta \mathrm{E}$, in the NADA NEWT model with (left plot) and without (right plot) neutrino transport. Note that the scale for $\Delta E$ is on the right side of each panel.}
    \label{fig:test_ccsn_energy_conservation}
\end{figure*}

\begin{table}
  \centering
  \caption{Neutrino opacities used for the 1D CCSN simulations discussed in Sect.~\ref{sec:CCSN}. ``$N$'' denotes nucleons and ``$A$'' and ``$A'$ '' denote nuclei. The $\nu \bar{\nu}$ pair processes are taken into account only for $\nu_x$ (for $\nu_e$ and $\bar \nu_e$ the $\beta$-processes are  by far dominant).}
	\begin{tabular}{cc}
		\hline
		Reaction & Neutrino  \\
		\hline
        $\nu + \mathrm{A} \leftrightarrow \nu + \mathrm{A}$ & $\nu_\mathrm{e}, \bar{\nu}_\mathrm{e}, \nu_\mathrm{x}$  \\
        $\nu + \mathrm{N} \leftrightarrow \nu + \mathrm{N}$ & $\nu_\mathrm{e}, \bar{\nu}_\mathrm{e}, \nu_\mathrm{x}$  \\
        $\nu_\mathrm{e} + \mathrm{n} \leftrightarrow \mathrm{e}^- + \mathrm{p}$ & $\nu_\mathrm{e}$  \\
        $\nu_\mathrm{e} + \mathrm{A} \leftrightarrow \mathrm{e}^- + \mathrm{A}'$ & $\nu_\mathrm{e}$  \\
        $\bar{\nu}_\mathrm{e} + \mathrm{p} \leftrightarrow \mathrm{e}^+ + \mathrm{n}$ & $\bar{\nu}_\mathrm{e}$ \\
        $\nu + \bar{\nu} \leftrightarrow \mathrm{e}^- + \mathrm{e}^+$ & $\nu_\mathrm{x}$  \\
        $\nu + \bar{\nu} + \mathrm{N} + \mathrm{N} \leftrightarrow \mathrm{N} + \mathrm{N}$ & $\nu_\mathrm{x}$ \\
        \hline
	\end{tabular}
	\label{tab:neutrino_opacity}
\end{table}
        
In this section, we discuss spherically symmetric simulations with more realistic microphysics of the collapse and post-bounce evolution of a 20\,$\mathrm{M}_{\odot}$ stellar progenitor with solar metallicity \citep{2007PhR...442..269W}. We employ the SFHo nuclear equation of state \citep{2012ApJ...748...70H,2013ApJ...774...17S}. The radial extent of our simulation domain is 10000\,km. We use 600 grid points with 170 equidistant points within $r<20~\mathrm{km}$, and 300 grid cells in the region $20~\mathrm{km}<r<200~\mathrm{km}$ (with size increasing by 1\% from cell to cell) and additional 130 grid points outside of $r=200~\mathrm{km}$ (size increasing by 3\% per cell). In order to save computation time, we reduce the number of transport steps relative to the GR-Hydro steps. During the collapse phase we apply neutrino transport every 50 hydrodynamics time steps (the hydrodynamics time step is given by equation \eqref{eq:gr_time_step} with CFL of 0.6). We apply neutrino transport every hydro time step between the time when the central density rises to $10^{12}~\mathrm{g/cm^3}$ and 20 ms post-bounce and subsequently every 10 hydrodynamics time steps. The energy grid is logarithmic, with 15 points covering energies from 0 to 400\,MeV, where 400 MeV is the upper boundary of the last energy bin. We evolve electron neutrinos ($\nu_e$), electron anti-neutrinos ($\bar{\nu}_e$), and $\nu_x$ neutrinos that are representative of the four heavy-lepton neutrinos. The neutrino reactions taken into account are listed in Table~\ref{tab:neutrino_opacity}. Their formulation is mostly based on \citet{1985ApJS...58..771B} and~\citet{2002A&A...396..361R}, but additionally includes corrections due to weak magnetism and recoil \citep{2002PhRvD..65d3001H}. We also take into account nucleon-nucleon bremsstrahlung. Following the recipe suggested by \citet{2015ApJS..219...24O}, we neglect pair-processes for electron-type neutrinos and treat pair-processes for $\nu_x$ neutrinos with a prescription that is formally equivalent to emission/absorption. 

We perform simulations with fully general relativistic hydrodynamics and transport, denoted by NADA GR, using each of the two flux-limiters, LP and Wilson (cf. eqs.~(\ref{eq:limiter})). However, in order to compare our code with a reference solution, we first discuss a simulation, called NADA NEWT, that is  identical to NADA GR with the LP limiter, but that is conducted with a Newtonian treatment of gravity and special relativistic hydrodynamics. We compare this model to model ALCAR NEWT which is performed with the ALCAR code using the Minerbo closure \citep{2015MNRAS.453.3386J, 2018MNRAS.481.4786J}. Model ALCAR NEWT contains exactly the same input physics, but it employs the M1 approximation for the neutrino transport and assumes non-relativistic hydrodynamics\footnote{Note that in the considered simulations the velocities are rather low
and, hence,  generic special relativistic effects should be small.} with 600 grid cells spanning a region of 10000 km.

In the left column of Fig.~\ref{fig:test_ccsn_nada_alcar_comp_1}, we compare characteristic properties of the collapse between models NADA NEWT and ALCAR NEWT, namely the electron fraction, $Y_e$, lepton fraction, $Y_{\mathrm{lep}}$, and entropy per baryon at the stellar center as function of the central density, $\rho$. Neutrino trapping sets in once the central density reaches $\sim 2 \times 10^{12}$\,g\,cm$^{-3}$. After the onset of neutrino trapping, the lepton fraction remains constant with a value around $0.37$ for both codes. The electron fraction roughly asymptotes at a central density of $\sim 2 \times 10^{13}$\,g\,cm$^{-3}$. The deleptonization slows down around $5\times10^{10}-10^{11}$\,g\,cm$^{-3}$ in both models due to neutron shell blocking and a low abundance of free protons \citep[e.g.][]{1985ApJS...58..771B}. After the onset of trapping, the entropy per baryon of the gas increases to $\approx 1.15\,k_\mathrm{b}$/baryon because of a growing number of free nucleons and $\alpha$-particles. Overall, both simulations agree very well in their deleptonization behavior. 

In the right column of Fig.~\ref{fig:test_ccsn_nada_alcar_comp_1}, we show the neutrino luminosity as well as the shock- and PNS-radii as functions of time until 20\,ms post bounce. For the present spherically symmetric case we define the comoving-frame luminosity as
\begin{eqnarray}
L_{\nu}(r) \equiv 4 \pi \eul^{4\phi} r^2  \int \mathcal{H}_{\nu}(r,\epsilon)~\mathrm{d}\epsilon~,
\end{eqnarray}
where $\phi=0$ for the case of Newtonian gravity. The luminosity obtained with the NADA code agrees well with that of the ALCAR code. The time-integrated energy loss associated with the neutrino burst (i.e. within the first 20\,ms of post-bounce evolution) is $\sim$5\,\% higher in the NADA model than that in the ALCAR model. The peak in the neutrino luminosities around 10-15\,ms post-bounce time is due to early expansion and subsequent compression of matter behind the shock as shown in the bottom-right plot of Fig.~\ref{fig:test_ccsn_nada_alcar_comp_1}. We also see that the luminosity of $\bar \nu_e$ rises earlier than $\nu_x$, which is different from existing earlier work where the $\nu_x$ luminosity rises earlier than the $\bar \nu_e$ luminosity (see, e.g. \citealt{2003ApJ...592..434T,2005PhRvD..71f3003K}). This difference, which is shared with the comparative ALCAR model, might be a consequence of the use of an analytic closure relation, or it might be linked to different sets of neutrino reactions employed by our scheme and previous models (e.g. the current NADA version ignores neutrino-electron scattering). Indeed, the rapid rise and first peak of the $\bar \nu_e$ luminosity disappears in ALCAR simulations when neutrino electron scattering is taken into account.

The left column of Fig.~\ref{fig:test_ccsn_nada_alcar_comp_2} provides various quantities as functions of time for the NADA NEWT and ALCAR NEWT simulations, namely the neutrino luminosities, $L_\nu$, the neutrino mean energies,
\begin{eqnarray}\label{eq:ccsn_mean_en}
\langle \epsilon_{\nu} \rangle (r) \equiv \frac{\int \mathcal{J}(r,\epsilon)~\mathrm{d}\epsilon}{\int \mathcal{J} (r,\epsilon)~\epsilon^{-1}~\mathrm{d}\epsilon}~,
\end{eqnarray}
the mass-accretion rate at 500 km, $\dot{M}$, the mass of the PNS, $M_\mathrm{ns}$, the total mass in the gain layer, $M_\mathrm{gain}$, and the total neutrino-heating rates, $Q_{\mathrm{gain}}$. The luminosities and mean energies, as well as almost all other quantities agree remarkably well between both codes. In Fig.~\ref{fig:test_ccsn_cumulative_energy_loss}, we notice that the cumulative energy radiated in neutrinos, $E^\mathrm{loss}_\nu$, after core bounce and measured at $r=$500\,km is $\sim$6\,\% higher in the NADA model than in the ALCAR model until 500\,ms post-bounce time. Moreover, the absolute difference in $E^\mathrm{loss}_\nu$ between NADA model and ALCAR model grows monotonically and continuously with time because the neutrino luminosities measured at $r=$500\,km are around 5\% higher in the NADA model compared to the ALCAR model. We also notice a secular drift towards higher mean energies in the NADA NEWT model, particularly at late times. We speculate that this difference might be related to what we see in Fig.~\ref{fig:test_ccsn_nada_alcar_comp_3}, where radial profiles of the mean flux factor,
\begin{eqnarray}\label{eq:ccsn_mean_ff}
\langle f_{\nu} \rangle(r) \equiv \frac{\int \mathcal{H}_{\nu} (r,\epsilon)~\epsilon^{-1}~\mathrm{d}\epsilon}{\int \mathcal{J}_{\nu} (r,\epsilon)~\epsilon^{-1}~\mathrm{d}\epsilon}~,
\end{eqnarray}
are plotted: In the NADA simulation, the flux factors rise at slightly smaller radii than in the ALCAR simulation, which means that neutrinos are effectively released from deeper within the PNS and therefore at higher temperatures. As a result,  the neutrino mean energies, $\langle \epsilon_{\nu} \rangle$, have higher values in the NADA simulation compared to the ALCAR simulation (see below for further discussion of Fig.~\ref{fig:test_ccsn_nada_alcar_comp_3}). The higher neutrino luminosities in model NADA NEWT might be linked to slightly higher mass-accretion rates compared to model ALCAR NEWT and to the fact that the flux factor in the NADA NEWT model rises to unity at a lower radial distance compared to the ALCAR NEWT model (see Fig.~\ref{fig:test_ccsn_nada_alcar_comp_3}). This rise in the flux factor at a smaller radial distance reflects the fact that the neutrinos become free-streaming at a higher optical depth in NADA NEWT model compared to the ALCAR NEWT model and hence can support higher neutrino luminosities. During the phase of high mass accretion ($t_\mathrm{pb} \lesssim 0.2 \mathrm{s}$: see Fig. \ref{fig:test_ccsn_nada_alcar_comp_2}, left column), we notice that model NADA NEWT produces slightly larger PNS radii (by $\sim$ 1$-$3km) and shock radii (by $\sim$5km) compared to model ALCAR NEWT, while afterwards both model agrees almost perfectly.  Nevertheless, apart from the aforementioned differences the overall very good agreement between ALCAR and NADA is encouraging and suggests that the combined neutrino-hydro solver functions well and that the equation of state and the neutrino interactions are implemented properly.

In the middle column of Fig.~\ref{fig:test_ccsn_nada_alcar_comp_2}, we compare the fully relativistic NADA GR simulation with the NADA NEWT model, using for both cases the LP limiter. The main impact of GR is to produce an effectively steeper gravitational potential. Hence, the core bounces $\approx 40\,$ms earlier in the GR case compared to the Newtonian case. Subsequently, the GR treatment produces a considerably more compact PNS and post-shock configuration. As a consequence of the higher compactness, the temperatures at the PNS surface are increased, which results in significantly enhanced neutrino luminosities and mean energies. The enhancement is even strong enough to overcompensate for the lower masses in the gain layer and to yield considerably higher total neutrino-heating rates compared to the Newtonian model. The qualitative differences found here between Newtonian and general relativistic CCSN models are in good agreement with previous studies \citep[e.g.][]{2001ApJ...560..326B, 2006A&A...445..273M, 2012ApJ...756...84M, OConnor2018a}. We conclude that the coupling of the neutrino-hydrodynamics components of the code to the Einstein solver is working well, at least in spherical symmetry.

In order to test the sensitivity with respect to the chosen flux-limiter, we also compare the NADA GR simulation that uses the LP limiter against a similar simulation that employs the Wilson limiter; see the right column of Fig.~\ref{fig:test_ccsn_nada_alcar_comp_2} for the corresponding quantities as functions of time. Using the Wilson limiter instead of the LP limiter results in an overall less compact configuration, i.e. in higher values of the shock-, PNS-, and gain-radii, particularly at earlier times, $t_{\mathrm{pb}}\la0.3\,$s, while later on the differences become smaller. The most likely reason is found when comparing the luminosities, which for electron-type neutrinos are significantly reduced during the first $\sim$0.2$-$0.3\,s of post-bounce evolution in the case of using the WILSON limiter. The lower neutrino-cooling rates explain the larger PNS radii, and those also cause \citep[see e.g.][]{2012ARNPS..62..407J} larger gain- and shock-radii in the case of using the WILSON limiter. The more powerful neutrino heating in the gain layer with the WILSON limiter is thus mainly a result of the increased mass in the gain layer compared to the case with the LP limiter. 

It is interesting that the differences between NADA GR (LP) and NADA GR (WILSON) are bigger than those between NADA NEWT (LP) and ALCAR NEWT. In this respect it is worth pointing out that ALCAR uses an M1 scheme with Minerbo closure \citep{2015MNRAS.453.3386J} and that \citet{1992A&A...256..452J} found the LP limiter to show better agreement with the limiter belonging to the Minerbo closure than with the Wilson limiter in the optically thick and semi-transparent regimes. We suspect that this fact explains why there is good agreement between models NADA NEWT (LP) and ALCAR NEWT but comparatively large deviations between model NADA GR (LP) and NADA GR (WILSON).

In Fig.~\ref{fig:test_ccsn_nada_alcar_comp_3}, we show the radial profile of the mean flux factor, eq.~(\ref{eq:ccsn_mean_ff}), for models NADA NEWT (with LP limiter) and ALCAR NEWT, at a time when the central density is $2 \times 10^{12}$\,g\,cm$^{-3}$ (left plot) and at 300\,ms post bounce (right plot). Although the M1 scheme used in ALCAR is not a fully accurate solution of the Boltzmann equation either, it is likely somewhat more reliable than the FLD solution (see \citealt{2015MNRAS.453.3386J} for a comparison of FLD and M1 with a Boltzmann solver for static CCSN-related configurations). In both cases, we see that the FLD solution makes the transition to free-streaming conditions at smaller radii compared to the M1-based ALCAR solution. Furthermore, in the FLD scheme, the flux factor jumps to high values artificially strongly near sharp drops in the transport opacity (see \citealt{1992A&A...256..452J} for a detailed discussion). As a result, the mean flux factor abruptly becomes $\approx 1$ already close to the PNS surface, i.e. well behind the shock, which lies at $r\approx 80-90\,$km in the right panel of Fig.~\ref{fig:test_ccsn_nada_alcar_comp_3}. The results concerning the flux factor are consistent with previous investigations of the FLD scheme: \citet{1992A&A...256..428D} identify a ``missing opacity problem'' of FLD, which originates from neglecting the time and spatial derivatives of the flux factor and the Eddington tensor as well as velocity-dependent terms and energy-derivative terms in the first moment equation (see Appendix~\ref{app:der_fld_flux_eqn} for details). This problem in principle can be circumvented by introducing an ``artificial opacity'' (\citealt{1992A&A...256..428D,1993A&A...273..338D}), however, no time-dependent multi-dimensional implementation exists so far. Nevertheless, the otherwise good agreement between NADA NEWT and ALCAR NEWT suggests that the aforementioned deficiencies are small enough to affect the 1D dynamics at most on the few-percent level.

As a final point we discuss the time-integration accuracy of a (future) multi-dimensional CCSN simulation based on our 1D simulation data. As we recall from Sect.~\ref{sec:numer-treatm-transp}, the time integration of the transport equations is done implicitly for the source terms as well as the radial fluxes and energy derivatives, and explicitly for the lateral fluxes. We consider for the case of an axisymmetric simulation the resulting characteristic time-step parameter,
\begin{eqnarray}
  r_\mathrm{diff}=\langle D_\nu \rangle \Delta t / (r \Delta \theta)^2 \, ,
  \label{eq:ccsn_rdiff}
\end{eqnarray}
i.e. the ratio of the integration time step employed for all explicit terms, $\Delta t$, and the characteristic timescale associated with the lateral diffusion terms, $(r \Delta \theta)^2/\langle D_\nu \rangle$. We assign $\Delta t$ the value of the hydrodynamics time step employed in the 1D simulation, given by equation \eqref{eq:gr_time_step}, and assume a suitable value of 1.4 degrees for $\Delta \theta$. The energy-averaged diffusion coefficient, $\langle D_{\nu} \rangle$, is estimated as
\begin{eqnarray}
    \langle D_{\nu} \rangle (r) = \frac{\int D_{\nu} (r,\epsilon)~\mathcal{J}_{\nu} (r,\epsilon)~\epsilon^{-1}~ \mathrm{d}\epsilon}{\int \mathcal{J}_{\nu} (r,\epsilon)~\epsilon^{-1}~\mathrm{d}\epsilon}~.
    \label{eq:ccsn_rdiff_D}
\end{eqnarray}
For this setup, the estimates of $r_\mathrm{diff}$, shown in Fig.~\ref{fig:test_ccsn_r_diff} for an early and a late post-bounce time, allow us to identify regions, $r_\mathrm{diff}\ga 1$, in which the explicit Allen-Cheng method is potentially less accurate and the RKL2 method will become unstable in describing the lateral neutrino propagation. We find, however, that high values, $r_\mathrm{diff}\ga 1$, are reached only near the very center of the PNS and close to the shock. This is reassuring, because deep inside the PNS ($r \la 2~\mathrm{km}$), neutrinos are trapped and neutrino fluxes are strongly dominated by radial advection fluxes, while at large radii in the vicinity of the shock lateral neutrino fluxes are anyway small compared to radial fluxes. Hence, our estimate indicates that the explicit treatment of lateral terms in multi-dimensional simulations will only have minor consequences on the dynamical evolution. In future 2D simulations, we will apply the lateral transport sweep at every hydrodynamics time step, but the radial transport sweep will be applied only at every few, say 10$-$50, hydrodynamics time steps. Furthermore, in order to improve the accuracy of the lateral transport, we may apply the RKL2 method instead of the Allen-Cheng method.

\subsubsection{Resolution dependence}\label{sec:CCSN_resolution_dependence}
In order to test the resolution dependence of our results, we run a low-resolution NADA Newtonian model, NADA NEWT LR (LP), with 400 radial grid cells (the widths of which are constant up to 4\,km and afterwards increase by 3\,\% from cell to cell), using the LP flux limiter. The left column of Fig.~\ref{fig:test_ccsn_nada_alcar_comp_2} also shows the results of model NADA NEWT LR (LP) (dashed lines). The biggest impact of the resolution is seen among the displayed quantities for the shock, gain radii, and neutron star radii, for which the agreement between the ALCAR model and the NADA model is clearly improved with higher resolution. In Fig.~\ref{fig:test_ccsn_nada_alcar_comp_3}, the mean flux factors for the NADA NEWT LR (LP) model are shown along with the NADA NEWT (LP) and the ALCAR NEWT model. Again we notice that the flux factors in the NADA NEWT LR (LP) model, similar to the NADA NEWT (LP) model, rise at a smaller radius than in the ALCAR model. As a result, both the high- and low-resolution NADA models have higher neutrino mean energies compared to the ALCAR model.

\subsubsection{Energy conservation}\label{sec:CCSN_energy_conservation}
To assess the energy conservation error of our code we show in Fig. \ref{fig:test_ccsn_energy_conservation} different components of the energy for our NADA NEWT model with neutrino transport (left plot) and without neutrino transport (right plot). We also evaluate the magnitude of energy violation (red lines), $\Delta E=E_\mathrm{tot}-E_\mathrm{tot,0}$ (where $E_\mathrm{tot,0}$ is the total energy at the beginning of the simulation, $\sim$318\,ms before core bounce), in our models. For the calculation of the total energy, $E_\mathrm{tot}$, we have taken into account the energy drain due to neutrino escape from the computational grid and energy gain due to the mass inflow through the outer boundary (not shown in Fig. \ref{fig:test_ccsn_energy_conservation}, because it is tiny). The gravitational potential energy, $E_\mathrm{grav}$, is evaluated from the Newtonian potential, $\Phi$, by:
\begin{eqnarray}
    E_\mathrm{grav} = - \frac{1}{8 \pi \mathrm{G}} \int \mathrm{d}V\,|\nabla\Phi|^2,
    \label{eq:ccsn_Egrav}
\end{eqnarray}
where the integration is carried out over the computational domain and we chose the normalization constant being equal to zero, meaning that we ignore the potential energy produced by the constant mass exterior of the computational domain. This method of evaluating the gravitational potential energy ensures better accuracy because the potential gradient, $\nabla\Phi$, is directly used in the hydrodynamics equations to describe the Newtonian gravitational force.

The energy violation in the purely hydrodynamical case is about $-0.5\times10^{51}$\,erg at\,$\sim$\,$450\,\mathrm{ms}$ post-bounce time. With neutrino transport included we find a total energy violation of about $+1.85\times10^{51}$\,erg  at\,$\sim$\,60\,ms post-bounce time and $+2\times10^{51}$\,erg at\,$\sim$525\,ms after bounce (left plot of Fig. \ref{fig:test_ccsn_energy_conservation}). However, referring this to the relevant energy scale of the problem, this is  $\sim$1.2$\%$ ($\sim$1.7$\%$) energy violation with respect to the gravitational (internal) energy at $\sim$60\,ms post-bounce time and $\sim$0.65$\%$ ($\sim$1$\%$) energy violation at $\sim$525\,ms post-bounce time relative to the gravitational (internal) energy (or again $\sim$0.65$\%$ when compared to the sum of internal energy and energies stored and escaping in neutrinos). 

Since we use an exponential extrapolation for the neutrino energy density at the upper boundary of the neutrino energy grid, it is possible for neutrinos to be lost through the upper energy boundary. To estimate how much neutrino energy is contained outside the upper energy boundary, we calculated the following ratio:
\begin{eqnarray}
    \frac{\sum\limits_{\nu} \int_{\epsilon_\mathrm{ob}}^{\infty} \mathrm{d}\epsilon \mathcal{J}}{\sum\limits_{\nu} \int_{0}^{\epsilon_\mathrm{ob}} \mathrm{d}\epsilon \mathcal{J}}~,
    \label{eq:ccsn_neutrino_energy_ratio}
\end{eqnarray}
where $\epsilon_\mathrm{ob}$=400\,MeV is the neutrino energy at the upper boundary and the sum is taken over all neutrino species. At around 525\,ms post-bounce time, when neutrinos have hard spectra, i.e. the highest mean energies, the mentioned ratio peaks around $r=$12\,km and has a value of less than 1\%. The total neutrino energy outside the energy grid integrated over the whole simulation volume is only 0.03\% of the total neutrino energy inside the energy grid at the mentioned time. Therefore, the neutrino leakage through the upper energy boundary cannot account for the entire energy violation. 

We also evaluated the total energy contained by the neutrino luminosity burst associated with shock breakout as it propagates radially outward and reaches the outer radial boundary at around\,$\sim$\,30\,ms. As the luminosity peak propagates through the optically thin material the total energy under the neutrino luminosity peak should be conserved. However, we observed energy violation of about +1.6\,$\times10^{51}$\,erg. Apparently, the FLD scheme seems to produce energy gain during the propagation of the radiation front through the optically thin region. 

For comparison, the Fornax code is mentioned to conserve the total energy on an excellent level of 0.05\,$\times 10^{51}$ erg until 1s after bounce for a purely hydrodynamical Newtonian 1D simulation with 608 radial zones (see section\,8.9 of \citealt{2019ApJS..241....7S}) and the total energy conservation is fulfilled on the level of $\sim$$10^{51}$\,erg (within 0.5\% of gravitational energy) during 500\,ms post-bounce time for a 2D simulation with neutrino transport and Newtonian monopolar gravity (see Fig.\,21 of \citealt{2019ApJS..241....7S}).

\section{Summary}\label{sec:summary}

In this paper, we presented a new code to solve multi-dimensional neutrino transport in spherical polar coordinates coupled to the GR-hydro code NADA \citep{2013PhRvD..87d4026B,2014PhRvD..89h4043M}. The transport solver assumes the flux-limited diffusion approximation and evolves the neutrino energy densities as measured in the frame comoving with the fluid. In order to improve the computational efficiency and parallel scalability compared to a scheme that solves the multi-dimensional FLD equations in a single, unsplit step, we employ operator splitting such that different parts of the equations (and different coordinate directions) are dealt with in separate, consecutive steps. The source terms as well as the radial- and energy-derivatives are integrated implicitly, while the non-radial derivatives are integrated explicitly using the Allen-Cheng method \citep{Allen1970} or the Runge-Kutta-Legendre method \citep{2012MNRAS.422.2102M} which remain stable in optically thin conditions.

We tested the algorithm and its implementation by conducting several problems in 1D and 2D and comparing to reference solutions. The tests demonstrate that the code runs stably and it robustly handles diffusion, transition to free-streaming, energy-bin coupling, multi-dimensional transport, microphysical neutrino interactions, and the coupling to GR-hydro. We confirmed that the Allen-Cheng method is, in contrast to conventional explicit schemes, unconditionally stable even if the diffusion timescale of a grid cell is shorter than the time step used for integration. However, estimates indicate that in multidimensional CCSN simulations, the diffusion timescale is typically longer than the integration time step except close to the coordinate center and the shock locations where lateral neutrino fluxes are strongly subdominant.

In terms of physics ingredients the most sophisticated tests performed here consider the core collapse and post-bounce evolution of a massive star in spherical symmetry. We compared a Newtonian version of this configuration with the results of the M1 code ALCAR \citep{2015MNRAS.453.3386J,2018MNRAS.481.4786J} and found that most global properties agree remarkably well, namely within $5-10\,\%$. We also compared the Newtonian simulation with its GR counterpart and were able to confirm the tendency of GR \citep[e.g.][]{2001ApJ...560..326B, 2006A&A...445..273M, 2012ApJ...756...84M, OConnor2018a} to lead to an overall more compact post-bounce configuration along with higher neutrino luminosities and mean energies. Another comparison of the GR simulation using the Levermore-Pomraning (LP) flux-limiter with another GR simulation using the Wilson limiter revealed notable differences, which, given the good agreement of the LP simulation with the ALCAR simulation, suggests that the LP limiter may be a better choice for
CCSN simulations than the Wilson limiter. Finally, we tested the resolution dependence and found that the impact of GR corrections to the FLD flux is relatively small.

We mainly chose the FLD method for its computational simplicity and its complementarity compared to M1, which is already employed by some existing GR codes. However, drawbacks of the FLD scheme are the potentially less accurate closure and the fact that FLD is not relativistically covariant and parabolic (in contrast to the hyperbolic GR-hydro solver). The accuracy of the closure may be improved in the future by employing a flux-limiter that is constructed to account for velocity effects and time-dependent terms in the 1st-moment equation of radiation transport.

\section*{Acknowledgements}
We thank Thomas W. Baumgarte, Pedro J. Montero, Bernhard M\"uller, Ewald M\"uller, Robert Bollig and Robert Glas for helpful discussions. We are also grateful to the anonymous referee for their helpful comments and for pointing us to the RKL2 method. NR and HTJ are grateful for support by the European Research Council through grant ERC-AdG No. 341157-COCO2CASA and the  Deutsche  Forschungsgemeinschaft  through  Sonderforschungbereich SFB 1258 ``Neutrinos and Dark Matter in Astro- and Particle Physics'' (NDM) and the Excellence Cluster Universe (EXC 153; http://www.universe-cluster.de/). OJ acknowledges support by the Special Postdoctoral Researchers (SPDR) program and iTHEMS cluster of RIKEN.




\bibliographystyle{mnras}
\bibliography{ref.bib}



\appendix
\onecolumn

\section{Derivation of energy equation in comoving frame}\label{app:der_ene_eqn}

\begin{table}
	\centering
	\caption{Meaning of various quantities used in Appendix~\ref{app:der_ene_eqn} and where to find their computation. The quantities $G$,$I^{a}$,$P^{ab}$,$Q^{abc}$ used here are denoted by $\mathcal{G}$,$\mathcal{I}^{a}$,$\mathcal{P}^{ab}$,$\mathcal{Q}^{abc}$ in \citet{2012arXiv1212.4064E} and by $\mathcal{Z}$,$\mathcal{Y}^{a}$,$\mathcal{X}^{ab}$,$\mathcal{W}^{abc}$ in \citet{2013PhRvD..87j3004C}, respectively. Also, note that all angular moments (projections of $\mathcal{U}^{abc}$) in \citet{2012arXiv1212.4064E} and \citet{2013PhRvD..87j3004C} are defined with a factor $\epsilon^{-2}$ ($\epsilon^{-1}$) compared to ours.}
	\begin{tabular}{lcc} 
		\hline
		\hline
		$\mathcal{J}$      & 0th angular moment of distribution function in comoving frame & eq.~\eqref{eq:tr_moment}               \\
                $\mathcal{H}^i$    & 1st angular moment of distribution function in comoving frame & eq.~\eqref{eq:tr_moment}                \\
                $\mathcal{K}^{ij}$ & 2nd angular moment of distribution function in comoving frame & eq.~\eqref{eq:tr_moment}                \\
		$\mathcal{E}$      & 0th angular moment of distribution function in lab frame      & eq.~\eqref{eq:tr_moment_lab}           \\
                $\mathcal{F}^i$              & 1st angular moment of distribution function in lab frame              & eq.~\eqref{eq:tr_moment_lab}            \\
                $\mathcal{S}^{ij}$           & 2nd angular moment of distribution function in lab frame              & eq.~\eqref{eq:tr_moment_lab}             \\
		$G$                & 0th projection of $\mathcal{U}^{abc}$ in lab frame            & eq.~(125) of \citet{2013PhRvD..87j3004C} \\
		$I^i$              & 1st projection of $\mathcal{U}^{abc}$ in lab frame            & eq.~(126) of \citet{2013PhRvD..87j3004C}  \\
		$P^{ij}$           & 2nd projection of $\mathcal{U}^{abc}$ in lab frame            & eq.~(127) of \citet{2013PhRvD..87j3004C}  \\
		$Q^{ijk}$          & 3rd projection of $\mathcal{U}^{abc}$ in lab frame            & eq.~(128) of \citet{2013PhRvD..87j3004C}  \\
		\hline
	\end{tabular}
	\label{tab:rad_tr_moments}
\end{table}

The transport equation used in our code evolves the energy density, $\mathcal{J}$, measured in an orthonormal comoving frame. This Appendix shows how this evolution equation can be obtained from corresponding equations evolving the lab-frame moments, $E$ and $F^i$. The lab-frame equations are derived and discussed in \citet{2011PThPh.125.1255S, 2012arXiv1212.4064E, 2013PhRvD..87j3004C}. In Table~\ref{tab:rad_tr_moments} we summarize the meaning of various quantities used here and where to find more information about them. The lab-frame equations contain the quantities, $G, I^a, P^{ab}, Q^{abc}$, which are related to the third-moment tensor,
\begin{eqnarray}
	\mathcal{U}^{a b c} \equiv \epsilon^3 \int l^a l^b l^c f(x^{\mu},p^{\hat \mu}) \mathrm{d}\Omega  ~,
	\label{eq:tr_third_moment_1}
\end{eqnarray}
by
\begin{eqnarray}
	\mathcal{U}^{a b c} = G n^a n^b n^c + I^a n^b n^c + I^b n^a n^c + I^c n^a n^b + P^{ab} n^c + P^{bc} n^a + P^{ac} n^b + Q^{abc}. 
	\label{eq:tr_third_moment_2}
\end{eqnarray}


We start from the lab-frame neutrino-energy equation as given in conservative form by equations (171),~(91-93),~(146),~(147), and~(173) of \citet{2013PhRvD..87j3004C}:
\begin{eqnarray}\label{eq:1.1} 
 &&\frac{1}{\sqrt{-g}}\frac{\partial}{\partial t}(\sqrt{\gamma} \mathcal{E}) 
 + \frac{1}{\sqrt{-g}}\frac{\partial}{\partial x^j}\{\sqrt{\gamma}(\alpha \mathcal{F}^j-{\beta}^j \mathcal{E})\}
 + \mathcal{F}^j \frac{\partial \ln \alpha}{\partial x^j}
 - \mathcal{S}^{jk}K_{jk}
 - \frac{1}{\epsilon^2}\frac{\partial}{\partial \epsilon}(\epsilon^2 F^\epsilon)
 = -n_\mu \frac{1}{\epsilon} \int p^\mu C \mathrm{d}\Omega~.
\end{eqnarray}
The definition of different symbols can be found in Table \ref{tab:rad_tr_moments}. Here,
\begin{eqnarray}\label{eq:1.2}
 F^\epsilon & \equiv & W\{I_j \frac{\partial v^j}{\partial \tau} + {P_j}^k \frac{\partial v^j}{\partial x^k}
 + \frac{1}{2} P^{jk}v^l\frac{\partial \gamma_{jk}}{\partial x^l}
 + (I^j - G v^j)\frac{\partial \ln \alpha}{\partial x^j}
 - P^{jk}K_{jk} + v^j I_k \frac{1}{\alpha} \frac{\partial \beta^k}{\partial x^j}\} \nonumber\\
 &&+ (I_j v^j - G)\frac{\partial W}{\partial \tau}
 + ({P_k}^j v^k - I^j)\frac{\partial W}{\partial x^j}~,
\end{eqnarray}
with $\frac{\partial}{\partial \tau}\equiv n^a\frac{\partial}{\partial x^a}$. Equations~(\ref{eq:1.1}) and~(\ref{eq:1.2}) are copied directly from \citet{2013PhRvD..87j3004C}, who employs slightly different definitions than us concerning the power of $\epsilon$ in the prefactor of the angular moments and third-moment projections. We now switch to our notation by doing the replacements $\epsilon^2\{\mathcal{E},\mathcal{F}^{i},\mathcal{S}^{ij},\mathcal{L}^{ijk}\} \rightarrow \{\mathcal{E},\mathcal{F}^{i},\mathcal{S}^{ij},\mathcal{L}^{ijk}\}$ and $\epsilon\{G,I^{i},P^{ij},Q^{ijk}\}\rightarrow \{G,I^{i},P^{ij},Q^{ijk}\}$. Moreover, we multiply eq.~(\ref{eq:1.1}) by $\epsilon^2\sqrt{\gamma}$ and introduce the notation $\hat X \equiv \sqrt{\gamma} X$ for any quantity $X$ to obtain:
\begin{eqnarray}\label{eq:1.4}
 \frac{1}{\alpha}\frac{\partial}{\partial t}(\hat{\mathcal{E}}) 
 + \frac{1}{\alpha}\frac{\partial}{\partial x^j}(\alpha \hat{\mathcal{F}}^j-{\beta}^j \hat{\mathcal{E}})
 + \hat{\mathcal{F}}^j \frac{\partial \ln \alpha}{\partial x^j}
 - \hat{\mathcal{S}}^{jk}K_{jk}
 - \frac{\partial}{\partial \epsilon}(\epsilon \hat{F}^\epsilon)
 = -n_\mu \sqrt{\gamma} \epsilon \int p^\mu C \mathrm{d}\Omega~,
\end{eqnarray}
where the formal definition of $F^\epsilon$ is still given by eq.~(\ref{eq:1.2}). Multiplying eq.~\eqref{eq:1.4} by $W$ and 
using the product rule (i.e. $W \frac{\partial f}{\partial x} = \frac{\partial (Wf)}{\partial x} - f \frac{\partial W}{\partial x}$) one finds
\begin{eqnarray}\label{eq:1.5}
 && \frac{1}{\alpha}\frac{\partial}{\partial t}(W \hat{\mathcal{E}})
 + \frac{1}{\alpha}\frac{\partial}{\partial x^j}\{W (\alpha \hat{\mathcal{F}}^j-{\beta}^j \hat{\mathcal{E}})\}
 + W \hat{\mathcal{F}}^j \frac{\partial \ln \alpha}{\partial x^j}
 - W \hat{\mathcal{S}}^{jk}K_{jk}
 - \frac{\partial}{\partial \epsilon}(W \epsilon \hat{F}^\epsilon) \nonumber\\
 &&- \frac{1}{\alpha} \hat{\mathcal{E}} \frac{\partial W}{\partial t}
 - \frac{1}{\alpha} (\alpha \hat{\mathcal{F}}^j-{\beta}^j \hat{\mathcal{E}}) \frac{\partial W}{\partial x^j}
 = -n_\mu W \sqrt{\gamma} \epsilon \int p^\mu C \mathrm{d}\Omega~.
\end{eqnarray}

We now extract the lab-frame energy-momentum equation from eqs.~(172),~(95-97),~(149),~(150), and~(174) of \citet{2013PhRvD..87j3004C}, again keeping their notation first:
\begin{eqnarray}\label{eq:2.1}
 &&\frac{1}{\sqrt{-g}}\frac{\partial}{\partial t}(\sqrt{\gamma} \mathcal{F}_i) 
 + \frac{1}{\sqrt{-g}}\frac{\partial}{\partial x^j}\{\sqrt{\gamma}(\alpha {\mathcal{S}^j}_i-{\beta}^j \mathcal{F}_i)\}
 + \mathcal{E} \frac{\partial \ln \alpha}{\partial x^i}
 - \mathcal{F}_j \frac{1}{\alpha} \frac{\partial \beta^j}{\partial x^i} \nonumber\\
 &&- \frac{1}{2} \mathcal{S}^{jk}\frac{\partial \gamma_{jk}}{\partial x^i}
 - \frac{1}{\epsilon^2}\frac{\partial}{\partial \epsilon}(\epsilon^2 S_i^\epsilon)
 = \gamma_{i \mu} \frac{1}{\epsilon} \int p^\mu C \mathrm{d}\Omega~,
\end{eqnarray}
where
\begin{eqnarray}\label{eq:2.2}
 S_i^\epsilon &\equiv& W\{P_{ij} \frac{\partial v^j}{\partial \tau} + {Q_{ij}}^k \frac{\partial v^j}{\partial x^k}
 + \frac{1}{2} {Q_i}^{jk}v^l\frac{\partial \gamma_{jk}}{\partial x^l}
 + ({P_i}^j - I_i v^j)\frac{\partial \ln\alpha}{\partial x^j}
 - {Q_i}^{jk}K_{jk} + v^j P_{ik} \frac{1}{\alpha} \frac{\partial \beta^k}{\partial x^j}\} \nonumber \\
 && + (P_{ik} v^k - I_i)\frac{\partial W}{\partial \tau}
 + ({Q_{ik}}^j v^k - {P_i}^j)\frac{\partial W}{\partial x^j}~.
\end{eqnarray}
Switching to our notation by doing the same replacements as for the energy equation above, we obtain:
\begin{eqnarray}\label{eq:2.4}
 &&\frac{1}{\alpha}\frac{\partial}{\partial t}(\hat{\mathcal{F}}_i) 
 + \frac{1}{\alpha}\frac{\partial}{\partial x^j}(\alpha {\hat{\mathcal{S}}^j}_i-{\beta}^j \hat{\mathcal{F}}_i)
 + \hat{\mathcal{E}} \frac{\partial \ln \alpha}{\partial x^i}
 - \hat{\mathcal{F}}_j \frac{1}{\alpha} \frac{\partial \beta^j}{\partial x^i} 
 - \frac{1}{2} \hat{\mathcal{S}}^{jk}\frac{\partial \gamma_{jk}}{\partial x^i}
 - \frac{\partial}{\partial \epsilon}(\epsilon \hat{S}_i^\epsilon)
 = \gamma_{i \mu} \sqrt{\gamma} \epsilon \int p^\mu C \mathrm{d}\Omega~,
\end{eqnarray}
where the formal definition of $S_i^\epsilon$ is still given by eq.~(\ref{eq:2.2}).
We multiply eq.~\eqref{eq:2.4} by $W$ and contract with $v^i$, to end up with
\begin{eqnarray}\label{eq:2.6}
 &&\frac{1}{\alpha}\frac{\partial}{\partial t}(W v^i \hat{\mathcal{F}}_i) 
 + \frac{1}{\alpha}\frac{\partial}{\partial x^j}\{W v^i (\alpha {\hat{\mathcal{S}}^j}_i-{\beta}^j \hat{\mathcal{F}}_i)\}
 + W v^i \hat{\mathcal{E}} \frac{\partial \ln \alpha}{\partial x^i}
 - W v^i \hat{\mathcal{F}}_j \frac{1}{\alpha} \frac{\partial \beta^j}{\partial x^i} 
 - W \frac{1}{2} \hat{\mathcal{S}}^{jk} v^i \frac{\partial \gamma_{jk}}{\partial x^i}
 - W \frac{\partial}{\partial \epsilon}(v^i \epsilon \hat{S}_i^\epsilon) \nonumber \\
 &&- v^i \hat{\mathcal{F}}_i \frac{1}{\alpha}\frac{\partial W}{\partial t} 
 - v^i (\alpha {\hat{\mathcal{S}}^j}_i-{\beta}^j \hat{\mathcal{F}}_i) \frac{1}{\alpha}\frac{\partial W}{\partial x^j}
 - W \hat{\mathcal{F}}_i \frac{1}{\alpha}\frac{\partial v^i}{\partial t} 
 - W (\alpha {\hat{\mathcal{S}}^j}_i-{\beta}^j \hat{\mathcal{F}}_i) \frac{1}{\alpha}\frac{\partial v^i}{\partial x^j}
 = \gamma_{i \mu} v^i W \sqrt{\gamma} \epsilon \int p^\mu C \mathrm{d}\Omega~.
\end{eqnarray}
Subtracting eq.~\eqref{eq:2.6} from eq.~\eqref{eq:1.5} we get
\begin{eqnarray}\label{eq:3.1}
 &&\frac{1}{\alpha}\frac{\partial}{\partial t}(W (\hat{\mathcal{E}}-v^i \hat{\mathcal{F}}_i)) 
 + \frac{1}{\alpha}\frac{\partial}{\partial x^j}
 [W \{(\alpha \hat{\mathcal{F}}^j-{\beta}^j \hat{\mathcal{E}})-v^i (\alpha {\hat{\mathcal{S}}^j}_i-{\beta}^j \hat{F}_i)\}]
 + W (\hat{\mathcal{F}}^j - v^j \hat{\mathcal{E}}) \frac{\partial \ln \alpha}{\partial x^j} 
 + W v^i \hat{\mathcal{F}}_j \frac{1}{\alpha} \frac{\partial \beta^j}{\partial x^i} \nonumber \\
 &&- W \hat{\mathcal{S}}^{jk}(K_{jk} - \frac{1}{2} v^i \frac{\partial \gamma_{jk}}{\partial x^i})
 - \frac{1}{\alpha} (\hat{\mathcal{E}} - v^i \hat{\mathcal{F}}_i) \frac{\partial W}{\partial t}
 - \frac{1}{\alpha} \{(\alpha \hat{\mathcal{F}}^j-{\beta}^j \hat{\mathcal{E}}) 
 - v^i(\alpha {\hat{\mathcal{S}}^j}_i-{\beta}^j \hat{\mathcal{F}}_i)\} \frac{\partial W}{\partial x^j}
 + W \hat{\mathcal{F}}_i \frac{1}{\alpha}\frac{\partial v^i}{\partial t} \nonumber \\
 &&+ W (\alpha {\hat{\mathcal{S}}^j}_i-{\beta}^j \hat{\mathcal{F}}_i) \frac{1}{\alpha}\frac{\partial v^i}{\partial x^j}
 - \frac{\partial}{\partial \epsilon}\{W \epsilon (\hat{F}^\epsilon - v^i \hat{S}_i^\epsilon)\}
 = -W \sqrt{\gamma} \epsilon (n_\mu + \gamma_{i \mu} v^i)  \int p^\mu C \mathrm{d}\Omega~.
\end{eqnarray}
We further rewrite $n^\mu \frac{\partial}{\partial x^\mu}=\frac{\partial}{\partial \tau}$ with $n^\mu=(1/\alpha,-\beta^i/\alpha)$, $W(n_\mu + \gamma_{i \mu} v^i) = u_\mu$, $-u_\mu p^\mu = \epsilon$, and redefine $\epsilon^2C \rightarrow C$:
\begin{eqnarray}\label{eq:3.5}
 &&\frac{1}{\alpha} \frac{\partial}{\partial t}(W (\hat{\mathcal{E}}-v^i \hat{\mathcal{F}}_i)) 
 + \frac{1}{\alpha} \frac{\partial}{\partial x^j}
 [W \{\alpha(\hat{\mathcal{F}}^j - v^i {\hat{\mathcal{S}}^j}_i) - {\beta}^j(\hat{\mathcal{E}} - v^i\hat{\mathcal{F}}_i)\}]
 + W (\hat{\mathcal{F}}^j - v^j \hat{\mathcal{E}}) \frac{1}{\alpha} \frac{\partial \alpha}{\partial x^j} 
 + W v^i \hat{\mathcal{F}}_j \frac{1}{\alpha} \frac{\partial \beta^j}{\partial x^i} \nonumber \\
 &&- W \hat{\mathcal{S}}^{jk}(K_{jk} - \frac{1}{2} v^i \frac{\partial \gamma_{jk}}{\partial x^i})
 - (\hat{\mathcal{E}} - v^i \hat{\mathcal{F}}_i) \frac{\partial W}{\partial \tau}
 - (\hat{\mathcal{F}}^j - v^i {\hat{\mathcal{S}}^j}_i) \frac{\partial W}{\partial x^j}
 + W \hat{\mathcal{F}}_i \frac{\partial v^i}{\partial \tau} 
 + W {\hat{\mathcal{S}}^j}_i \frac{\partial v^i}{\partial x^j} \nonumber \\
 &&- \frac{\partial}{\partial \epsilon}\{W \epsilon (\hat{F}^\epsilon - v^i \hat{S}_i^\epsilon)\}
 =  \sqrt{\gamma} \int C \mathrm{d}\Omega~.
\end{eqnarray}
The term inside the energy derivative of eq.~\eqref{eq:3.5} is given by:
\begin{eqnarray}\label{eq:5.1}
\epsilon W (F^\epsilon - v^i S_i^\epsilon)
 &=& \epsilon W[W\{I_j \frac{\partial v^j}{\partial \tau} + {P_j}^k \frac{\partial v^j}{\partial x^k}
 + \frac{1}{2} P^{jk}v^l\frac{\partial \gamma_{jk}}{\partial x^l}
 + (I^j - G v^j)\frac{\partial \ln\alpha}{\partial x^j}
 - P^{jk}K_{jk} + v^j I_k \frac{1}{\alpha} \frac{\partial \beta^k}{\partial x^j}\} \nonumber \\
 &&+ (I_j v^j - G)\frac{\partial W}{\partial \tau}
 + ({P_k}^j v^k - I^j)\frac{\partial W}{\partial x^j}
 - v^i W\{P_{ij} \frac{\partial v^j}{\partial \tau} + {Q_{ij}}^k \frac{\partial v^j}{\partial x^k}
 + \frac{1}{2} {Q_i}^{jk}v^l\frac{\partial \gamma_{jk}}{\partial x^l} \nonumber \\
 &&+ ({P_i}^j - I_i v^j)\frac{\partial \ln\alpha}{\partial x^j}
 - {Q_i}^{jk}K_{jk} + v^j P_{ik} \frac{1}{\alpha} \frac{\partial \beta^k}{\partial x^j}\}
 - v^i(P_{ik} v^k - I_i)\frac{\partial W}{\partial \tau}
 - v^i({Q_{ik}}^j v^k - {P_i}^j)\frac{\partial W}{\partial x^j}] \nonumber \\
 & =& \epsilon \Bigg\{ W^2 [(I_j - v^i P_{ij}) \frac{\partial v^j}{\partial \tau}  
 + ({P_j}^k - v^i {Q_{ij}}^k) \frac{\partial v^j}{\partial x^k}
 + \{ (I^j - G v^j) - v^i ({P_i}^j - I_i v^j) \} \frac{\partial \ln\alpha}{\partial x^j}  \nonumber \\
 && + v^j (I_k - v^i P_{ik}) \frac{1}{\alpha} \frac{\partial \beta^k}{\partial x^j}
 + (P^{jk} - v^i {Q_i}^{jk}) (\frac{1}{2} v^l\frac{\partial \gamma_{jk}}{\partial x^l} - K_{jk}) ]
 + W \{ (I_j v^j - G) - v^i(P_{ik} v^k - I_i) \} \frac{\partial W}{\partial \tau} \nonumber \\
 &&+ W \{ ({P_k}^j v^k - I^j) - v^i({Q_{ik}}^j v^k - {P_i}^j) \} \frac{\partial W}{\partial x^j} \Bigg\}~.
\end{eqnarray}
In order to express $\{G,I^{i},P^{ij},Q^{ijk}\}$ in terms of the lab-frame moments, $\{\mathcal{E},\mathcal{F}^i,\mathcal{S}^{ij}\}$, we use eqs.~(74)-(80) of \citet{2012arXiv1212.4064E}, however with $\epsilon$ set to 1 in their equations to account for the different definitions:
\begin{eqnarray}\label{eq:5.3}
\epsilon W (F^\epsilon - v^i S_i^\epsilon) &=& \epsilon \Bigg\{W \Big[ \mathcal{F}_j \frac{\partial v^j}{\partial \tau}  
 + {\mathcal{S}_j}^k \frac{\partial v^j}{\partial x^k}
 + (\mathcal{F}^j - \mathcal{E} v^j) \frac{\partial \ln\alpha}{\partial x^j} 
 + v^j \mathcal{F}_k \frac{1}{\alpha} \frac{\partial \beta^k}{\partial x^j}
 + \mathcal{S}^{jk} (\frac{1}{2} v^l\frac{\partial \gamma_{jk}}{\partial x^l} - K_{jk}) \Big] \nonumber \\
 &&- (\mathcal{E} - v^i \mathcal{F}_i) \frac{\partial W}{\partial \tau} 
 - (\mathcal{F}^j - {\mathcal{S}_k}^j v^k) \frac{\partial W}{\partial x^j}\Bigg\}~.
\nonumber \\
 & \equiv & \epsilon R_\epsilon~,
\end{eqnarray}
where in the last line we defined $R_\epsilon$ used in the main text of this paper. While $R_\epsilon$ is expressed in eq.~(\ref{eq:5.3}) in terms of the lab-frame moments, it can be re-expressed in terms of the comoving-frame moments using
\begin{eqnarray}\label{eq:lab_com_trafo_red}
  \mathcal{E} - v^i \mathcal{F}_i &=& \mathcal{J} + \bar v_{\hat i} \mathcal{H}^{\hat i}, \nonumber \\
  \mathcal{F}^i - {\mathcal{S}_j}^i v^j &=& \frac{1}{W} e^{i}_{\hat i} \mathcal{H}^{\hat i} + v^i \mathcal{J} + W v^i \bar v_{\hat i} \mathcal{H}^{\hat i}, \nonumber \\
  \mathcal{F}^i - \mathcal{E} v^i &=& W \Big(e^i_{\hat i} - \frac{W}{W+1} v^i \hat v_{\hat i} \Big)\Big(\mathcal{H}^{\hat i} +  \hat v_{\hat j} \mathcal{K}^{\hat i \hat j} \Big)   \end{eqnarray}
as well as eq.~(\ref{eq:tr_moment_lab}). Using the same transformations also for the remaining terms of eq.~(\ref{eq:3.5}), we finally obtain the neutrino energy equation in terms of the comoving-frame neutrino moments as:
\begin{eqnarray}\label{eq:comframeeq}
	&&\frac{1}{\alpha} \frac{\partial}{\partial t} [W(\mathcal{\hat J} + \bar v_{\hat i}\mathcal{\hat H}^{\hat i})]
    + \frac{1}{\alpha} \frac{\partial}{\partial x^j} [\alpha W (v^j-\beta^j/\alpha) \mathcal{\hat J}] +
    \frac{1}{\alpha} \frac{\partial}{\partial x^j} [\alpha e^j_{\hat i} \mathcal{\hat H}^{\hat i}] \nonumber \\ 
    &&+ \frac{1}{\alpha} \frac{\partial}{\partial x^j} \Big[ \alpha W \Big(\frac{W}{W+1}v^j-\beta^j/\alpha \Big) 
    \bar v_{\hat i}\mathcal{\hat H}^{\hat i} \Big] +
    \hat R_\epsilon - \frac{\partial}{\partial \epsilon} (\epsilon \hat R_\epsilon)
    = \sqrt{\gamma} \int C \mathrm{d}\Omega ~.
\end{eqnarray}

\section{Derivation of FLD equations and test of GR corrections}\label{app:der_fld_flux_eqn}
\begin{figure*}
    \includegraphics[width=\textwidth]{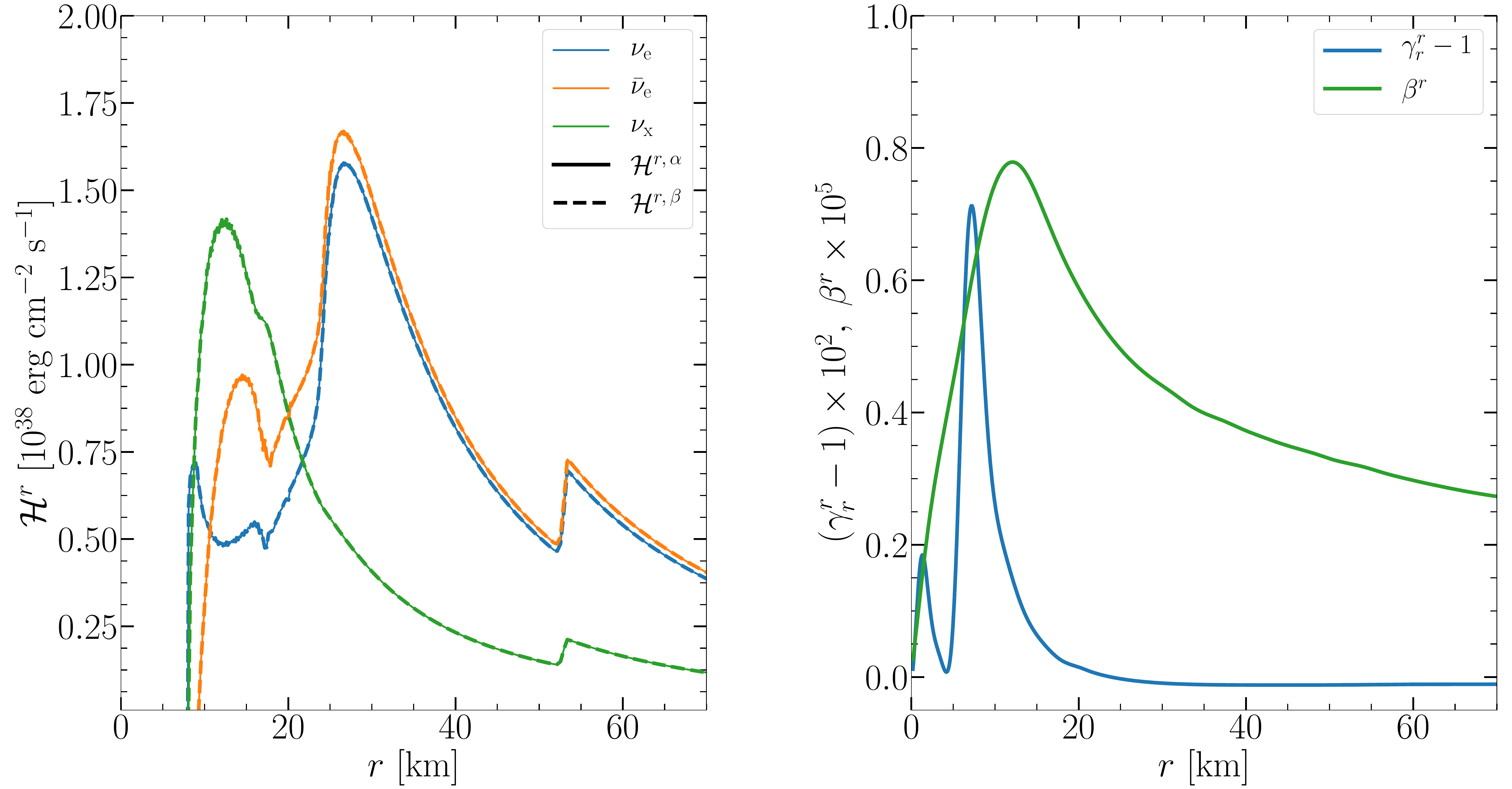}
    \vspace*{-6mm}
    \caption{GR corrected radiation fluxes $\mathcal{H}^{\mathrm{r},\,\alpha}$ and $\mathcal{H}^{\mathrm{r},\,\beta}$, given by equations~(\ref{eq:B.7}) and ~(\ref{eq:B.8}), respectively, are shown in the left plot. These radiation fluxes are post-processed from the NADA GR (LP) model, which is discussed in section~\ref{sec:CCSN}, at a post-bounce time of 500\,ms. The solid and dashed lines lie on top of each other, indicating negligibly small effects associated with the $\beta$-derivatives. In the right plot we show the value $\gamma^r_r-1$ and $\beta^r$ multiplied by $10^2$ and $10^5$, respectively, from the mentioned model.}
    \label{fig:test_ccsn_flux_correction}
\end{figure*}
In this section, we outline the derivation of the FLD radiation flux and test several GR corrections to the FLD flux. We start from eq.~(\ref{eq:2.4}). As is usually done in FLD implementations (see, e.g., \citealt{1981ApJ...248..321L}), we ignore all velocity dependent terms, the time derivatives of the comoving flux $\mathcal{H}_i$ and the energy derivative terms $\partial_\epsilon(\hat{S}_i^\epsilon)$ in equation~(\ref{eq:2.4}). Applying the aforementioned assumption in equation~(\ref{eq:2.4}), we get
\begin{eqnarray}\label{eq:B.1}
 && 
 \frac{1}{\alpha}\frac{\partial}{\partial x^j}(\alpha {\hat{\mathcal{S}}^j}_i)
 + \hat{\mathcal{E}} \frac{\partial \ln \alpha}{\partial x^i}
 - \frac{\beta^j}{\alpha} \frac{\partial \hat{\mathcal{F}}_i}{\partial x^j}
 - \hat{\mathcal{F}}_i \frac{1}{\alpha} \frac{\partial \beta^j}{\partial x^j} 
 - \hat{\mathcal{F}}_j \frac{1}{\alpha} \frac{\partial \beta^j}{\partial x^i}
 - \frac{1}{2} \hat{\mathcal{S}}^{jk}\frac{\partial \gamma_{jk}}{\partial x^i}
 - \hat{\mathcal{S}}^j_i \frac{\partial \ln \alpha}{\partial x^j}
 = - \sqrt{\gamma} \kappa_\mathrm{t}\mathcal{H}_i.
\end{eqnarray}
In equation~(\ref{eq:B.1}), the only term survived from $\partial_\epsilon(\epsilon \hat{S}_i^\epsilon)$ (see, equation~(\ref{eq:2.2}) for definition of $\hat{S}_i^\epsilon$) after taking the limit $v \rightarrow 0$ is $- \hat{\mathcal{S}}^j_i \partial_j \ln \alpha$ which comes from $\hat{P}^j_i \partial_j (\ln \alpha)$ term of $\hat{S}_i^\epsilon$. In the limit $v \rightarrow 0$, $\mathcal{E}$ and $S^j_i$ reduce to
\begin{eqnarray}\label{eq:B.2}
    \mathcal{E} &=& \mathcal{J}, \nonumber \\
    \mathcal{F}_i &=& {e_i}^{\hat{i}} \mathcal{H}_{\hat{i}}, \nonumber \\
	S^j_i &=& e^i_{\hat i} e_{j\hat j} \mathcal{K}^{\hat i \hat j}.
\end{eqnarray}
And in the diffusion limit $\mathcal{K}^{\hat i \hat j}$ reduces to
\begin{eqnarray}\label{eq:B.3}
	\mathcal{K}^{\hat i \hat j} &=& \frac{1}{3}\delta^{\hat i \hat j} \mathcal{J}.
\end{eqnarray}
Using equation~(\ref{eq:B.2}) and (\ref{eq:B.3}) in equation~(\ref{eq:B.1}) we obtain:
\begin{eqnarray}
 && 
 \frac{1}{3 \alpha}\frac{\partial}{\partial x^j}(\alpha \sqrt{\gamma} \gamma^j_i \mathcal{J})
 + \sqrt{\gamma} \mathcal{J} \frac{\partial \ln \alpha}{\partial x^i}
 - \frac{\beta^j}{\alpha} \frac{\partial (e_i^{\hat i} \hat{\mathcal{H}}_{\hat i})}{\partial x^j}
 - \sqrt{\gamma} {e_i}^{\hat i} \mathcal{H}_{\hat i} \frac{1}{\alpha} \frac{\partial \beta^j}{\partial x^j}
 - \sqrt{\gamma} {e_j}^{\hat i} \mathcal{H}_{\hat i} \frac{1}{\alpha} \frac{\partial \beta^j}{\partial x^i}
 - \sqrt{\gamma} \frac{\mathcal{J}}{6} \gamma^{jk}\frac{\partial \gamma_{jk}}{\partial x^i} \nonumber \\
 && - \sqrt{\gamma} \frac{\mathcal{J}}{3} \gamma^j_i \frac{\partial \ln \alpha}{\partial x^j}
 = - \sqrt{\gamma} \kappa_\mathrm{t}\mathcal{H}_i\,, \nonumber
\end{eqnarray}
and further:
\begin{eqnarray}\label{eq:B.4}
 &&
 \frac{1}{3 \alpha} \frac{\partial}{\partial x^j}(\alpha \gamma^j_i \mathcal{J})
 + \frac{1}{3} \gamma^j_i \mathcal{J} \frac{1}{\sqrt{\gamma}} \frac{\partial}{\partial x^j}(\sqrt{\gamma})
 + \mathcal{J} \frac{\partial \ln \alpha}{\partial x^i}
 - \frac{\mathcal{J}}{3} \gamma^j_i \frac{\partial \ln \alpha}{\partial x^j}
 - \frac{\beta^j}{\alpha} \frac{1}{\sqrt{\gamma}} \frac{\partial (e_i^{\hat i} \hat{\mathcal{H}}_{\hat i})}{\partial x^j}
 - {e_j}^{\hat i} \mathcal{H}_{\hat i} \frac{1}{\alpha} \frac{\partial \beta^j}{\partial x^j}
 - {e_j}^{\hat i} \mathcal{H}_{\hat i} \frac{1}{\alpha} \frac{\partial \beta^j}{\partial x^i} \nonumber \\
 && - \frac{\mathcal{J}}{6} \gamma^{jk}\frac{\partial \gamma_{jk}}{\partial x^i}
 = -\kappa_\mathrm{t}\mathcal{H}_i\,.
\end{eqnarray}
Now using the identity $\partial_i(\sqrt{\gamma}) = 1/2 \sqrt{\gamma} \gamma^{jk} \partial_i \gamma_{jk}$ and  assuming $\gamma^i_j \rightarrow \delta^i_j$, $\beta^j \rightarrow 0$, and $\partial_j \gamma^j_i = 0$ (while keeping $\partial_i \beta^j \neq 0$) in equation~(\ref{eq:B.4}), we get
\begin{eqnarray}\label{eq:B.5}
 && 
 \frac{1}{\alpha} \frac{\partial}{\partial x^j}(\alpha \mathcal{J})
 + 2 \mathcal{J} \frac{\partial \ln \alpha}{\partial x^i}
 = -3\kappa_\mathrm{t}\mathcal{H}_i
  + 3{e_i}^{\hat i} \mathcal{H}_{\hat i} \frac{1}{\alpha} \frac{\partial \beta^j}{\partial x^j}
  + 3{e_j}^{\hat i} \mathcal{H}_{\hat i} \frac{1}{\alpha} \frac{\partial \beta^j}{\partial x^i}\,.
\end{eqnarray}
Based on eq.~(\ref{eq:B.5}) we can now compute different versions of the diffusive flux, each accounting for different GR related corrections. First, assuming $\partial_i \beta^j = 0$ in the equation~(\ref{eq:B.5}) and after a little bit of algebra we obtain the following GR corrected diffusion flux ($\alpha$ correction):
\begin{eqnarray}\label{eq:B.6}
 && 
 \mathcal{H}^{\hat i\,,\,\alpha}_\mathrm{diff} = - \frac{e^{i\hat i}}{3\kappa_\mathrm{t} \alpha^3}
 \frac{\partial}{\partial x^i}(\alpha^3 \mathcal{J})\,.
\end{eqnarray}
This $\alpha$ correction has been derived by \citet{1989ApJ...346..350S} and is related to gravitational redshifting, time dilation, and ray-bending effects. Now, introducing the flux-limiter $\lambda$, evaluated using $\mathcal{H}^{\hat i\,,\,\alpha}_\mathrm{diff}$ (cf. Appendix~\ref{app:evl_R_lambda_D}) and the flux-limited diffusion coefficient $D=\lambda/\kappa_\mathrm{t}$, we obtain the FLD flux
\begin{eqnarray}\label{eq:B.7}
 \mathcal{H}^{\hat i\,,\,\alpha}
 &=& - \frac{\lambda e^{i \hat i}}{\kappa_\mathrm{t} \alpha^3} \frac{\partial}{\partial x^i}(\alpha^3 \mathcal{J})\,, \nonumber \\
 &=& - D \frac{e^{i \hat i}}{\alpha^3} \frac{\partial}{\partial x^i}(\alpha^3 \mathcal{J})\,.
\end{eqnarray}
Next, relaxing the assumption $\partial_i \beta^j = 0$ in equation~(\ref{eq:B.5}), one finds the following correction due to the $\beta$-dependent terms:
\begin{eqnarray}\label{eq:B.8}
 \mathcal{H}^{\hat i\,,\,\beta}_\mathrm{diff} &=& - \frac{1}{3 \kappa_\mathrm{t}  \alpha^3} \frac{1}{\{e_{i \hat i} 
 + (3 \kappa_\mathrm{t})^{-1}(e_{i \hat i} \partial_j \beta^j
 + e_{j \hat i} \partial_i \beta^j)\}}
 \frac{\partial}{\partial x^i}(\alpha^3 \mathcal{J})\,, \nonumber \\
 \mathcal{H}^{\hat i\,,\,\beta} &=& - \frac{D}{\alpha^3} \frac{1}{\{e_{i \hat i}
 + (3 \kappa_\mathrm{t})^{-1}(e_{i \hat i} \partial_j \beta^j
 + e_{j \hat i} \partial_i \beta^j)\}}
 \frac{\partial}{\partial x^i}(\alpha^3 \mathcal{J})\,.
\end{eqnarray}
Finally, to derive the Newtonian diffusion flux and FLD flux from equations~(\ref{eq:B.6}) and (\ref{eq:B.7}), one needs to further assume $\alpha \rightarrow 1$ and a tetrad given in the spherical polar coordinates by $e^{k \hat i} = \mathrm{diag}(1,1/r,1/(r\sin\theta))$ corresponding to the flat spatial metric:
\begin{eqnarray}\label{eq:B.9}
 && 
 \mathcal{H}^{\hat i,\,\mathrm{NEWT}}_\mathrm{diff} = - \frac{e^{i\hat i}}{3\kappa_\mathrm{t}}
 \frac{\partial \mathcal{J}}{\partial x^i}\,.
\end{eqnarray}
\begin{eqnarray}\label{eq:B.10}
 \mathcal{H}^{\hat i,\,\mathrm{NEWT}}
 &=& - \frac{\lambda e^{i \hat i}}{3 \kappa_\mathrm{t}} \frac{\partial}{\partial x^i}(\mathcal{J})~, \nonumber\\
 &=& - D e^{i \hat i} \frac{\partial}{\partial x^i}(\mathcal{J})~.
\end{eqnarray}
Using the Newtonian FLD flux in equation \eqref{eq:comframeeq} we obtain the FLD transport equation:
\begin{eqnarray}\label{eq:B.11}
	&&\frac{1}{\alpha} \frac{\partial}{\partial t} [W(\mathcal{\hat J} + \bar v_{\hat i}\mathcal{\hat H}^{\hat i})]
    + \frac{1}{\alpha} \frac{\partial}{\partial x^j} [\alpha W (v^j-\beta^j/\alpha) \mathcal{\hat J}] -
    \frac{1}{\alpha} \frac{\partial}{\partial x^j} [\alpha \gamma^{i j} D \partial_i \mathcal{J}] \nonumber \\ 
    &&- \frac{1}{\alpha} \frac{\partial}{\partial x^j} \Big[ \alpha W \Big(\frac{W}{W+1}v^j-\beta^j/\alpha \Big) 
    v^i D \partial_i \mathcal{J} \Big] +
    R_\epsilon - \frac{\partial}{\partial \epsilon} (\epsilon R_\epsilon)
    = \sqrt{\gamma} \int C \mathrm{d}\Omega~.
\end{eqnarray}
Alternatively using the GR corrected flux given by the equation~(\ref{eq:B.7}), one obtains:
\begin{eqnarray}\label{eq:B.12}
	&&\frac{1}{\alpha} \frac{\partial}{\partial t} [W(\mathcal{\hat J} + \bar v_{\hat i}\mathcal{\hat H}^{\hat i})]
    + \frac{1}{\alpha} \frac{\partial}{\partial x^j} [\alpha W (v^j-\beta^j/\alpha) \mathcal{\hat J}] -
    \frac{1}{\alpha} \frac{\partial}{\partial x^j} [\alpha^{-2} \gamma^{i j} D \partial_i (\alpha^3 \mathcal{J})] \nonumber \\ 
    &&- \frac{1}{\alpha} \frac{\partial}{\partial x^j} \Big[ \alpha^{-2} W \Big(\frac{W}{W+1}v^j-\beta^j/\alpha \Big) 
    v^i D \partial_i (\alpha^3 \mathcal{J}) \Big] +
    R_\epsilon - \frac{\partial}{\partial \epsilon} (\epsilon R_\epsilon)
    = \sqrt{\gamma} \int C \mathrm{d}\Omega~.
\end{eqnarray}
In Fig.~\ref{fig:test_ccsn_flux_correction}, the left plot shows the radial component of the GR corrected radiation fluxes $\mathcal{H}^{\mathrm{r},\,\alpha}$ and $\mathcal{H}^{\mathrm{r},\,\beta}$ given by equations~(\ref{eq:B.7}) and (\ref{eq:B.8}), respectively. These radiation fluxes are post-processed from the results of the NADA GR (LP) model, which is discussed in section~\ref{sec:CCSN}, at a post-bounce time of 500\,ms, when the neutron star has a very compact structure and general relativity has the strongest impact on the neutrino emission during the simulation. Note that model NADA GR (LP) uses equations~(\ref{eq:B.7}) and (\ref{eq:B.12}) for neutrino transport. It is evident that the inclusion of spatial derivatives of the shift vector $\beta^i$ in the radiation fluxes (see equation~\ref{eq:B.8}) does not make any visible changes in the values of the radiation fluxes. Moreover, in the right plot of Fig.~\ref{fig:test_ccsn_flux_correction}, we show the values of $\gamma^r_r-1$ and $\beta^r$ from the mentioned NADA GR (LP) model multiplied by $10^2$ and $10^5$, respectively, and we see that the deviations from $\gamma^r_r \rightarrow 1$ and $\beta^r \rightarrow 0$ are less than one percent and less than $10^{-5}$, respectively. Therefore, we conclude that the assumptions made to derive the neutrino transport equations~(\ref{eq:B.7}) and (\ref{eq:B.12}) are justifiable in the context of CCSNe, and the mentioned neutrino transport equations already include the dominant general relativistic effects. However, the impact of the various GR corrections on the FLD flux in a more dynamic configuration, e.g. in a rapidly rotating neutron star, may be different and needs to be examined in future works.

\section{Evaluation of Knudsen number, flux limiter and diffusion coefficient}\label{app:evl_R_lambda_D}
In this section, we describe the procedure to calculate the Knudsen number, flux limiter, and diffusion coefficient. Our evaluation of the flux limiter ensures that causality is not violated by both the individual radiation flux components ($|\mathcal{H}^{\hat i}| \le \mathcal{J}$) and the total radiation flux ($|\mathcal{H}|=\sqrt{(\mathcal{H}^r)^2+(\mathcal{H}^\theta)^2+(\mathcal{H}^\phi)^2} \le \mathcal{J}$). In contrast, the method described in  \citet{2009ApJS..181....1S} ensures causality only for the individual fluxes but not for the total radiation flux. In the hemispheric difference test described in Section~\ref{sec:hemisph-diff-test}, we obtained a maximum flux factor, $f=|\mathcal{H}|/\mathcal{J}$, of 1.2 with the \citet{2009ApJS..181....1S} method, whereas with our alternative method the maximum value of the flux factor dropped to 1.0. In this section, we describe the numerical scheme to calculate the Knudsen number, flux limiter, and diffusion coefficient flat metric tetrad for spherical for coordinate, given by $e^{k \hat i} = \mathrm{diag}(1,1/r,1/(r\sin\theta))$. We use $i,j,k$ to indicate the cell center in the radial, $r$, polar, $\theta$, and azimuthal, $\phi$ direction, respectively. First, we calculate the absolute value of the gradient of the radiation energy density at cell center $(i,j,k)$, $|\nabla \mathcal{J}|_{i,j,k}$, by:
\begin{eqnarray}\label{eq:C.1}
    |\nabla \mathcal{J}|_{i,j,k} = |e^{i \hat i} \partial_i \mathcal{J}|_{i,j,k}
    = \sqrt{\Bigg(\frac{\mathcal{J}_{i+1,j,k}-\mathcal{J}_{i-1,j,k}}{r_{i+1}-r_{i-1}}\Bigg)^2
    +\Bigg(\frac{\mathcal{J}_{i,j+1,k}-\mathcal{J}_{i,j-1,k}}{r_i(\theta_{j+1}-\theta_{j-1})}\Bigg)^2
    +\Bigg(\frac{\mathcal{J}_{i,j,k+1}-\mathcal{J}_{i,j,k-1}}{r_i\sin{\theta_j}(\phi_{k+1}-\phi_{k-1})}\Bigg)^2}.
\end{eqnarray}
Using this gradient, we calculate the Knudsen number, $R_{i,j,k}$, as follows:
\begin{eqnarray}\label{eq:C.2}
    R_{i,j,k} = \frac{|\nabla \mathcal{J}|_{i,j,k}}{(\kappa_\mathrm{t})_{i,j,k} \mathcal{J}_{i,j,k}}
    = \frac{3 |\mathcal{H}_\mathrm{diff}|_{i,j,k}}{\mathcal{J}_{i,j,k}},
\end{eqnarray}
where $({\kappa_\mathrm{t}})_{i,j,k}$ is the transport opacity at the cell center $(i,j,k)$. Afterwards, we calculate the flux limiter, $\lambda_{i,j,k}(R_{i,j,k})$ using one of the closures given by equation~(\ref{eq:limiter}). Finally, the diffusion coefficient is given by:
\begin{eqnarray}\label{eq:C.3}
    D_{i,j,k} = \frac{\lambda_{i,j,k}}{(\kappa_\mathrm{t})_{i,j,k}}.
\end{eqnarray}
The values of the diffusion coefficients at the cell interfaces are obtained by linear interpolation of cell-centered values. Furthermore, the cell centered values of the Knudsen number and the flux limiter are used to calculate the Eddington factor and the Eddington tensor with equations~(\ref{eq:edd_factor_fld}) and (\ref{eq:edd_tensor_fld}), respectively. 

We can assess the capability of this method to maintain causality by evaluating the flux factor in the free-streaming limit. In the free-streaming limit, the Knudsen number goes to $R \rightarrow \infty$ and the flux limiter asymptotes to $\lambda \rightarrow 1/R$. The flux factor in the free-streaming limit is then given by:
\begin{eqnarray}\label{eq:C.4}
    f_{i,j,k} = \frac{|H|}{\mathcal{J}_{i,j,k}}
    = \frac{\lambda_{i,j,k}|\nabla \mathcal{J}|_{i,j,k}}{\kappa_{\mathrm{t}~i,j,k}\mathcal{J}_{i,j,k}}
    \rightarrow \frac{|\nabla \mathcal{J}|_{i,j,k}}{R_{i,j,k}\kappa_{\mathrm{t}~i,j,k}\mathcal{J}_{i,j,k}} = 1.
\end{eqnarray}
In the last step, we used equation~(\ref{eq:C.2}). As we can see from the above derivation, the new method of evaluating the diffusion coefficient ensures causality in the free-streaming limit.


\bsp	
\label{lastpage}
\end{document}